\documentclass[onecolumn,authoryear]{els-mrw} 

\newcommand{\En}{Encyclopedia}
\newcommand{\Enp}{Encyclopedia of Particle Physics}
\newcommand{\Enn}{Encyclopedia of Nuclear Physics}
\usepackage[utf8]{inputenc}

\usepackage{array}[=2016-10-06]

\usepackage[utf8]{inputenc}

\usepackage{amsmath}
\usepackage{bbm}
\usepackage{mathtools}
\usepackage{amsfonts}
\usepackage{mathrsfs}
\usepackage{bm}
\usepackage{slashed}
\usepackage{tensor}
\usepackage{bbold}
\usepackage{MnSymbol}
\usepackage{textcomp}
\usepackage[export]{adjustbox}

\usepackage{float}
\usepackage{graphicx}
\usepackage{color}
\usepackage[abs]{overpic}

\usepackage{placeins}

\usepackage{makecell}
\usepackage{subcaption}

\usepackage{xspace}
\usepackage{siunitx}
\usepackage{xfrac}
\usepackage{hyperref}
\usepackage[nameinlink]{cleveref}
\usepackage{appendix}

\Crefname{appsec}{Appendix}{Appendices}

\usepackage{tikz}
\usetikzlibrary{shapes, arrows}
\tikzstyle{roundbox} = [rectangle, draw, text centered, rounded corners, 
minimum height=2em, minimum width=2em, draw=black, fill=blue!10]
\tikzstyle{process} = [rectangle, draw, minimum height=1em, 
minimum width=3em, text centered, draw=black, fill=green!10]
\tikzstyle{integration} = [ellipse, draw, text centered, minimum height=1em, 
minimum width=3em, draw=black, fill=red!10]

\newcommand*{\upB}{\textup{B}}

\newcommand*{\muB}{\mu_{\upB}}

\def\Slash#1{\setbox0=\hbox{$#1$} 
	\dimen0=\wd0 
	\setbox1=\hbox{/} \dimen1=\wd1 
	\ifdim\dimen0>\dimen1 
	\rlap{\hbox to \dimen0{\hfil/\hfil}} 
	#1 
	\else 
	\rlap{\hbox to \dimen1{\hfil$#1$\hfil}} 
	/ 
	\fi}

\newcommand{\pslash}{\Slash{p}}

\usepackage{xifthen}
\usepackage{xcolor}
\hypersetup{
	colorlinks,
	linkcolor={red!75!black},
	citecolor={blue!75!black},
	urlcolor={blue!75!black}
}

\usepackage{booktabs}
\usepackage{multirow}

\newcolumntype{C}{>{$}c<{$}}
\AtBeginDocument{
	\heavyrulewidth=.08em
	\lightrulewidth=.05em
	\cmidrulewidth=.03em
	\belowrulesep=.65ex
	\belowbottomsep=0pt
	\aboverulesep=.4ex
	\abovetopsep=0pt
	\cmidrulesep=\doublerulesep
	\cmidrulekern=.5em
	\defaultaddspace=.5em
}

\makeatletter
\def\l@subsubsection#1#2{}
\makeatother

\graphicspath{{./figures/}{./figures/equations/}{./figures/results/}}

\graphicspath{{./figures/}{./figures/equations/}{./figures/results/}}
 
\begin{document}

\chapter{Phase structure of strong interaction matter from Functional QCD}\label{chap1}

\author[1,2]{Christian~S.~Fischer}%
\author[3,4]{Jan~M.~Pawlowski}%

\address[1]{\orgname{Institut f{\"u}r Theoretische Physik}, \orgdiv{Justus-Liebig-Universit{\"a}t Gießen}, \orgaddress{Heinrich-Buff-Ring 16, 35392 Gießen, Germany}}
\address[2]{\orgname{Helmholtz Forschungsakademie Hessen für FAIR (HFHF)}, \orgdiv{GSI Helmholtzzentrum für Schwerionenforschung, Campus Gießen}, \orgaddress{Heinrich-Buff-Ring 16, 35392 Gießen, Germany}}
\address[3]{\orgname{Institut f{\"u}r Theoretische Physik}, \orgdiv{Universit{\"a}t Heidelberg}, \orgaddress{Philosophenweg 16,
		69120 Heidelberg, Germany}}
\address[4]{\orgname{ExtreMe Matter Institute EMMI}, \orgdiv{GSI Helmholtzzentrum für Schwerionenforschung}, \orgaddress{Planckstr. 1,
		64291 Darmstadt, Germany}}


\maketitle

\begin{glossary}[Glossary]
\term{Functional QCD} is an approach to QCD using functional methods, i.e.~methods working with integral/differential equations for QCD correlation functions that can be derived from 
the QCD path integral.

\end{glossary}

\begin{glossary}[Nomenclature]
\begin{tabular}{@{}lp{34pc}@{}}
CEP &Critical End Point\\
DSE &Dyson-Schwinger Equations\\
EoM &Equation of Motion\\
fRG &functional Renormalisation Group Equations\\
ONP &Onset of New Phases\\
QCD & Quantum Chromo Dynamics\\
\end{tabular}
\end{glossary}

\begin{abstract}[Abstract]
In this contribution to the \Enn, we aim to provide a pedagogical introduction to the functional approach to QCD at 
finite temperature and chemical potential. We briefly outline the general framework and address 
its complementarity to other first-principle approaches to non-perturbative QCD. We discuss selected results  
obtained with Dyson-Schwinger equations (DSE) and the functional renormalisation group (fRG) in the context of
a general physics perspective on the QCD phase diagram. This article is specifically aimed at students and 
non-practitioners of functional methods alike and may serve as a short guide to further literature.   
\end{abstract}

\section{Introduction}\label{sec:intro}

Exploring the properties of strongly interacting matter in different regions of the QCD phase diagram, 
unravelling the potential existence of a critical end point (CEP), or more generally mapping out the high density region 
of QCD with the potential onset of new phases (ONP) are major goals of current and future experimental 
programs at the Relativistic Heavy Ion Collider (RHIC) in Brookhaven, USA, at the Large Hadron Collider 
at CERN in Geneva, Switzerland, the FAIR facility in Darmstadt, Germany and HIAF in Huizhou, China. Experiments like STAR, ALICE, 
HADES and CBM, and CEE+ seek to probe signals for continuous and non-continuous phase transitions from the hadronic 
state of matter at low temperatures and densities to exotic and new phases of QCD at high temperatures 
and/or large densities. 

In a broader context, these regions are of relevance for the understanding of the universe. Microseconds 
after the Big Bang, our universe cooled along a trajectory at rather small baryon and isospin chemical 
potentials and underwent a (probably continuous) transition from the QCD high temperature phase to today's 
low temperature phase. On the other hand, in the high density region of large baryon (and non-zero isospin)
chemical potential we find conditions which we believe are present in the interior of neutron stars
and during neutron star collisions. Having in mind, that the latter may be responsible for the creation
of heavy elements in various types of rapid neutron capture processes ('r-process') it is mandatory to 
understand the nature of strongly interacting matter under these conditions.  
  
Providing solid theoretical predictions and interpretations of experimental and astrophysical data 
is of utmost relevance. For most relevant aspects of this task, the underlying dynamics builds on the 
non-perturbative low-energy regime of QCD, the theory of strong interaction. Therefore, genuine non-perturbative 
approaches are mandatory, and in this Chapter we focus mainly on the functional approach to QCD. In the past decade, 
functional QCD at finite temperature and density has developed from a qualitative level, well-suited for 
exploratory studies, to a first principles quantitative QCD approach.

At zero and small chemical potential, functional QCD complements another non-perturbative approach, lattice QCD. 
In order to understand the complementary nature of these two approaches we need to dive a little bit into 
their technical foundations. This is done in \Cref{sec:technics}. We would like to note that
lattice QCD is our prime source of information at zero and imaginary baryon chemical potential $\mu_B$, 
whereas at real baryon chemical (the region probed by experiment) lattice data provide indirect evidence 
within extrapolations. This allows for reliable (approximate) 
results in the chemical potential regime $\mu_B/T \lesssim  3$. Beyond this region, the extrapolation errors accumulate rapidly 
and solid predictions are increasingly difficult. Explicit computations with functional methods, on the other hand, 
require approximations from the start (again, see \Cref{sec:technics} for details).å However, they allow 
for direct computations everywhere in the QCD phase diagram. Key to their qualitative or even quantitative 
reliability is the systematic error control within a given approximation scheme. The latter has developed 
substantially in the past decade. Therefore, the functional approach provides excellent opportunities to explore 
the exciting and physics-rich large chemical potential region of QCD.  

In this Chapter of the \Enn{} we aim at a pedagogical overview of the functional approach to the phase structure of QCD
and the properties of QCD matter at large temperatures and/or chemical potentials. We start in \Cref{sec:general} 
with brief introductions to the phase diagram and the two main functional methods, Dyson-Schwinger equations and the functional renormalisation group. Many more details can be found in recent reviews, see e.g.~\cite{Fischer:2018sdj, Dupuis:2020fhh, Fu:2022gou, Rennecke:2025bcw, Huber:2025cbd, Fischer:2026uni}. We also address 
the extraction of order parameters for the chiral and confinement-deconfinement transition of QCD and provide
a short overview on the calculation of fluctuation observables with functional methods. In section \Cref{sec:results} we discuss
selected results on the phase structure of QCD under variation of essential QCD parameters. We explore the interplay
of explicit vs dynamical chiral symmetry breaking and explain the rich physics of the Columbia plot that serves as a key quality 
check for the power of the functional approach to QCD. In sections \Cref{sec:cep} and \Cref{sec:fluc-Exp-Theo} we discuss
recent results on the location of a chiral critical end point in the QCD phase diagram and the potential onset of new
phases at large densities and connect these results with freeze-out data extracted from heavy ion collisions. We conclude
in \Cref{sec:conclusion}.

\section{Generalities}
\label{sec:general}

\subsection{Regions of the phase diagram}
\label{sec:regions}

\begin{figure}[t]
	\centering
	\includegraphics[scale=0.42]{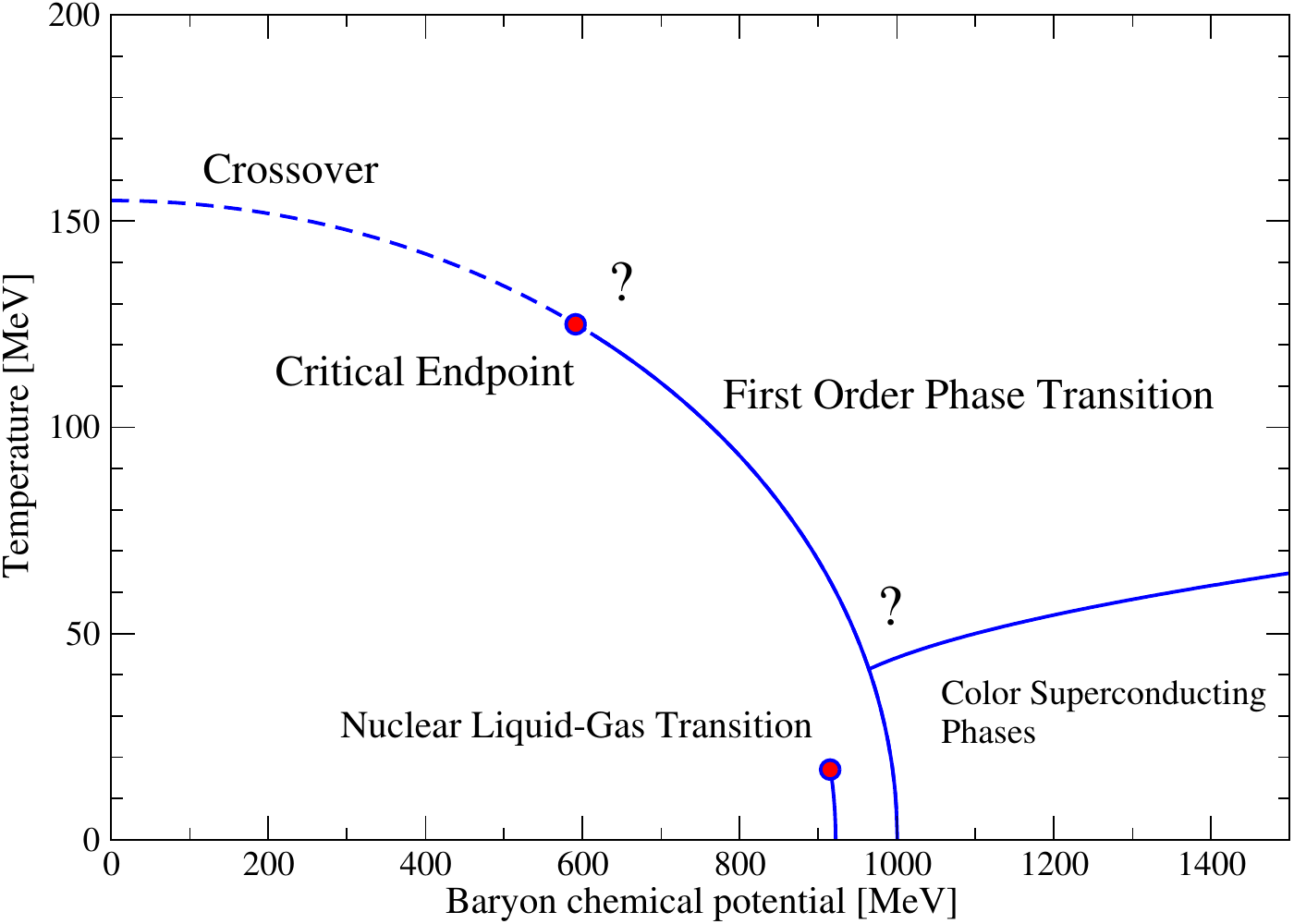}    
	\caption{Sketch of the QCD phase diagram in the temperature and baryon chemical potential plane.
		Figure taken from~\cite{Fischer:2018sdj}. \label{fig:phase}}
\end{figure}

Before we discuss the merits of the functional approach to QCD, let us start with a couple of general considerations that 
will become useful in the following. Consider the sketch of the QCD phase diagram shown in \Cref{fig:phase}. 
There is widespread agreement that results from lattice QCD, see e.g.~\cite{Aoki:2006we, Borsanyi:2010bp, Bazavov:2011nk, Bhattacharya:2014ara, Bazavov:2018mes, Borsanyi:2020fev},  
demonstrate an analytic crossover at zero chemical potential from a low-temperature phase characterised 
by confinement and chiral symmetry breaking to one or more (see the Chapter on 'Chiral Spin Symmetry' in this \En) high-temperature phases. 
The corresponding pseudo-critical temperature for the chiral transition has been localised at $T_c \approx 157$ MeV with 
an error margin below two MeV, \cite{Bazavov:2018mes,Borsanyi:2020fev}. Further observables include chiral condensates, thermodynamic observables, fluctuations of conserved charges and the slope or curvature of the chiral crossover line within an expansion in $\mu_B^2$. These results have been corroborated quantitatively within the functional approach to QCD, see in particular \cite{Fu:2019hdw, Gao:2020fbl, Gunkel:2021oya, Pawlowski:2025jpg, Fu:2026qnl}, and the QCD phase structure at vanishing baryon chemical potential still serves as one of the crucial benchmark tests for functional QCD. 

At non-zero baryon chemical potential $\mu_B$, lattice QCD is obstructed by the sign problem (explained in detail in other Chapters of this \En). However, various methods like Taylor 
expansion, re-weighting schemes or extrapolation from imaginary chemical potential have been developed and refined over the years, and provide a reliable access to the region $\mu_B/T \lesssim 3$. Specifically, the chiral crossover line at sufficiently small baryon chemical potential can be determined within an expansion in $\mu^2_B/T^2$. It is found that the linear order is dominating by far, 
\begin{align}
\frac{T_c(\mu_B)}{T_c} = 1-\kappa_2\left(\frac{\mu_B}{T_c}\right)^2 - \kappa_4 \left(\frac{\mu_B}{T_c}\right)^4 - \cdots\,,
\label{eq:kappa}
\end{align}
with pseudo-critical temperature $T_c(\mu_B)$ and $T_c=T_c(0)$.\footnote{A word of 
	caution is
	in order here: this expansion has been used in the literature in different forms, sometimes it is formulated not
	in baryon but in quark chemical potential and sometimes factors of $\pi^2$ are included, resulting in trivial 
	changes of the values of the expansion coefficients $\kappa_2$ and $\kappa_4$. Sometimes on the right hand side 
	$T_c(\mu_B)$ is used instead of $T_c(0)$. The latter change
	is immaterial for small chemical potential but has some impact on extrapolations at larger $\mu_B$. In this review
	we will stick to the formulation \labelcref{eq:kappa}.} The expansion is quadratic, since the grand canonical 
QCD partition function $Z$ is symmetric with respect to a change of sign in $\mu_B/T$, 
$Z\left(\frac{\mu_B}{T}\right) = Z\left(-\frac{\mu_B}{T}\right)$, see e.g.~\cite{deForcrand:2002hgr}, 
and therefore all odd powers of $\mu_B/T$ in the expansion have to vanish. 

If we assume for the moment that no new physics emerges at large chemical potential, then \labelcref{eq:kappa} can be generalised to
a {\bf baseline expectation} for the chiral transition line across the whole phase diagram, see \cite{Fischer:2018sdj}. 
To this end we consider the ellipse 
\begin{align}
	\left(\frac{T_c(\mu_B)}{T_c}\right)^2 = 1 - 2\kappa\left(\frac{\mu_B}{T_c}\right)^2\,.
	\label{eq:ellipse}
\end{align}
which automatically satisfies the boundary conditions $Z\left({\mu_B}\right) = Z\left(-{\mu_B}\right)$ and 
$Z\left(T \right) = Z\left(-T \right)$. This expression can be expanded for small baryon chemical potential and then matched
to \labelcref{eq:kappa} leading to the identification $\kappa_2 = \kappa$, $\kappa_4 = \frac{1}{2}\kappa^2$ (and further ones 
for $\kappa_{2n}$ with $n=3,\dots,\infty$). 
The transition temperature at zero chemical potential, $T_c\equiv T_c(\mu_B=0) \approx 157$ MeV is well 
known from lattice QCD, as discussed above. 
At zero temperature but finite baryon chemical potential $\mu_B$ 
the nuclear liquid-gas transition happens at $\mu_B^{(\textrm{lg})} \approx 923$ see \Cref{fig:phase}. This 
value arises from 
difference between the nucleon mass in vacuum, $m_N\approx 939$\,MeV, and the nucleon binding energy in nuclear matter, 
$\epsilon_b\approx 16$\,MeV, for a functional perspective see e.g.~\cite{Fukushima:2023wnl}. It is safe to assume that the chiral transition at $T=0$ cannot happen for chemical 
potentials smaller than $\mu_B^{(\textrm{lg})}$,  since all nuclei are composed of nuclear matter in the chirally broken phase. 
This directly results in an estimate for the upper bound of the curvature  
\begin{align}
	\kappa \lesssim 0.0145\,. 
	\label{eq:kappaguess}
\end{align}
Using the above identifications $\kappa_2 = \kappa$ and $\kappa_4 = \frac{1}{2}\kappa^2$ this value is compatible with contemporary 
extractions of $\kappa_{2,4}$ from lattice QCD and functional QCD, see \cite{Fischer:2026uni} for a recent overview. 

However, we do expect genuine density fluctuations to kick in at sufficiently large chemical potentials, changing the very nature of the QCD at high densities. This will {\bf modify the baseline} \labelcref{eq:ellipse}. It furthermore offers the exciting possibility of the onset of new phases in hot and dense QCD. A fairly comprehensive list of different scenarios is: {\bf (i)} the possible appearance of 
a critical end point, i.e.~the termination of the crossover line by a second order phase transition point followed by a first order 
transition line; {\bf (ii)} the possible appearance of a so-called 'moat' regime and/or a phase with spatially varying (inhomogeneous) order 
parameter; {\bf (iii)} the possible appearance of a colour-superconducting regime with homogenous or 
inhomogeneous diquark condensates and various pairing scenarios at relatively low temperature. 

All these scenarios {\bf (i-iii)} require the control of emergent density-dependent QCD dynamics at larger densities and in particular in the regime $\mu_B/T\gtrsim 4.5$, see \cite{Pawlowski:2025jpg}. Consequently, any prediction or even full resolution of the high density necessitates \textit{direct} QCD computations in this regime. While this is evident in the scenarios {\bf (ii,iii)}, seemingly {\bf (i)} escapes this necessity by not requiring new density-dependent dynamics in the first place. However, its presence or absence can only be validated by a direct high density QCD computations. In conclusion, \textit{any} extrapolation of QCD results at low densities carries with it implicit assumptions about the nature of the high density regime of QCD: Direct extrapolations of QCD results at small chemical potentials fall into the category {\bf (i)} or variations thereof, and hence cannot be viewed as a prediction of the CEP. Extrapolations of low density QCD in terms of low energy effective theories whose parameters are fixed with vacuum QCD or QCD at finite temperature,  carry the high density effects of the specific low energy effective theory at hand. Again this cannot be viewed as predictive at large densities. Still, direct and low energy effective theory extrapolations offer a collection of baseline computations and are chiefly important in view of the mainly uncharted high density QCD regime. 

To date, functional QCD is the only framework that can be used for direct non-perturbative QCD computations at high densities. This is particularly relevant in the regime $\mu_B/T \gtrsim 4.5$ which shoes signals of emergent new dynamics (see below). However, it is fair to say, that we are currently only at the very beginning of this exciting endeavour. In the following we give an overview on important results and properties of this approach.

\subsection{Technical foundations}
\label{sec:technics}

In this subsection we detail technical aspects of functional methods. Readers with primary interest in an overview of results may
safely omit this section and the following \Cref{sec:observ} on first reading. An introductory overview on functional
methods in general is given in the Chapter \cite{Huber:2025cbd} and an introduction to results for
hadron physics is given in the Chapter \cite{Eichmann:2025wgs} within the \Enp. For a more comprehensive list of reviews see \cite{Fischer:2026uni}. 

Experiments and non-perturbative theoretical approaches to QCD unravel the complicated and highly exciting infrared quantum, thermal 
and density dynamics of QCD with the simple classical action 
\begin{align} 
	S_\textrm{QCD}[A_\mu, q,\bar q] = \frac14 \int_x F_{\mu\nu}^a F_{\mu\nu}^a -\int_x \bar q \left( \slashed{D} + m_q+\gamma_0 \mu_q\right) q \,.
	\label{eq:SQCD}
\end{align}  
The pure glue part, $F_{\mu\nu}^a F_{\mu\nu}^a$ contains a quadratic non-interacting term such as in QED and three- and four-point 
self-interactions of the 
gluons $A_\mu^a$ with $a=1,...,8$ for the strong gauge group $SU(3)$.  The quark part contains the standard kinetic term for all quarks $q=(u,d,s,...)$ with (current) 
quark masses $m_{q}=\textrm{diag}(m_u,m_d,m_s,...)$, a quark-gluon interaction hidden in the Dirac operator $\slashed{D}$ with the covariant derivative $D_\mu=\partial_\mu - i\, g_s A_\mu$ with the strong coupling $g_s$. The quark densities $\bar q_i\gamma_0 q_i$ are coupled to the quark chemical 
potentials $\mu_{q_i}$ with $\mu_q =\textrm{diag}(\mu_u,mu_d,\mu_s,...)$. 

The functional approach to QCD is based on the free energy or grand potential $\Gamma[A_\mu, q, \bar q]$ of QCD with a general glue and quark background, $A_\mu, q, \bar q$. Diagrammatically, $\Gamma$ is the (1PI) one-particle irreducible effective action, that can be obtained
from the general functional including \Cref{eq:SQCD} and additional source terms via the usual Legendre transform.  
A general background is arranged for by external sources with the equations of motion 
\begin{align} 
 \frac{\delta \Gamma[\Phi]}{\delta \Phi} = J(\Phi)\,,\qquad \textrm{with} \qquad \Phi=(A_\mu, c,\bar c, q,\bar q)\,,\qquad J=(J_\mu, -\bar \eta,\eta )\,. 
	\label{eq:EoMQCD}
\end{align}
The external sources or backgrounds such as an external (chromo-) magnetic or (chromo-) electric field are encoded in the source term $\int_x J\cdot \Phi$, and the dot stands for the contraction of all Lorentz and internal indices. The field  $\Phi$ collects all the fundamental fields in QCD, including the auxiliary ghost fields that come with the gauge fixing. The latter is commonplace in functional approaches as explicit computations are done within a diagrammatic framework that uses the propagators of the dynamical fields including the gluons. The common choice is the Landau gauge, $\partial_\mu A_\mu=0$ for both conceptual and practical reasons.  Moreover, most 
applications of functional QCD work with the \textit{Euclidean} version of the QCD generating functional that describes 
strongly interacting matter in thermodynamic equilibrium as a grand-canonical ensemble. The temperature $T$ is hidden 
in the four dimensional integral $\int_x = \int_0^T d\tau \int_{\vec{x}}$ in \labelcref{eq:SQCD}. This is the Euclidean Matsubara 
or imaginary time formalism, see e.g.~\cite{Landsman:1986uw} for a general review. 

Finally, the grand potential or 1PI effective action reduces to the standard grand potential in the absence of external sources or background by solving \labelcref{eq:EoMQCD} for $J=0$, 
leading to $\Phi_\textrm{\tiny{EoM}}$. For a given spatial volume, temperatures and chemical potentials we obtain the grand potential  
\begin{align} 
	\Omega(V,T,\mu_B,\mu_Q,\mu_S) =  \Gamma\left[\Phi_\textrm{\tiny{EoM}}\right]\,, 
\label{eq:GrandPotential} 
	\end{align} 
with the \textit{baryon} chemical potential $\mu_B$, the \textit{electric charge} chemical potential $\mu_Q$ and the \textit{strangeness} chemical potential $\mu_S$. The volume, temperature and chemical potential derivatives of the grand potential $\Omega$ provide access to the thermodynamics of QCD across its phase structure. These derivatives and their combinations provide us with the pressure, entropy, baryon charge, strangeness densities, and  fluctuations observables (higher order derivatives), see \Cref{sec:observ} and \Cref{sec:fluc}. 

In the Functional Approach to QCD these observables are computed from the right hand side of \labelcref{eq:GrandPotential}, evaluated on the equations of motion. The computation of the Effective Action $\Gamma[\Phi]$ is typically done in terms of an expansion in powers of the fields. The expansion coefficients are the full correlation functions of the respective fields and carry considerably more information than the thermodynamic observables extracted from the grand potential $\Omega$. The common denominator of all functional methods is the resolution of the effective action in terms of these correlations functions, i.e.~(inverse) propagators, the three- and four-point
vertices and higher Green's functions. 

We would like to note that both functional methods follow from specific equations of motions \labelcref{eq:EoMQCD} or rather specific choices of $J$.\footnote{Roughly speaking, the functional DSE is the quantum field theory version of Ehrenfest's theorem: The quantum equation of motion (EoM) is the expectation value of the classical EoMs: \labelcref{eq:EoMQCD} with $J=\langle \delta S[\hat\Phi]/\Delta\hat \Phi\rangle[\Phi]$ with the quantum field $\hat \Phi$ with $\Phi=\langle \hat\Phi\rangle$ and the classical gauge-fixed action $S$.} In the case of the flow equation one may start with a variant of \labelcref{eq:EoMQCD} that is obtained by taking a derivative w.r.t.~the propagator, linking it to yet another functional methods based on two-particle (or $n$-particle) (2PI and nPI) irreducible effective actions.\footnote{The 2PI effective action $\Gamma[\Phi, G_\Phi]$ with the propagator $G_\Phi$ is obtained by also coupling a source $J_G$ to two fields $\hat\Phi(x)\,\hat\Phi(y)$. Then, the flow equation can be written as an equation of motion in the presence of an external current: $\delta\Gamma/\delta G= R$ where $J_G=R$ is the infrared regulator in the flow equation.}  The latter, and specifically the three-particle effective action and ensuing resummation schemes are also used within the DSE approach. The above already emphasises the multitude of direct connections between functional methods and the practical possibility of combined application which is indeed often used.  

\begin{figure*}[t]
	\centering
	\begin{minipage}[t]{.55\linewidth}
		\begin{subfigure}{\linewidth}
			\centering
            \includegraphics[width=\linewidth]{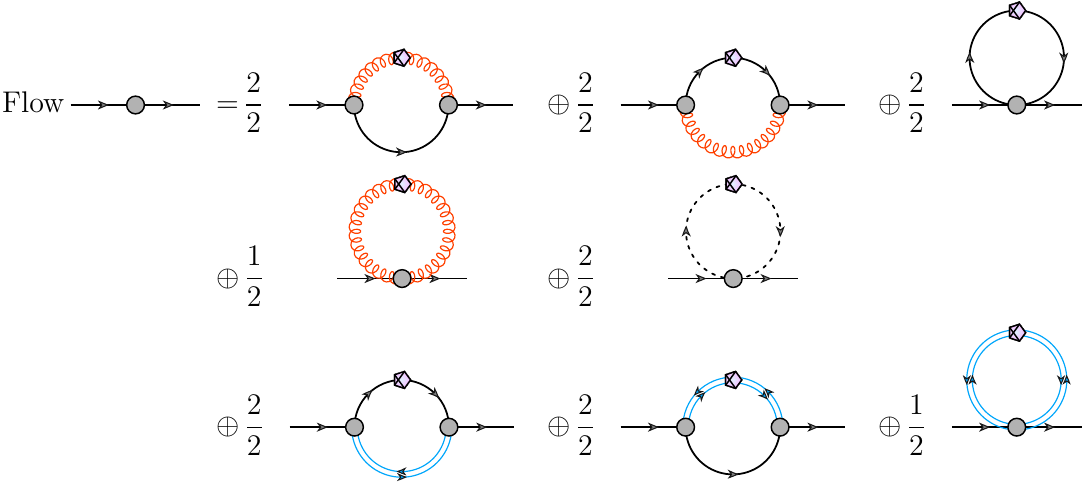} 
            \subcaption{Diagrammatic part of the flow equation for the quark propagator. \hspace*{\fill}}
            \label{fig:QuarkGapfRG} 
		\end{subfigure}%
	\end{minipage}
	\hspace{0.04\linewidth}%
	\begin{minipage}[t]{.38\linewidth}
		\begin{subfigure}{\linewidth}
			\centering 
			\includegraphics[width=\linewidth]{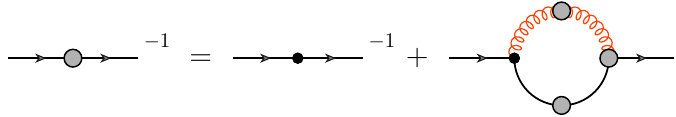} \vspace{1cm}
			\subcaption{Quark gap equation.\hspace*{\fill}}
			\label{fig:QuarkGapDSE} 
		\end{subfigure}
	\end{minipage}
	\vspace{.1cm}
	\caption{Full propagators and vertices are indicated by grey blobs, the classical vertices are indicated by small black blobs. In the fRG all ingredients are dressed, it only depends on full propagators and vertices, and hence we have dropped the grey blobs in the propagators for the sake of notational simplicity.  
	Gluons are represented by red spiral lines, ghosts by back dotted ones, and quark by straight black ones. 
	The $\bigoplus$ accommodate symmetry factors and relative minus signs. The crossed symbols in the fRG indicate
	cutoff insertions, see \cite{Ihssen:2024miv} for a detailed account of the flow.	
	In \Cref{fig:QuarkGapfRG} we depict the flow equation for the quark propagator, in \Cref{fig:QuarkGapDSE} the quark DSE. \hspace*{\fill}}
	\label{fig:FunDSE+QuarkDSE}
\end{figure*}

If expanded in terms of the amputated 1PI parts $\Gamma^{(n)}$ of $n$-point functions, the functional DSE and flow equations for QCD lead to infinite towers of coupled one- (fRG) or two-loop (DSE) exact relations for these $\Gamma^{(n)}$, schematically written as 
\begin{align} 
	\Gamma^{(n)} = \textrm{FunRel}_n\left[ \left\{ \Gamma^{ (3\leq m\leq n+2) }  \right\}\,,\, G_\Phi \right]\,.
	\label{eq:GenFun}
\end{align}
The tower of DSEs also involves the classical vertices $S^{(m)}$ and starts with $n=1$ while that of the flow equations starts with $n=0$ and only involves the full vertices $\Gamma^{(m)}$. 
\begin{align} 
	G_\Phi(x,y)=\langle \Phi(x) \Phi(y)\rangle =\frac{1}{\Gamma^{(2)}_\Phi}(x,y)\,,
		\label{eq:Props}
\end{align}
for their key importance in diagrammatic relations. We illustrate the generic structure of the relations with the quark flow equation \Cref{fig:QuarkGapfRG} and the quark gap equation \Cref{fig:QuarkGapDSE} in \Cref{fig:FunDSE+QuarkDSE}. All structural features of \labelcref{eq:GenFun} can be seen at work. To begin with, both relations are one-loop exact. While this is a generic feature of flow equations, DSEs in the glue sector also contain non-perturbative two-loop diagrams but, importantly, not more. 
We continue with a brief technical discussion of both equations, starting with the DSE for the quark propagator, \Cref{fig:QuarkGapDSE}. The first diagram of 
the right hand side stands for the free, renormalised propagator as extracted from the QCD action. The following non-perturbative one-loop 
diagram generates the quark self energy and encodes all interactions of the quark with itself via gluon emission and absorption. It contains 
the dressed (exact) gluon propagator and one bare and one dressed quark-gluon vertex. It also contains the fully dressed quark 
propagator, which appears also in its inverse form on the left hand side of the equation. Given an exact expression for the 
gluon and the quark-gluon vertex, this equation can be solved exactly at least within numerical accuracy) by standard iteration procedures.
Obviously, the task is then to provide for exact expression for the gluon and the quark-gluon vertex. These satisfy their own 
DSEs, which in principle have a similar structure, i.e.~non-perturbative loop diagrams on the right hand side involving other Green's functions. 
In general, every DSE for an $n$-point function contains at least $(n+1)$-functions on the respective right hand side. 

Now we proceed to the quark flow equation shown in \Cref{fig:QuarkGapfRG}. The diagrammatic part consists out of more diagrams than its DSE counterpart: three versus one quark-gluon diagram. The diagrams in the second line take into account resonant parts of the  quark-gluon and four-quark scattering vertex in the first line in terms of emergent composites. In most cases this treatment is restricted to the softest modes in QCD, the pions and the $\sigma$-mode, for more details see \cite{Fischer:2026uni} and references therein. In the DSE all these contributions are included in the dressed quark-gluon vertex. As in the DSE, given exact expression for the gluon and quark-gluon vertex (in the fRG split into its different parts), the quark flow equation can be solved by starting at a large cutoff scale with the perturbative quark propagator. In comparison to the DSE this amounts to solving a integral-differential equation instead of an integral equation. 
and is solved with standard PDE solvers, a compilation of respective codes can be found here:  \url{https://fqcd-collaboration.github.io/}. For a more direct comparison with the DSE one may even integrate the fRG over the flow time which leads to a five-dimensional DSE-type hierarchy of integral equations that can be solved similarly to the DSE hierarchy. 

In summary, the diagrammatic structure of DSEs and fRGs are manifestly different although the global structure is the same: 
They both form an infinite tower of coupled integral of integral-differential equations which can be solved exactly only in very special momentum limits in QCD. A specifically interesting one is the infrared limit, see \cite{Fischer:2006vf, Fischer:2009tn} for a combined application of DSE and fRG hierarchies in QCD, for related work using DSE and nPI (skeleton expansion) hierarchies see \cite{Alkofer:2008jy}. In practical applications to QCD one has to resort to 'truncations' of the full hierarchies of equations, i.e.~approximations of the full vertices as well as Ansätze. In this context we also would like to emphasise the intriguing property of functional QCD, that one can systematically include results for correlation functions obtained within other approaches: In our example \Cref{fig:FunDSE+QuarkDSE} one may take the required input of the gluon propagator and/or the quark-gluon vertex from other functional approaches or from lattice QCD. In the latter case this mostly concerns the propagators as to date, quantitative results for vertices are still rare as they require a very high statistics. Finally, the mere existence of several  functional  approaches (fRG, DSE, 2PI, 3PI,...) can (and has been) turned into an advantage: given a certain level of truncation, the structure of both towers of equations is sufficiently different that in combination, results from both approaches provide non-trivial cross-checks for functional QCD: if a sufficiently rich truncation leads to similar results, this suggests that the combined error from both approaches is small. Note that this does not suffice to provide comprehensive systematic error estimates but is is an important ingredient. This is discussed in 
detail in the review \cite{Fischer:2026uni} and we will come back to this point frequently in the results section of this article.

\subsection{Order parameters from functional methods}
\label{sec:observ}

The phase structure of QCD is mapped out with the help of order parameters for the chiral and the  confinement-deconfinement phase transition. In the chiral limit and vanishing baryon chemical potential, the chiral transition connects a high energy phase with massless quarks to a low energy phase with massive (constituent) quarks. In physical QCD quarks have finite current quark masses owing to the Higgs mechanism. The light up and down quarks have current quark masses of $2-5$\,MeV which is small in comparison to the dynamical infrared scale of QCD of a few 100 MeV. In the presence of these small current quark masses, the chiral phase transition turns into a -relatively sharp- crossover which is still well described by the thermal evolution of the chiral order parameters, for more details see \Cref{sec:chiral}. 

The labelling 'constituent' already refers to the second dynamical infrared phenomenon in QCD, namely confinement. At low energies quarks and gluons are bound in colourless hadrons and the coloured quarks experience a linear potential between them. In the heavy-quark limit this phenomenon is related to center symmetry breaking in QCD. In turn, for dynamical quarks, center symmetry is broken and since the light quark masses are so small, order parameters for center symmetry are not capturing any more the intricate details of the confinement-deconfinement transition sufficiently well. Consequently, the interpretation of the behaviour of these order parameters in physical QCD has been the subject of a long and ongoing debate. We refrain from entering this debate here but only introduce the respective order parameters in \Cref{sec:deconf}. Later we also discuss further observables that are related to the transition from the high energy phase with perturbative quarks to the hadronic one, most prominently the kurtosis of baryon number.

\subsubsection{Chiral transition}
\label{sec:chiral}

Let us first discuss the chiral transition. The prime candidate for a suitable 
order parameter is the quark condensate $\langle\bar{q} q  \rangle_f$ which is the expectation value of the mass operator in \labelcref{eq:SQCD}. For a quark with flavour $f$ it
is given by the trace of the fully dressed quark propagator $G_f(p)$ via
\begin{align} 
\langle\bar{q}q\rangle_f \simeq -\sumint_p \textrm{Tr}\, \bigl[ G_f (p) \bigr]\,, 
\label{eq:condensate}
\end{align}
where the trace sums over Dirac and colour indices, and the frequency sum is over fermion Matsubara frequencies $\omega_n=(2n+1)\pi T$ (and 
we omitted multiplicative renormalisation factors for brevity). The momentum four-vector is given by $p = (p_0,\vec{p})= (\omega_p ,\vec{p})$. 
The expression \labelcref{eq:condensate} requires further additive renormalisation leading to the \textit{reduced} condensate $\Delta_{l,h}$ or the \textit{renormalised} condensate $\Delta_{f,R}$ (both with similar physics content):
\begin{align}
\Delta_{l,h} = \langle\bar q q\rangle_l - \frac{m_l}{m_h}\langle\bar q q\rangle_h\,, \qquad \textrm{and} \qquad \Delta_{f,R}= \langle\bar{q}q\rangle_f(\mu_B,T) - \langle\bar{q}q\rangle_f(0,0)\,.   
\label{eq:cond_renorm}
\end{align}
In $\Delta_{l,h}$, the divergent part of the light-quark condensate ($l \in \{u,d\}$) 
is cancelled by the divergent part of the heavy ($h$) quark condensate. In $\Delta_{f,R}$, the divergent part of the condensates cancel each other due to the subtraction.  In most applications either the reduced strange condensate $\Delta_{l,s}$ or the light quark renormalised condensate $\Delta_{l,R}$ is considered. For more details see \cite{Fischer:2026uni} and references therein. 

The reduced condensate vanishes if the 'light' quark mass $m_l$ and the 'heavy' quark mass $m_h$ agree. Hence, in order to extract a clean signal for the chiral transition of the light quark flavour, the mass of the 
heavy quark needs to be sufficiently larger than the one of the light quark. For the extraction of the
light up/down-quark condensate a heavy quark mass of the order of the strange quark is already sufficient. In turn, the renormalised condensate vanishes at $T,\mu_B=0$ and only carries the thermal and density dependence of the condensate. For heavy quarks it is dominated by the explicit mass and dynamical chiral symmetry breaking is a subleading effect while for the light quarks $l=(u,d)$  dynamical chiral symmetry breaking is dominant. 

For physical quark masses and small chemical potential the chiral transition is a crossover which leads to 
ambiguities in the definition of a pseudo-critical temperature. Frequently used quantities to determine 
$T_c$ the maximum of the chiral magnetic susceptibility 
\begin{align}
\chi^{\ }_{\langle\bar{q}q\rangle} = \frac{\partial \langle\bar{q}q\rangle_l}{\partial m_{l}}\,, 
 \label{eq:chisusz}
 \end{align}
or the thermal susceptibility, i.e.~the maximum of $\partial \langle\bar{q}q\rangle_l/\partial T$ which 
determines the inflection point of the condensate. All these variants have been computed within functional QCD, see \cite{Fischer:2026uni} and references therein.

\subsubsection{Confinement-deconfinement transition} 
\label{sec:deconf}

Even after decades, a comprehensive description of confinement and its dynamics still eludes us\footnote{A concise and detailed discussion of many aspects of confinement 
	can be found e.g.~in~\cite{Greensite:2003bk, Alkofer:2006fu, Greensite:2011zz}.}. It is precisely for this lack of a full understanding that its description here or rather that within functional QCD is a bit more detailed than that of chiral symmetry breaking. 

We start this discussion with the part which is understood, namely confinement in Yang-Mills theory. To begin with, in Yang-Mills theory there is a well-defined notion of the confinement-deconfinement phase transition in terms of the breaking of center symmetry: In the absence of dynamical quarks, QCD is center-symmetric as the gluon, living in the adjoint representation of the $SU(3)$ gauge group, is insensitive to center symmetry. The center group $Z_3\in SU(3)$ is its subgroup that consists of the elements of $SU(3)$ that commute with all other group elements. A respective order parameter is the expectation value of the traced Polyakov loop in the fundamental representation, where the Polyakov loop is the Wilson loop winding around the full time direction at finite temperature, 
\begin{align} 
	{L}(\boldsymbol{x})= \frac{1}{3} \textrm{tr}_\textrm{f}\, P(\boldsymbol{x})\,,\qquad \textrm{with} \qquad P(\boldsymbol{x}) =  \mathcal{P}\, e^{ -i\, g_s \int\limits_0^\beta d \tau \,A_0(\tau,\boldsymbol{x}) }= e^{2\pi i \hat \varphi}\,.
	\label{eq:PolloopP+L}
\end{align} 
Under a center transformation with $z\in Z_3$ the trace Polyakov loop $L$ in the fundamental representation transforms with $L\to z\,L$. 
The expectation value of \labelcref{eq:PolloopP+L} can be understood as the free energy of a static quark, for more details and references we refer to \cite{Fischer:2026uni}. For large temperatures, Yang-Mills theory is in the center-broken phase. The theory only allows for temporal gauge field fluctuations about $A_0=0$ and the expectation value of the traced Polyakov loop approaches unity, $\langle L\rangle \to 1$, for asymptotically large temperatures. This entails that QCD with heavy quarks enters the perturbative phase at large temperatures or large energy, and the free energy of the quark 'state' is finite. More precisely, while  quarks cannot be defined as asymptotic states even at high temperatures or energies, they can still be observed indirectly for example in terms of gluon jets at the Large Hadron Collider (LHC). 

In turn, for $T\to 0$, large fluctuations of the gauge field are present and we have to sum the Polyakov loop over all $z\in Z_3$. Then, center symmetry is restored, leading to $\langle L\rangle =0$ below the phase transition. The latter is a first order one and the symmetry is that of a three-state Potts model. 
The vanishing of the expectation value of the Polyakov loop entails that the free energy of a quark 'state' is infinite. More precisely, quarks are bound in colourless states, the hadrons. Moreover, quarks and antiquarks at large distance $r$ exhibit a linear potential $V_{q\bar q}(r)$ between them, whose strength is called the string tension $\sigma$, 
\begin{align} 
	V_{q\bar q}(r\to \infty) \propto \sigma r\,.
\end{align}
In functional approaches to QCD, the theory is formulated in gluonic degrees of freedom and $\langle L[A_0]\rangle $ is an infinite order correlation function of the gauge field. While this is still accessible, see \cite{Herbst:2015ona}, it is not a natural observable within functional approaches. Indeed, the expectation value of the algebra element $\hat\varphi$ of the Polyakov loop in terms of its eigenvalues is an alternative order parameter that is far easier accessible in functional approaches: Moreover, it has the direct physics interpretation of the dynamical gluonic background in which the functional relations have to be solved. Constant fields $\hat\varphi$ can be rotated into the Cartan and we yield,   
\begin{align} 
\bar A_0 := g_s \beta \varphi \,,\qquad\qquad  \qquad  \varphi = \sum_{n=1}^{N_c^2-1} \nu_n \varphi_n \,,\qquad \textrm{with }\qquad  \nu_n=\langle \hat\nu_n\rangle\,,\quad \textrm{and} \quad \hat \varphi |\psi_n\rangle =  \hat\nu_n  |\psi_n\rangle
\label{eq:OrderParametervarphi}
\end{align} 
with $\beta=1/T$ and $n=1,...,N_c^2-1$ with the colour $N_c=3$ in QCD. The eigenvalues $\nu_n$ are gauge invariant as a gauge transformation of $\hat\varphi$ is a unitary rotation with the gauge group element. For more details see e.g.~\cite{Lu:2025cls} and references therein. Hence, the expectation value of the algebra field $\hat\varphi$ in \labelcref{eq:PolloopP+L}, or the respective \textit{gauge-invariant} background $\bar A_0$, are alternative order parameter, see \cite{Braun:2007bx, Marhauser:2008fz, Fister:2013bh, Fischer:2013eca}. The order parameter $\varphi$ or $\bar A_0$  can be readily computed with the fRG flow or the DSE of the effective action from its equation of motion 
\begin{align} 
	\frac{\partial V(\varphi)}{\partial \varphi_n} = \frac{1}{\beta {\cal V}_3} \frac{\partial \Gamma[\bar A_0(\varphi)]}{\partial \varphi_n}=  \frac{\partial \bar A^a_0(\varphi)}{\partial \varphi} \left[\frac{1}{\beta {\cal V}_3}\frac{\partial \Gamma[\bar A_0]}{\partial \bar A_0^a } \right]= 0 \,, \qquad \textrm{with}\qquad {\cal V}_3 = \int_\mathbb{R} d^3 x\,\quad \textrm{and} \quad \beta=\frac{1}{T}\,.
	\label{eq:GluonicEoMfininteT}
\end{align} 
This provides a direct functional access to the confinement-deconfinement phase transition, for applications in Yang-Mills theory and QCD see \cite{Braun:2007bx, Marhauser:2008fz, Braun:2009gm, Braun:2010cy, Fister:2013bh, Herbst:2015ona} (fRG) and \cite{Fukushima:2012qa, Kashiwa:2012td, Fischer:2013eca, Fischer:2014vxa, Fischer:2014ata, Lu:2025cls, Lu:2026ezr} (DSE) and \cite{Reinhardt:2012qe, Quandt:2022lhe} (Hamiltonian functional QCD). For related perturbative works in the Curci-Ferrari model and the refined Gribov-Zwanziger action see e.g.~\cite{Reinosa:2014ooa, Reinosa:2014zta, Dudal:2023nbt, MariSurkau:2026irs} and the lecture notes \cite{Reinosa:2024njc}. For its use in low energy effective models we refer to the review \cite{Fukushima:2017csk}. Early foundational work in perturbation theory has been initiated in the seminal work \cite{Gross:1980br} and \cite{Weiss:1980rj}.  

The functional order parameter $\varphi$ can be cast in a more familiar form as the traced Polyakov loop, evaluated for the algebra field $\varphi$, or rather for the gluonic background $\bar A_0(\varphi)$, 
\begin{align} 
	L[\bar A_0] = 
	\left\{ \begin{array}{rl} 0 & \textrm{for} \ T< T_\textrm{conf} \\[1ex] 
	                                        \neq 0 & \textrm{for} \  T > T_\textrm{conf} 
	                                        \end{array} \right. \,, 
	\label{eq:FunPolLoop}  
\end{align} 
with the critical temperature $T_\textrm{conf}$ of the first order confinement-deconfinement transition in $SU(3)$ Yang-Mills theory. The order parameter 	\labelcref{eq:FunPolLoop} has the same properties as the standard expectation value of the traced Polyakov loop, but approaches the high temperature limit (unity) far quicker (for $T \approx  1.3 \,T_\textrm{conf}$). The respective critical temperature in Yang-Mills theory agrees quantitative, as it must, with that computed from the expectation value of the Polyakov loop, see \cite{Braun:2007bx, Marhauser:2008fz, Fister:2013bh}. It links the gluon mass gap in Landau gauge QCD directly to the confinement-deconfinement critical temperature and hence to the mass gap in QCD, see e.g.~\cite{Ferreira:2025tzo}. Moreover, it can be shown that the gluon mass gap directly enters in the computation of the physical spectrum of Yang-Mills theory and QCD, and in particular in that of the low lying glueball states, see e.g.~\cite{Huber:2020ngt,Huber:2021yfy,Pawlowski:2022zhh,Huber:2025kwy}. We close this brief review of confinement in QCD with heavy quarks with the remark, that the above example of the confinement-deconfinement phase transition in Yang-Mills theory or QCD with heavy quarks is an early and chiefly interesting example for an inverse phase transition. 

We now proceed with the discussion of confinement in full dynamical QCD. To begin with, we emphasise once more that $\bar A_0(\varphi)$ is the solution of the gluonic equation of motion \labelcref{eq:GluonicEoMfininteT}, and provides us with the dynamical glue background at finite temperature. General $n$-correlation functions of QCD, $\Gamma^{(n)}$, as well as general observables have to be evaluated on this background, see \labelcref{eq:EoMQCD} and \labelcref{eq:GrandPotential}. This important property goes far beyond its mere use as an order parameter of confinement and we shall come back to it in  \Cref{sec:fluc} about fluctuations of conserved charges, one of the experimentally important observables. 

We still have to discuss its fate as an order parameter of the confinement-deconfinement phase transition. In QCD with dynamical quarks center symmetry is explicitly broken and the lighter the quarks, the stronger is the explicit breaking. The dynamical light up and down quarks with their current quark masses of a few MeV drive us far away from the heavy quark limit and a classification of the confinement properties of QCD with center symmetry and its breaking may not be that revealing. This scenario is present for $T \gg T_c$ where $T_c$ is the chiral crossover temperature. As we shall see, chiral symmetry, as measured by the order parameter of chiral symmetry breaking, is indeed restored rather rapidly for $T> T_c$ (in physical QCD). In turn, for $T\ll T_c$, 
the constituent quark mass is rather large in comparison to all dynamical scales in the glue sector and the quark dynamics decouples quickly, effectively reducing dynamical QCD to the heavy quark limit of QCD. Note that this heuristic argument is not invalidated by the presence of colourless soft modes such as the pion. 

In summary, the rôle and relevance of the Polyakov loop expectation value \labelcref{eq:FunPolLoop} or similar observables for the confinement dynamics is intricate. However, in view of the strongly correlated QCD regime for temperatures roughly between 
\begin{align} 
	T_c\lesssim T \lesssim 2 \,T_c\,,
\label{eq:SC-QCD} 
\end{align} 
it should be considered as one of the many observables required to fully resolve this 
regime.\footnote{
We would like to add a word of caution for the experts in the field: The above also implies that any interpretation of one or even several of these observables in the context of one of the simple scenarios around comes with quite some caveats and should be taken with more than one grain of salt.
Keeping all the above intricacies in mind, we merely observe the following: in full QCD the Polyakov loop observable \labelcref{eq:FunPolLoop} shows a peak in the thermal susceptibility at about the chiral phase transition temperature, $T_\textrm{peak}\approx T_c$, see   
 \cite{Braun:2009gm, Fischer:2013eca, Fischer:2014vxa, Fischer:2014ata, Lu:2025cls, Lu:2026ezr} (DSE). Technically speaking, the  proximity of $T_\textrm{peak}$ to the chiral phase crossover temperature does not come as a surprise as the constituent quark mass rapidly melts around $T_c$ and the light quarks trigger close perturbative values for the gluonic EoM \labelcref{eq:GluonicEoMfininteT}. Seemingly this suggests that in full QCD the Polyakov loop \labelcref{eq:FunPolLoop} is basically sensitive to the chiral crossover and has lost its connection to confinement. Note however, that the pure glue part $V_\textrm{glue}$ of the potential $V(\varphi)$ also shows a transition from its confining from to the deconfining one in this temperature regime, $T_\textrm{melt} \approx 200$\,MeV, see \cite{Haas:2013qwp}. In the heavy quark limit this temperature is simply the confinement-deconfinement temperature $T_\textrm{conf} = T_\textrm{peak}$ and it can be shown that it is \textit{not} sensitive to the melting down of the constituent quark mass. As discussed above, the gluon mass gap is directly to $T_\textrm{conf}$ and the mass gap in Yang-Mills theory and also determines the (purely gluonic part) of physical spectrum of QCD. Finally, we shall see in the next Chapter, \Cref{sec:fluc}, that it is the gluonic background which is instrumental for the even qualitatively correct fluctuations of conserved charges. On of them, the kurtosis $\kappa_B$, \labelcref{eq:kurtosis} in \Cref{sec:fluc}, is commonly used as a measure for the transition from the baryonic phase (with $\kappa_B=1$ for a non-interacting gas of hadrons) at low temperatures to the perturbative quark-gluon phase (with $\kappa_B=1/9$ for a non-interacting gas of quarks) at high temperatures. This transition happens at about $T\approx T_c - 200$\,MeV. Again, technically it can be related to the melting of the constituent quark mass \textit{and} the melting of $\bar A_0$ also triggered by the transition of the pure glue potential $V_\textrm{glue}$. This entails that, as expected, the situation is less clear as it seems and further work is required to fully unravel the exciting dynamics in the strongly correlated phase \labelcref{eq:SC-QCD} of QCD. }

\subsection{Fluctuations from functional methods}
\label{sec:fluc}

It is one of the main goals of the experimental heavy ion programs to study the existence and location of the critical end 
point in the QCD phase diagram. To this end it is vital to identify observables that connect the theoretical properties 
of the CEP with experimental data, see e.g.~\cite{Luo:2017faz, Zhang:2026dny} for details. It has been suggested, see e.g.~ \cite{Stephanov:1998dy,Stephanov:1999zu,Asakawa:2000wh,Jeon:2000wg,Koch:2005vg,Ejiri:2005wq,Friman:2011pf}, 
that fluctuations of conserved charges may provide important information on the location of the CEP. In the experiments 
these appear as event-by-event fluctuations of the net baryon number $B$, the electric charge $Q$ or the strangeness $S$
of the heavy ion system. In particular, ratios of susceptibilities may provide the cleanest signals. 

In order to compute these quantities in QCD, one starts from the dimensionless pressure $p/T^4$. The thermodynamic pressure is nothing but the volume derivative of the free energy. In the absence of volume fluctuations, it can be extracted 
from the QCD partition function via
\begin{align}
\frac{p}{T^4} = \frac{1}{V T^3} \ln \frac{Z(V,T,\mu_B,\mu_Q,\mu_S)}{ Z(V,0,0,0,0)}\,. 
\label{eq:PressureDensity}
\end{align}    
In \labelcref{eq:PressureDensity} we have normalised the pressure with the vacuum pressure, hence only defining the thermal and density pressure difference. \Cref{eq:PressureDensity} depends on  Lagrange multipliers for the baryon chemical potential $\mu_B$, the charge $\mu_Q$ and the strangeness chemical
potential $\mu_S$, and it can be understood as an integrated density distribution with moments $\langle n_B^i\, n_Q^j  \,n_S^k\rangle $. In functional QCD one typically introduces quark chemical potentials, which are related to the baryon, charge and strangeness ones via 
\begin{align}
	\mu_u = \mu_B/3 + 2\mu_Q/3\,,      \qquad 
	\mu_d = \mu_B/3 -  \mu_Q/3\,,        \qquad 
	\mu_s = \mu_B/3 -  \mu_Q/3 - \mu_S\,.
\end{align}
Then, the irreducible parts of these moments, also called the normalised generalised susceptibilities, are given by 
\begin{align}
\chi^{BSQ}_{lmn} = \frac{\partial^{l+m+n}\left(p/T^4\right)}{\partial\left(\mu_B/T\right)^l \,\partial\left(\mu_S/T\right)^m \,\partial\left(\mu_Q/T\right)^n}\,, \qquad \textrm{and}\qquad C^{BSQ}_{lmn} = V T^3 \chi^{BSQ}_{lmn}\,, 
\label{eq:chiBSQ}
\end{align}  
and the first two susceptibilities are the mean and the variance of the distribution. 
\begin{align}
	\textrm{mean:}\hspace*{10mm} M_{B} = C_1^{B}\,,  \qquad \qquad 
	\textrm{variance:}\hspace*{10mm} \sigma^2_{B} &= C_2^{B}\,, 
	\label{eq:Moments1+2}
\end{align}
They define the Gau\ss ian part of the distribution and carry a very limited information about the dynamics of QCD. In turn, the higher susceptibilities (non-Gau\ss ianities) encode the dynamics of QCD and become trivial in the free limit. Typically, one concentrates in ratios of cumulants, 
as they do not depend explicitly on the volume (though there may be implicit dependencies) 
and can be directly compared with ratios of theoretical susceptibilities, see \cite{Luo:2017faz} for details. The first two higher order cumulants of baryon number are given by 
\begin{align}\nonumber 
\textrm{skewness:}\hspace*{10mm} S_{B} &= \frac{C_3^{B}}{(C_2^{B})^{3/2}}\,, \\[2ex]
\textrm{kurtosis:}\hspace*{10mm} \kappa_{B} &= \frac{C_4^{B}}{(C_2^{B})^2}\,. 
\label{eq:kurtosis}
\end{align}
Analogous expressions for charge and strangeness fluctuations are given by derivatives with respect to $\mu_Q,\mu_S$. The kurtosis is specifically interesting as in non-interacting limits it simply counts degrees of freedom. For $T\to \infty$, asymptotic freedom in QCD leads us to a weakly, asymptotically free gas of quarks and gluons. In turn, for $T\to 0$ we enter the hadronic phase with weakly interacting hadrons. Hence,  we conclude 
\begin{align} 
	\left(  \sigma_B \kappa_B\right)(T\to 0)=1 \,,\qquad \qquad  \left(  \sigma_B \kappa_B\right) (T\to \infty)=\frac{2}{3\pi^2}\,, 
	\label{eq:CountingPhases} 
\end{align}
as the baryon number kurtosis counts baryonic degrees of freedom. In the hadronic phase for $T\to 0$ the baryon number kurtosis approaches unity, while for $T\to \infty$ it counts quarks with the baryon number $1/3$.  The high temperature result in \labelcref{eq:CountingPhases} is that of one-loop thermal perturbation theory with $\chi_2^{B}(T\to \infty)=1/3$ and  $\chi_4^{B}(T\to \infty)=2/(9 \pi^2)$.  

Within the functional approach to QCD, thermodynamic and statistical quantities can be accessed via several methods.
The 'standard' way is to determine the quark number susceptibilities or their flow 
\begin{align} 
	n_q^f(T,\mu) \simeq -\sumint_p \textrm{Tr}\, \bigl[ \gamma_0 \,G_f (p) \bigr]\,, \qquad \qquad \partial_t n_q^f(T,\mu) = - \frac{d}{d \mu_B} \partial_t \Omega(V,T,\mu_B,\mu_Q,\mu_S) \,, 
	\label{eq:sus}
\end{align}
where the grand potential $\Omega$ is nothing but the effective action, evaluated on the equations of motion, \labelcref{eq:GrandPotential}. Hence, in functional methods the quark number susceptibilities can be computed 
on the same footing as the quark condensate, \labelcref{eq:condensate}. The pressure and subsequently the energy density and the entropy can then be determined from standard thermodynamic relations using the grand potential that either follows from the integration of the quark number density (DSE) or from its flow (fRG). For practical purposes it might be beneficial to use
lattice input for the trace anomaly at zero chemical potential as truncation gauge or even as input, see e.g.
\cite{Lu:2023mkn, Lu:2025cls, Lu:2026ezr}. A variant of this method relying on Maxwell relations has been outlined in \cite{Isserstedt:2020qll}.
In the traditional approach, less used today due to its large systematic error, the pressure is determined directly from the 2PI effective action or the flow of the grand potential, see
e.g.~\cite{Roberts:2000aa} for an early review of this method. 

Higher order susceptibilities and their ratios (hyper fluctuations) are also accessible but require a rapidly increasing computational effort: For sampling methods such as lattice QCD simulations the required statistics increases dramatically and rapidly beyond currently available resources. Note that a similar rise of the required statistics also applies to experimental measurements of hyper fluctuations and the impressively high interaction rate is an important advantage of the future CBM experiment at the FAIR facility. In turn, for functional methods the numerical costs increase significantly as taking numerical higher order derivatives with respect to $\mu_B,\mu_Q,\mu_S$ requires a sizeable increase of numerical precision. Still, even high order hyper fluctuations are accessible, see e.g.~\cite{Fu:2023lcm}.  

In order to take into account the situation of heavy-ion collisions, these need to be adjusted appropriately.
Strangeness conservation in the colliding nuclei implies that the mean density of strange quarks vanishes, i.e.~ 
$\langle n_s \rangle = \chi^S_1 = 0$. On the other hand, typical ratios of the number of baryons to protons
in Au-Au and Pb-Pb collisions imply that $\langle n_Q \rangle = Z/A \,\langle n_B \rangle$ with $Z/A \approx 0.4$.
Thus the dependence of $\mu_Q$ and $\mu_S$ on $\mu_B$ or alternatively $\mu_u$, $\mu_d$ and $\mu_s$ need to be 
defined such that these conditions are satisfied.

\section{Discussion of results}
\label{sec:results}

In the preceding sections we have provided a brief review of the conceptual and technical foundations of the 
functional approach to QCD. In the following we will walk the reader through a selection of the main results 
of the functional approach to the QCD phase structure and observables at finite temperature and chemical potential. 

As has been discussed before, one of the most promising strategies for testing and benchmarking functional methods is to compare functional results with that from  
lattice QCD as systematically as possible in regions of QCD parameter space, where lattice QCD is applicable. This is
important and interesting for two reasons:\\[-1ex]

(i) Non-perturbative QCD in general is a tricky business 
and it is always advisable to cross-check physics results obtained from as many approaches as possible. An application where 
this strategy has been followed very successfully is the anomalous magnetic moment of the muon, where a large theory 
cooperation of several dozen colleagues working with distinctly different methods such as dispersion theory using 
input from various experiments, lattice QCD as well as holographic and functional approaches managed to provide a 
theory prediction with small systematic error, see \cite{Aoyama:2020ynm,Aliberti:2025beg} for details and corresponding 
Chapters in the Encyclopedia of Particle Physics for a pedagogical overview. \\[-2ex]

(ii) These cross-checks
serve as a gauging procedure for the quality of the truncation used in the functional approaches. Only truncations rich
enough to pass these cross-checks in all relevant areas should be trusted as potential candidates for reliable results
in those areas of the QCD phase diagram which cannot be explored by lattice QCD. As discussed above, this is the 
highly interesting region of the onset of new phases in the dense realm of QCD at large chemical potential. \\[-1ex]

\begin{figure}[t]
	\centering
	\includegraphics[scale=0.40]{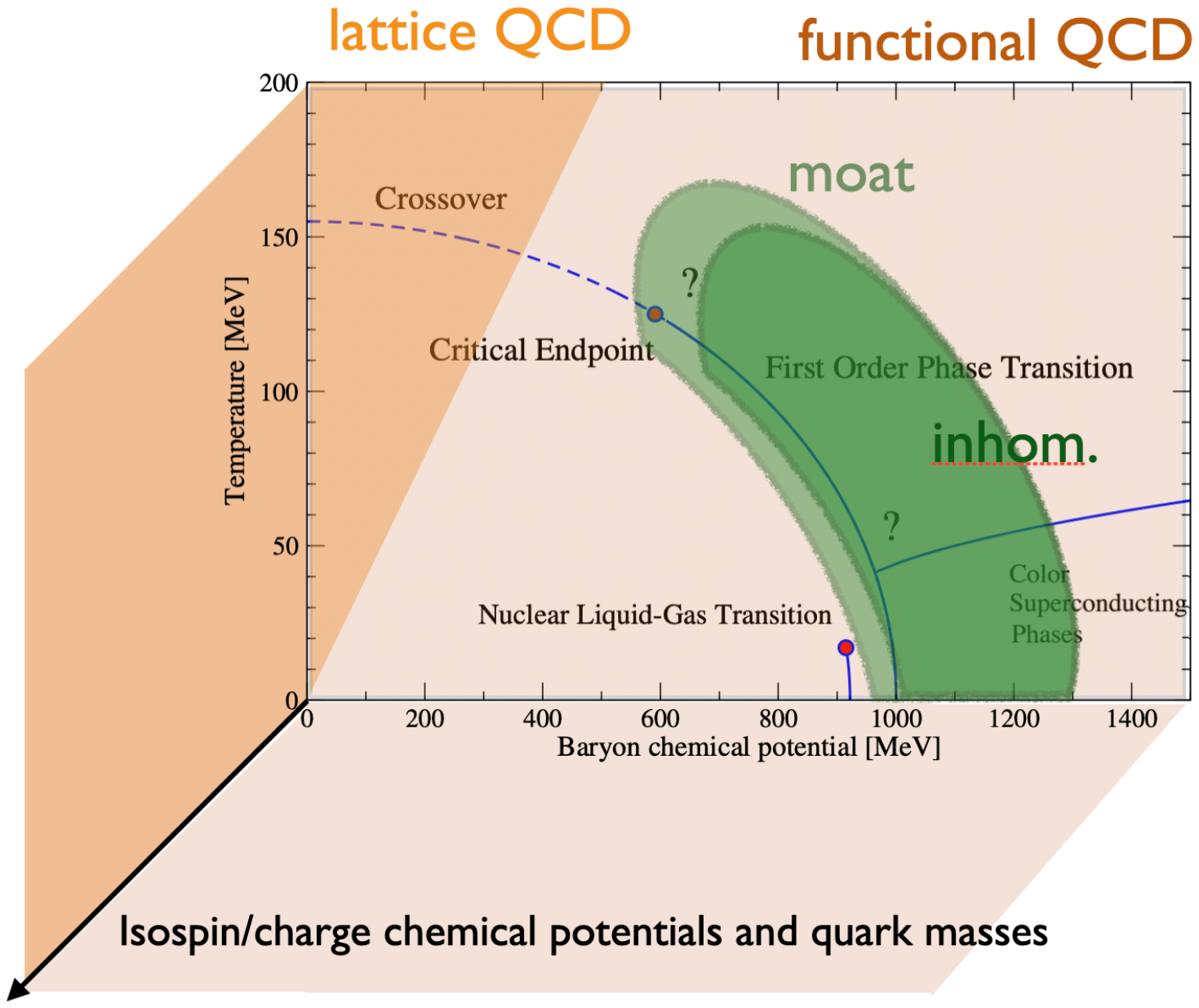}    
	\caption{Sketch of the QCD phase diagram in the temperature and baryon chemical potential plane augmented with 
		further axis to symbolize the QCD parameter space. Also sketched is the potential appearance of new
		inhomogeneous phases and/or precursor moat phenomena, see text for details. \hfill  \label{fig:phase2}}
\end{figure}
In order to be prepared for the discussion in this Chapter we need to briefly recapitulate, what is the parameter space 
of QCD and which regions or accessible by lattice gauge theory. To this end we refer to \Cref{fig:phase2} for a 
schematic overview. QCD features six quark flavours, but only three are relevant for the structure of the QCD phase
diagram at temperatures and quark chemical potentials up to several hundred MeV. This is mainly due to scales; the top 
and bottom quarks are simply to heavy to play a dynamical role, whereas the charm quark is only of 
minor importance (see below). Consequently, almost all studies 
currently available work with up, down (light) and strange (heavy) quarks in $N_f=2+1$ flavour QCD. Up to a global scale, we therefore encounter seven relevant 
QCD parameters: the temperature $T$, the current quark masses $m_u,m_d,m_s$ of the three lightest quarks, and their chemical potentials $\mu_u,\mu_d,\mu_s$. The latter may be
re-expressed in terms of chemical potentials for conserved quantities (baryon number $\mu_B$, electric charge $\mu_Q$ 
and strangeness $\mu_S$ chemical potentials), as discussed above in section \Cref{sec:fluc}.

Four of these seven parameters, $T$ and $\mu_{q_i}$'s, reflect the physics of the environment where strongly interacting matter is probed. In the early universe, conventional 
analysis suggests that all three chemical potentials are close to zero thus probing only the temperature
direction of parameter space (although the potential occurrence of large lepton asymmetries may induce sizeable 
isospin/charge and even baryon chemical potentials, see e.g.~\cite{Schwarz:2009ii,Gao:2021nwz}). In heavy ion 
collisions, the charge chemical potential is non-zero and, depending on beam energies, sizeable baryon chemical 
potentials can be reached, including the large density regions with the potential onset of new phases (i.e.
the green regions in \Cref{fig:phase2}). Finally, the physics of neutron stars and neutron star
mergers probes the region of both, sizeable baryon and isospin/charge chemical potential. Here, new phenomena
like inhomogeneous or colour superconducting phases may be induced by $\mu_B$, whereas pion condensation 
is known to occur in the $\mu_Q$-direction of the phase diagram. The interplay between these different phenomena is
highly intricate and is largely unexplored to date. 

The other three of these seven parameters are the fundamental parameters of QCD fixed by nature: within QCD, the up, down and strange quark masses 
are input parameters that are uniquely determined by the details of electroweak symmetry breaking via the Higgs-mechanism 
at the electroweak scale or order $100$\,GeV. This is far above the 1\,GeV scale of strongly correlated infrared QCD.
Nevertheless, it is interesting to systematically explore the change of the location and the nature of the QCD 
phase transitions at 'unphysical' values of these quark masses. In principle, this allows for deep theoretical insights 
into the internal structure of the theory. This comes with a high potential for additional theoretical understanding that in turn 
helps to interpret and understand corresponding results at physical quark masses. This is also true for 
analytic continuations of chemical potentials, in particular $\mu_B$, into the imaginary direction. Since lattice
QCD at imaginary baryon chemical potential does not suffer from the sign problem, this direction has been extensively 
explored in the past. This not only offers excellent opportunities for systematic comparisons with functional approaches 
but QCD at complex chemical potential allows for a theoretical approach to the CEP away from the physical QCD plane with $\mu_B\in\mathbb{R}$. 

In the following we will first discuss results for variations of the 'theoretical' QCD parameters, i.e.~the current quark 
masses and imaginary chemical potential. Then we proceed to QCD at physical quark masses and discuss results for
the location of the critical end-point, its properties and the region of onset of new phases in the QCD phase diagram.

\subsection{Theoretical explorations at unphysical values of quark masses and chemical potential}
\label{sec:unphys}

\begin{figure}[t]
	\centering
	\includegraphics[width=0.4\columnwidth]{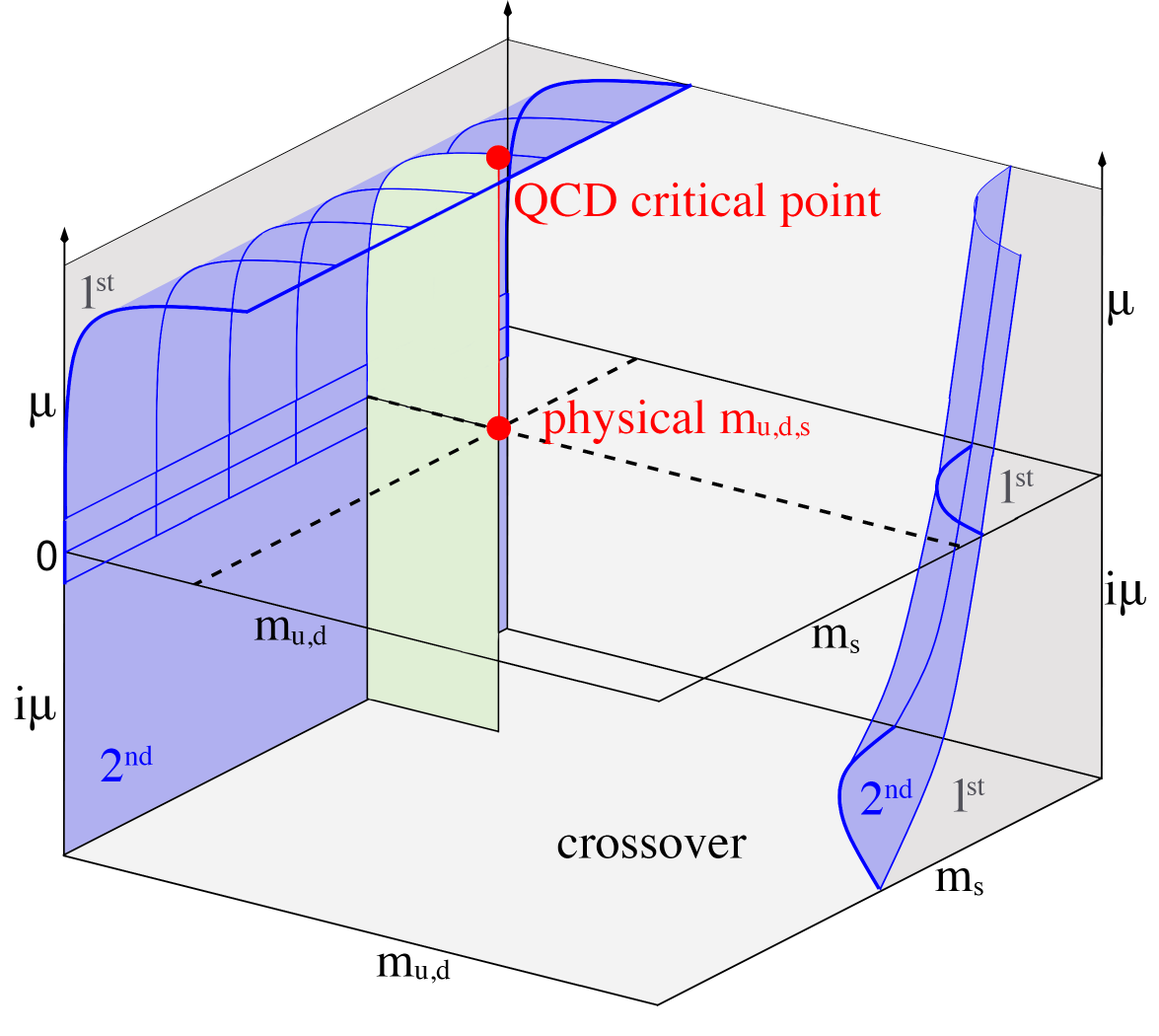}\hfill
	\includegraphics[width=0.4\columnwidth]{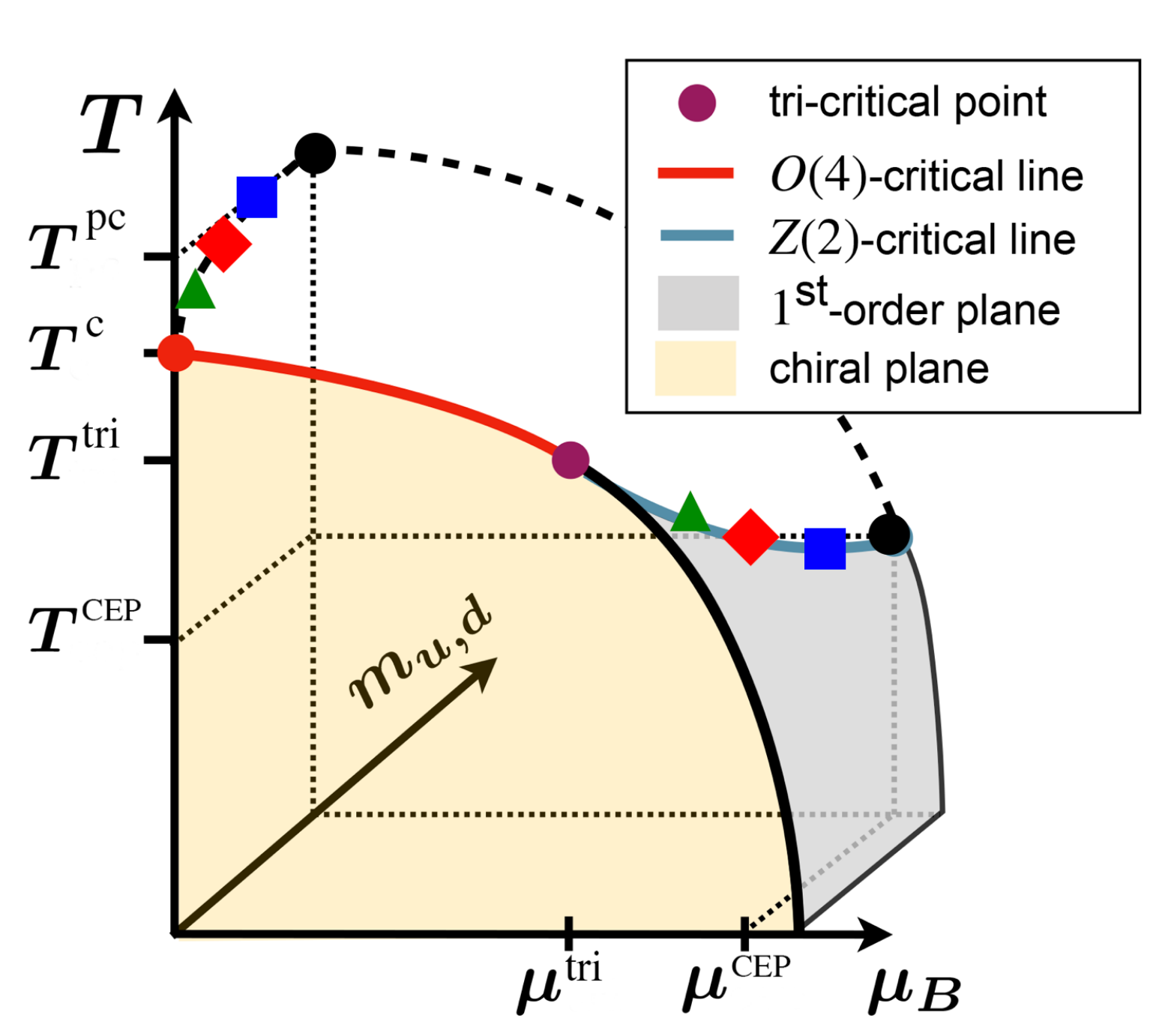}
	\caption{
		Left: 3d-Columbia plot with real and imaginary chemical potential as third axis. The second order critical surface
		of the confinement-deconfinement transition in the heavy quark limit has been explored with functional methods
		in ~\cite{Fischer:2014vxa}. The chiral critical surface connecting the QCD critical point discussed in 
		previous sections with the second order transition line at zero chemical potential has been studied in
		~\cite{Bernhardt:2025fvk}.
		Moreover the crossover nature of the green area in the imaginary chemical potential region is confirmed by both,
		lattice calculations \cite{DAmbrosio:2022kig, Cuteri:2022vwk, DAmbrosio:2025ldv} and functional methods \cite{Bernhardt:2025fvk}.
		Right: Sketch of the QCD phase diagram in temperature $T$ and baryon chemical
		potential $\muB$ for varying degenerate light quark masses. Adapted from~\cite{Ding:2024sux}. The symbols
		indicate the quantitative results from~\cite{Bernhardt:2025fvk} for $m_\pi =  140, 110, 80, 55$ MeV. 
		The qualitative behaviour under variation of the quark mass matches general expectations, 
		see text for details. Figures taken from~\cite{Bernhardt:2025fvk}. \hspace*{\fill}}
	\label{fig:columbia}
\end{figure}

We start with a short overview on the so-called Columbia plot (see the corresponding dedicated Chapter of this \En{} for
many more details). A three-dimensional version of this plot is given in the left figure of \Cref{fig:columbia}.
The light up and down quark masses are varied from zero to infinity along the x-axis, the strange quark mass is
varied from zero to infinity along the y-axis and the baryon chemical potential from imaginary values (negative z-axis)
to real values (positive z-axis) along the third dimension. The Columbia plot indicates the order of the QCD phase
transition with temperature, which may be an analytic crossover (light grey), second order (blue) or first order (dark grey). 
The physics of the Columbia plot is extremely interesting by itself and can be explored with various methods. 

In particular, the quark mass dependence of the QCD phase structure provides ample opportunity for systematic comparisons between 
different ab initio approaches to the QCD phase diagram such as lattice QCD and functional methods. Lattice QCD is fully 
operational at not too small quark masses and zero as well as imaginary chemical potential and provides extrapolations toward 
the various chiral limits, $m_{u,d} \rightarrow 0$ with $m_s \rightarrow \infty$, $m_{u,d} \rightarrow 0$ with $m_s$ fixed at 
the physical strange quark mass and $m_{u,d,s} \rightarrow 0$. Functional methods can be cross-checked in these regions with 
lattice QCD and, provided these are successful, may deliver reliable results in all regions inaccessible by lattice QCD.
In particular this is true for large baryon chemical potential. 

Technically, functional methods directly address the physics of the Columbia plot simply by varying the current quark masses $m_f$ that 
appear in the inverse bare quark propagators $ i \pslash + m_f$ in the corresponding quark DSEs in \Cref{fig:FunDSE+QuarkDSE}. For the fRG the current quark mass does not enter the flow itself but occurs as a parameter in the boundary condition of the flow equation, the initial effective action $\Gamma_\Lambda$ at the electroweak scale (or below). Each of the inverse quark propagators $G_f$ obeys its own DSE or fRG flow equation. In most practical computations in  $2+1$-flavour QCD the functional relations are solved with the assumption of (approximate) strong isospin symmetry, i.e.~two degenerate light quarks and a heavier strange quark, based on the smallness of the light current quark masses, $2-5$\,MeV and their difference in terms of the scale of dynamical chiral symmetry breaking, which leads to constituent quark masses of about $M_{u,d}\approx 350$\,MeV. In view of the latter scale the charm quark with a mass of the order of 1\,GeV may also be (borderline) relevant, but explicit calculations in a $N_f=2+1+1$ setup (two degenerate
light quarks, and one strange and one charm quark) show that effects on the location of the phase boundaries are of 
the order of only a few MeV \cite{Fischer:2014ata}. Although these might become relevant in the future,
given the level of today's precision requirements these can be safely neglected. 

Let us briefly walk through the Columbia plot. 
The physics of the pure gauge/heavy quark region of the Columbia plot is dominated by the $SU(3)$ gauge theory, the
associated first order confinement-deconfinement phase transition and the Roberge-Weiss (RW) point at imaginary chemical potential \cite{Roberge:1986mm}: its location depends crucially on the fact the QCD is periodic under shifts of the chemical potential with 
\begin{align} 
	\mu_B\to \muB + \frac{2 \pi}{3} i\,,  
\label{eq:RWsymmetry} 
\end{align} 
where the $1/3$ originates from center symmetry of the gauge field, underlying the first order confinement-deconfinement phase transition in QCD with heavy quarks. In the absence of center symmetry the periodicity would increase to $2\pi i$. In functional QCD this has been studied in~\cite{Fischer:2009wc, Fischer:2009gk, Fischer:2014vxa, Wan:2025wdg} (DSE), and in \cite{Braun:2009gm} (fRG). For an overview see \cite{Fischer:2018sdj}, including a graphical visualisation of the heavy-quark phase diagram at imaginary chemical potential capturing the Roberge-Weiss transitions. In the heavy quark limit the functional studies reproduce the second order critical surface emerging from the plane of the first Roberge-Weiss transition including the expected tricritical scaling from the Roberge-Weiss endpoint in agreement with previous and later findings from 
lattice gauge theory \cite{deForcrand:2010he, Saito:2011fs, Fromm:2011qi, Ejiri:2019csa, Cuteri:2020yke, Kiyohara:2021smr}. Moreover, the functional studies also provide a prediction for the location of the phase transition lines at imaginary chemical potential in the chiral limit, that awaits to be corroborated with lattice simulations. 

In this context we would like to pick up the discussion on the relevance of the traced Polyakov loop expectation value or temporal background $\bar A_0$ for the confinement properties of QCD with dynamical quarks: In functional approaches, the self-consistent dynamical background $\bar A_0$, obtained from the equation of motion \labelcref{eq:GluonicEoMfininteT}, is carrying the correct periodicity pattern \labelcref{eq:RWsymmetry} under shifts of the imaginary chemical potential. Accordingly, $\bar A_0$ is linked to the location of the RW-transition with its confinement-deconfinement correlation. This is discussed in detail in \cite{Braun:2009gm, Fischer:2014vxa, Wan:2025wdg}.

At the physical point (indicated by the lower red dot) and in the chiral limit, QCD at imaginary chemical potential has been explored within functional QCD in~\cite{Bernhardt:2023ezo} and \cite{Braun:2009gm} respectively. The corresponding pseudo-critical transition temperatures have been determined in the 
region between zero chemical potential and the first Roberge-Weiss transition. The purpose of the study in~\cite{Bernhardt:2023ezo} has been to provide a unified picture of the QCD transition from imaginary to real 
baryon chemical potential up to the critical end-point. In addition, it has provided two interesting cross-checks. 
On the one hand it served as a qualitative cross-check for the functional approach by direct comparison with the corresponding lattice 
results of~\cite{Borsanyi:2020fev} for $T_c(i \mu_B)$. On the other hand, it served as a cross-check for the extrapolation
method used in~\cite{Borsanyi:2020fev} to determine $T_c(\mu_B)$ at real baryon chemical potential from $T_c(i \mu_B)$.
Both cross-checks have been successful. The functional approach reproduced $T_c(i \mu_B)$ for a large range of $\mu_B$, with an increasing systematic error 
at larger $i \mu_B$, originating from neglecting the $A_0$-background discussed above. Moreover, using the same extrapolation method as in~\cite{Borsanyi:2020fev},
functional QCD results for $T_c(\mu_B)$ at real baryon chemical potential could be reproduced: up to large chemical potentials 
not very much smaller ($\approx 20 \%$) than the one of the critical endpoint, the extrapolation worked extremely well. 
For larger chemical potential, the extrapolated transition line undershoots the calculated one and, of course, the extrapolation cannot be used to indicate the location of the CEP. This underlines the (expected) intrinsic limits of any such extrapolation procedure. 

In \cite{Wan:2025wdg}, the functional QCD approach to imaginary baryon chemical potential has been extended to complex baryon chemical potentials $\mu_B\in\mathbb{C}$. This also accommodates Lee-Yang edge singularity (LYES) and their convergence towards the critical end point at real $\mu_B$, for first functional steps towards LYES and in particular their scaling see \cite{Connelly:2020gwa, Mukherjee:2021tyg, Rennecke:2022ohx, Johnson:2022cqv}. All these works confirm the potential of functional methods to access QCD at complex chemical potential and to provide further structural insights and quantitative results beyond that gained with lattice simulations. Amongst the results in the qualitative study \cite{Wan:2025wdg} is an estimate for the LYES and its approach towards the critical end point that awaits further studies and confirmation.  

A further interesting, and not yet fully resolved, situation emerged in the past years for the chiral limit, i.e.~the left hand
side of the zero chemical potential plane of \Cref{fig:columbia}. Starting from the physical point and approaching the chiral limit
$m_{u,d} \rightarrow 0$ with fixed strange quark mass, strong indications for the second-order nature of this point have 
been reported from lattice QCD \cite{HotQCD:2019xnw,Cuteri:2021ikv} and functional methods
\cite{Braun:2020ada, Gao:2021vsf, Bernhardt:2023hpr, Braun:2023qak}. The qualitative behaviour of the corresponding transition 
temperatures along this line is also shown in the right diagram of \Cref{fig:columbia} (the symbols in the upper left part of the 
diagram at $\mu_B=0$). Both, lattice QCD and the functional approach show decreasing chiral transition temperatures with decreasing 
light quark mass. This feature is readily understood; reducing the light quark mass reduces the amount of explicit breaking of 
chiral symmetry and consequently we find restoration at lower temperatures. 

Walking now along the left hand side of the Columbia plot, the 'zero' in the plot marks the SU(3)-symmetric limit $m_{u,d,s} \rightarrow 0$
at zero baryon chemical potential. It is not clear, and in fact highly debated, whether this point features a second order or a (weak) 
first order transition with a very small extension into the finite mass region of the Columbia plot. For recent work see e.g.~\cite{Resch:2017vjs, Fejos:2022mso, Pisarski:2024esv, Fejos:2024bgl}, for a detailed discussion see the corresponding dedicated 
article of this \En. 
 
Furthermore, we discuss the shape of the critical surface connecting the chiral limit line $m_{u,d} \rightarrow 0$ with varying strange
quark mass at zero chemical potential with the QCD critical point at finite $\mu_B$ and physical quark masses (the upper red dot in the 
diagram). A very recent work in the DSE-approach has addressed this issue, 
\cite{Bernhardt:2025fvk}. There, a qualitative study of the critical surface has been performed and the nature of the transition in the 
green area indicated in \Cref{fig:columbia}. Interestingly, a flat chiral critical surface at large
chemical potential has been found, that presumably ends in a line of tricritical points along the chiral left-hand side of the 3$d$-Columbia 
plot. Furthermore, inside the green area in the figure, a crossover down to the plane of the first 
Roberge-Weiss transition has been seen \cite{Bernhardt:2025fvk}. These findings are in agreement with and confirm 
previous notions from lattice gauge theory for the structure of the Columbia plot \cite{DAmbrosio:2022kig, Cuteri:2022vwk, DAmbrosio:2025ldv} 
and for temperature bounds for the transition temperature of the critical endpoint \cite{Halasz:1998qr, Karsch:2019mbv}. The latter scenario 
is summarised in the diagram on the right hand side of \Cref{fig:columbia} with explicit results from functional QCD denoted by the 
coloured symbols along the blue line on the right hand side of the sketch. These results support the existence of a tricritical point, 
whose precise location still needs to be determined.

\subsection{Physical values of quark masses and chemical potential: the quest for the CEP}
\label{sec:cep}

\begin{figure}[t]
	\centering
	\includegraphics[width=.65\textwidth]{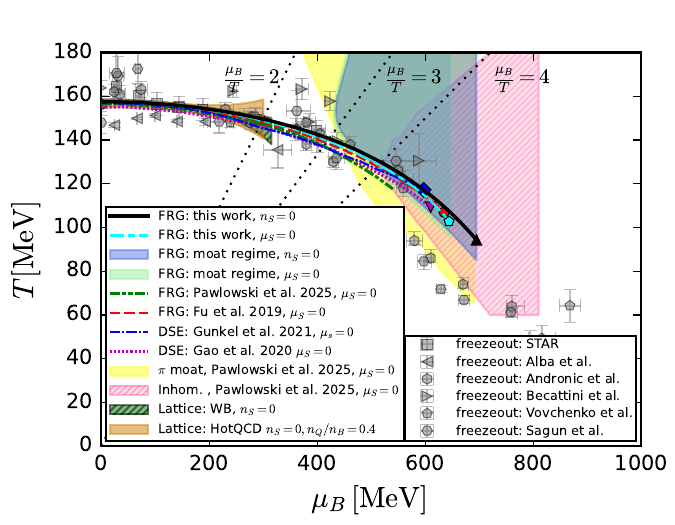}
	\caption{Comprehensive functional phase structure, including strangeness \cite{Fu:2026qnl}. Shown are 
		results from lattice QCD (the shaded 'trumpets'), crossover lines and CEPs from different functional 
		calculations, and the region with the onset of new phases (ONP) (large shaded regions), see text for 
		details and references. Shown are also free-out points extracted from experiment with statistical models. Figure taken from~\cite{Fu:2026qnl}. \hspace*{\fill}}
	\label{fig:FunQCDPhaseStructureAll}
\end{figure}

In this section we focus on the variation of the two physical parameters, temperature and baryon chemical potential, for QCD with physical quark masses, i.e.~we explore the line connecting the two red dots in the left diagram of \Cref{fig:columbia}. The 
corresponding two-dimensional QCD phase diagram has been sketched in \Cref{fig:phase} - here we discuss the explicit results 
shown in \Cref{fig:FunQCDPhaseStructureAll}. The plot is very busy, so let us walk through it step by step. 
First observe the region of small baryon chemical potential $\mu_B \lesssim  400$ MeV. This is the realm approachable by extrapolations 
of results from lattice QCD. In the figure we included the brown and green-shaded regions where two major lattice collaborations
locate the chiral crossover with increasing error (the 'opening of the tube') with chemical potential 
\cite{Bazavov:2018mes,Borsanyi:2020fev}. A collection of results from the functional approach in high quality truncations 
('DSE' and 'FRG')  \cite{Fu:2019hdw, Gao:2020fbl, Gunkel:2021oya, Fu:2026qnl}
is also shown and agrees nicely with the lattice results at small chemical potential. At large chemical potential, the chiral transition lines obtained from functional QCD using $\mu_s=0$ (zero strange quark chemical potential) or $\mu_S=0$ (zero strangeness chemical potential) show a critical 
end point in the region 
\begin{align}
\mu_S=0,\mu_s=0:\qquad \qquad 	(T, \mu_B)_\textrm{CEP} \in \bigl(115-105\,,\, 600-650\bigr)\textrm{MeV}\,. 
	\label{eq:FunEstimateCEP}
\end{align} 
The interested reader can find a comprehensive discussion of the (complementary) differences in the truncation schemes of 
functional results in general and \cite{Fu:2019hdw, Gao:2020fbl, Gunkel:2021oya} in particular in the review article 
\cite{Fischer:2026uni}. 

In \cite{Fu:2026qnl} the functional QCD analysis of the phase structure was extended to net strangeness density $n_s=0$ (which roughly corresponds to $\mu_S \approx \mu_B/5$), a physical condition which is realised in all heavy ion collision experiments. This condition shifts the CEP somewhat to even larger baryon chemical potential.
\begin{align} 
	(T, \mu_B)_\textrm{CEP} \approx  \bigl(92\,,\, 696\bigr)\textrm{MeV}\,. 
\label{eq:eq:FunEstimateCEP-ns} 
\end{align}
The truncation used in the fRG-study \cite{Fu:2026qnl} also improves on some aspects about that in \cite{Fu:2019hdw}, but overall is subject to the same systematic error margin as for $\mu_S=0, \mu_s=0$ in \labelcref{eq:FunEstimateCEP}. 

Importantly, all high quality functional approaches at $\mu_S,\mu_s,n_s=0$ see a CEP in the same region of the QCD phase diagram between $\mu_B=600-700$\,MeV. It has been shown in \cite{Fu:2023lcm} that a CEP in this regime is mirrored by a peak of the susceptibilities and specifically the kurtosis at the freeze-out line at collision  energies $\sqrt{s_{NN}}$ with 
\begin{align} 
	\sqrt{s_\textrm{peak}} \in (4.5\,,\,5.5)\, \textrm{GeV} \,. 
\label{eq:PkeakFreezeout} 
\end{align} 
Moreover, the location of the peak shows a very mild dependence on the location of the CEP, and curiously, the peak location $\sqrt{s_\textrm{peak}} $ even very mildly increases with increasing $\mu_B$. The best current estimate for the experimental location ($n_s=0$) that follows from \cite{Fu:2023lcm} as approximately 
\begin{align} 
	\sqrt{s_\textrm{peak}}(n_s=0) \approx 5\,\textrm{GeV}\,.
\label{eq:BestEstimate} 
\end{align} 
More details on the connection between the experimental observables and the current theoretical results is provided in \Cref{sec:fluc-Exp-Theo}. 

Is this the end of the story? Not by far. What is also shown in \Cref{fig:FunQCDPhaseStructureAll} are regions of potential onsets of new 
phenomena: for $\mu_B/T \approx 4$, the crossover line enters the 'moat regime' \cite{Pisarski:2021qof}. The moat is characterised by the appearance of a non-zero momentum scale 
in static meson correlation functions such as the one for the pseudoscalar pion ('$\pi$-moat') or the one of the scalar 
sigma meson. These may result in non-trivial dispersion relations and measurable signals in heavy ion collisions \cite{Pisarski:2021qof}. 
First theoretical indications of such a moat-regime have been found with functional QCD \cite{Fu:2019hdw, Fu:2024rto} and have 
been solidified recently in a self-consistent functional QCD computation \cite{Pawlowski:2025jpg}. In the latter work, 
indications for a true instability at even larger density have been found for $\mu_B/T \gtrsim 4.5$. 
This suggests the presence of an inhomogeneous phase in this regime. Such inhomogeneous phases are characterised by momentum-dependent order parameters (e.g.~the chiral condensate but also other realisations such as momentum-dependent diquark
condensates are possible). They have already been studied comprehensively in low-energy effective theories, for a review 
see \cite{Buballa:2014tba}. Moreover, the path towards applications within functional QCD has been prepared by now in
\cite{Motta:2023pks,Motta:2024rvk}. A first analysis of respective experimental signatures of the moat and inhomogeneous 
phases with Hanbury Brown-Twiss interferometry has been performed in \cite{Rennecke:2023xhc, Fukushima:2023tpv}. Finally, 
the physics of a potential CEP may be modified by the mixing of the critical scalar massless mode with density fluctuations 
inducing a sizeable glue component, see \cite{Haensch:2023sig}.

\subsection{Connecting theory and experiment: fluctuations and freezeout}
\label{sec:fluc-Exp-Theo}

The last point we wish to address in this pedagogical article is the connection with experiment. Let us reiterate some general remarks:\\[-2ex]

 (i) As explained in \Cref{sec:fluc} above, baryon number fluctuations are readily available in all approaches based on the 
Euclidean path integral such as lattice QCD and functional methods. However, (ratios of) event-by-event fluctuations extracted 
from experiment are that of proton number fluctuations. This is an important difference whose consequences have yet to be addressed comprehensively. Moreover, functional and lattice QCD are based on the grand potential of QCD. This is a good enough approximation for large collision energies, while at smaller collision energies the canonical ensemble is more appropriate. A first phenomenological account of this is included in \cite{Fu:2023lcm} based on the sub-ensemble acceptance method (SAM) introduced in \cite{Vovchenko:2020tsr}, for further discussions see also \cite{Lu:2026ezr}. The latter point already raises the question of the effective volume in a heavy ion collision, addressed in (ii).\\[-1ex]

(ii) Heavy ion experiments are always subject to finite and expanding volume effects due to the finite extent of the expanding fireball
in heavy ion experiments. Although major volume corrections are cancelled out in the ratios of susceptibilities that are generally
considered (cf. \Cref{sec:fluc}; see also \cite{Bernhardt:2021iql} for a direct study of the volume dependence of the CEP), 
subleading corrections may still play a role and ultimately have to be addressed. \\[-2ex] 

(iii) Non-equilibrium effects have 
to be taken into account, first steps towards functional QCD computations in the transport regime have been taken 
in \cite{Bluhm:2018qkf, Tan:2025bsv}. The final goal of this endeavour is the first principles access to the full timeline of a Heavy Ion Collision (HIC). On the QCD side, this requires the computation of real-time correlation functions in and out of equilibrium. Again to date, these seem only accessible with functional QCD due to the real-time sign problem in corresponding lattice simulations. This is ongoing work in functional QCD with a current concentration on real-time equilibrium QCD correlation functions, where the advantages at finite temperature and density gain a lot from respective ones in the area of hadron resonances and decays, see e.g.~\cite{Weil:2017knt,Williams:2018adr,Santowsky:2020pwd} and references therein. Real-time QCD correlation functions provide a QCD-access to the transport and hydrodynamical phase in a HIC, for first results on transport coefficients see \cite{Haas:2013hpa, Christiansen:2014ypa}, for the progress in real-time functional methods see the e.g.~\cite{Berges:2004yj, Gasenzer:2007za, Berges:2008sr, Kamikado:2013sia, Pawlowski:2015mia, Horak:2020eng, Roth:2021nrd, Tan:2021zid, Horak:2022myj, Braun:2022mgx, Horak:2022aza, Chen:2024lzz, Roth:2024hcu, Tan:2025bsv, Huang:2026zfv}. It goes without saying that far-from-equilibrium QCD is even more challenging.  \\[-1ex]

\begin{figure*}[t]
	\centering
	\begin{minipage}[t]{.48\linewidth}
		\begin{subfigure}{\linewidth}
			\centering
			\includegraphics[width=\linewidth]{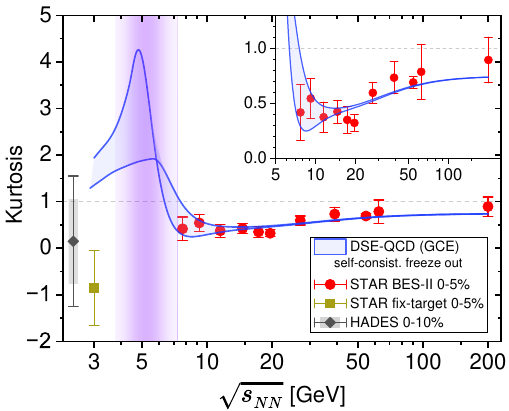} 
			\subcaption{Kurtosis as a function of collision energy $\sqrt{s_{\tiny{NN}}}$. \hspace*{\fill}}
			\label{fig:kurtosis} 
		\end{subfigure}%
	\end{minipage}
	\hspace{0.02\linewidth}%
	\begin{minipage}[t]{.48\linewidth}
		\begin{subfigure}{\linewidth}
			\centering
			\includegraphics[width=\linewidth]{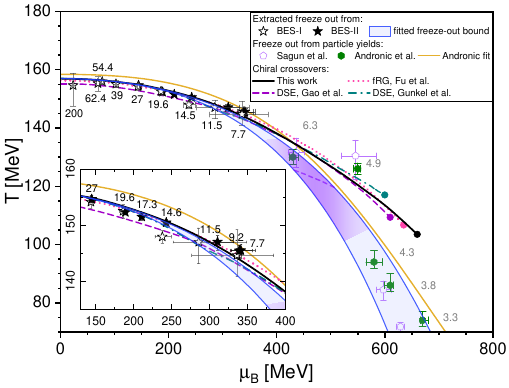} \vspace{-.20cm}
			\subcaption{Freezeout-line in the QCD phase diagram.\hspace*{\fill}}
			\label{fig:freezeout} 
		\end{subfigure}
	\end{minipage}
	\caption{Left: Results for the kurtosis extracted in \cite{Lu:2026ezr} using functional methods and compared to experimental 
		results from BES-II~\cite{STAR:2025zdq} and HADES \cite{HADES:2020wpc}. The blue band indicates the systematic error of the extraction. 
		Right: Freeze-out region (blue band) in the $T$-$\mu_B$ plane as extracted in \cite{Lu:2026ezr}, using functional methods and data 
		from BES-I~\cite{STAR:2021iop} and BES-II~\cite{STAR:2025zdq}. The collision energies are labelled in GeV units. 
		Also shown are the locations of the CEP extracted in \cite{Fu:2019hdw, Gao:2020fbl, Gunkel:2021oya, Lu:2025cls}. 
		The freeze-out points extracted from particle yields~\cite{Andronic:2017pug,Sagun:2017eye} are inside the blue band. 
		whose collision energies are indicated by the grey numbers in GeV unit. The peak position of the kurtosis (from the left diagram)
		along the freeze-out line is indicated by the purple region. Figures taken fromn\cite{Lu:2026ezr}.\hspace*{\fill}}
	\label{fig:fluct}
\end{figure*}

With these caveats (and further ones) in mind, equilibrium QCD calculations of baryon number fluctuations at infinite 
volume provide an equilibrium QCD baseline for the experimental proton number fluctuation. Confronted with the experimental data they provide information about the importance of the differences described in (i-iii). For sufficiently small densities they can be obtained by both, lattice and functional QCD, while for larger densities only functional QCD results are available. In the following we briefly discuss the latter mainly drawing from \cite{Fu:2023lcm, Lu:2026ezr}: 

We start with the kurtosis, defined in \labelcref{eq:kurtosis} and shown in \Cref{fig:kurtosis} as a function of the 
collision energy $\sqrt{s_{\tiny{NN}}}$. In \Cref{fig:freezeout} the respective collision energies are marked as black labels along the 
corresponding freeze-out line, i.e.~the temperature and chemical potential where the abundances of hadrons emerging from a 
fireball with collision energy $\sqrt{s_{\tiny{NN}}}$ are fixed. Note that there are still elastic re-scattering effects after chemical 
freeze-out, but these do not change the particle abundances. The point where these become negligible is called the kinematic 
freeze-out. From this translation, one finds that the kurtosis remains roughly constant as long as the freeze-out line is
on top of the chiral transition line. These start to substantially deviate around $\sqrt{s_{\tiny{NN}}} < 7$\,GeV. In this region,
the kurtosis starts to develop a peak indicated by the purple shaded bands in both figures. The properties of this peak
are tightly connected to the properties of the critical end point, see \cite{Fu:2023lcm} for a detailed 
discussion. Two specific, experimentally well-accessible properties are the location $\sqrt{s_\textrm{peak}}$ of the peak and its height. We have already discussed in \Cref{sec:fluc} that the location of the peak has a very mild dependence on the location of the CEP in the baryon chemical potential regime $600-700$\,MeV. The best current estimate for the experimental situation with strangeness neutrality is given by \labelcref{eq:BestEstimate}. In turn, the peak height is very sensitive to the location of the CEP, but at present it depends on many unknowns related to the unresolved physics (i-iii) discussed at the beginning of this Chapter.  

In summary, the precise connection of the experimental measurements and the theoretical predictions is still subject of a considerable amount of ongoing research in both experiment and theory. The authors share the solid expectation that functional QCD will play an important rôle in resolving these open challenges in the next decade. 

\section{Conclusion}
\label{sec:conclusion}

With the ongoing experimental program at ALICE/CERN, the beam energy scan at RHIC/BNL as well as 
the future dedicated programs at HADES/CBM/FAIR and HIAF, it remains a major task and challenge for 
the theoretical approaches of QCD to make solid qualitative and quantitative predictions for the
potentially rich structure of the QCD phase diagram and the associated physics of strongly interacting 
matter. 

In this Chapter to the \Enn{} we provided a pedagogical overview to the functional approach to QCD at finite 
temperature and chemical potential. 
The interested reader will find more information on functional methods in two corresponding Chapters of the 
\Enp, \cite{Huber:2025cbd,Eichmann:2025wgs}, as well as a number of renowned and recent review articles, partially or fully devoted to the functional approach to QCD, see e.g.
\cite{Roberts:1994dr, Litim:1998nf, Berges:2000ew, Alkofer:2000wg, Roberts:2000aa, Schaefer:2004en, Fischer:2006ub, Gies:2006wv, Braun:2011pp, Pawlowski:2014aha, Eichmann:2016yit, Fischer:2018sdj,  Huber:2025cbd, Dupuis:2020fhh, Fu:2022gou, Rennecke:2025bcw, Fischer:2026uni}. 
These contain discussions and references to many more results than we could discuss in the available space here.

In general we have seen that highly advanced contemporary truncation schemes allows us to address 
the physics of the Columbia plot and its extensions to real and imaginary chemical potential as well as
the structure of the QCD phase diagram at realistic quark masses in a systematic and quantitative manner.
The approach has been thoroughly cross-checked with results from lattice QCD in regions of the QCD parameter 
space, where direct lattice calculations are possible and naturally extends the reach of first-principles methods
into the realm of large baryon chemical potential. The combined evidence from the functional approach,
lattice QCD extrapolations based on Lee-Yang zeros and other methods such holographic QCD signal the 
appearance of a potential critical end point in this region. The properties and maybe even very existence of 
this CEP is potentially called into question by the possibility of moat regimes and more complicated phases
such as the ones featuring inhomogeneous condensates. An important task for the next years is to intensify 
contact with experimental heavy ion physics in all possible respects. Important interfaces are the calculation 
of fluctuations and ratios thereof, the calculation of thermodynamic quantities and transport coefficients and 
the access to static correlators and spectral functions of light and heavy mesons.

\begin{ack}[Acknowledgments]
{}We thank Julian Bernhardt, Szabolcs Borsanyi, Jens Braun, Michael Buballa, Yong-rui Chen, Wei-jie Fu, Kenji Fukushima, Fei Gao, Leonid Glozman, Jana Guenther, Pascal Gunkel, Chuang Huang, Friederike Ihssen, Philipp Isserstedt, Keiwan Jamaly, Yi Lu, Frithjof Karsch, Volker Koch, Konrad Kockler, Xiaofeng Luo, Larry McLerran, Theo Motta, Jorge Noronha, Owe Philipsen, Rob Pisarski, Fabian Rennecke, Dirk Rischke, Franz Sattler, Bernd-Jochen Schaefer, Lorenz von Smekal, Misha Stephanov, Yang-yang Tan, Shi Yin, Rui Wen, Nicolas Wink, Nu Xu and the members of the QCD collaboration \cite{fQCD} for discussions and work related to topics of this review. 

This work is funded by the Deutsche Forschungsgemeinschaft (DFG, German Research Foundation) under Germany’s Excellence Strategy EXC 2181/1 - 390900948 (the Heidelberg STRUCTURES Excellence Cluster) and the Collaborative Research Centre SFB 1225 - 273811115 (ISOQUANT) as well as the Collaborative Research Centre TransRegio CRC-TR 211 “Strong-interaction matter under extreme conditions” and the individual grant FI 970/16-1.
\end{ack}


\bibliographystyle{Harvard}
\bibliography{../ref-lib.bib}

\begin{thebibliography*}{153}
\providecommand{\bibtype}[1]{}
\providecommand{\natexlab}[1]{#1}
{\catcode`\|=0\catcode`\#=12\catcode`\@=11\catcode`\\=12
|immediate|write|@auxout{\expandafter\ifx\csname
  natexlab\endcsname\relax\gdef\natexlab#1{#1}\fi}}
\renewcommand{\url}[1]{{\tt #1}}
\providecommand{\urlprefix}{URL }
\expandafter\ifx\csname urlstyle\endcsname\relax
  \providecommand{\doi}[1]{doi:\discretionary{}{}{}#1}\else
  \providecommand{\doi}{doi:\discretionary{}{}{}\begingroup
  \urlstyle{rm}\Url}\fi
\providecommand{\bibinfo}[2]{#2}
\providecommand{\eprint}[2][]{\url{#2}}

\bibtype{Article}%
\bibitem[Abdallah et al.(2021)]{STAR:2021iop}
\bibinfo{author}{Abdallah M} and  et al. (\bibinfo{collaboration}{STAR})
  (\bibinfo{year}{2021}).
\bibinfo{title}{{Cumulants and correlation functions of net-proton, proton, and
  antiproton multiplicity distributions in Au+Au collisions at energies
  available at the BNL Relativistic Heavy Ion Collider}}.
\bibinfo{journal}{{\em Phys. Rev. C}} \bibinfo{volume}{104}
  (\bibinfo{number}{2}): \bibinfo{pages}{024902}.
  \bibinfo{doi}{\doi{10.1103/PhysRevC.104.024902}}.
\bibinfo{note}{[Erratum: Phys.Rev.C 111, 029902 (2025)]}, \eprint{2101.12413}.

\bibtype{Article}%
\bibitem[Aboona et al.(2025)]{STAR:2025zdq}
\bibinfo{author}{Aboona BE} and  et al. (\bibinfo{collaboration}{STAR})
  (\bibinfo{year}{2025}).
\bibinfo{title}{{Precision Measurement of Net-Proton-Number Fluctuations in
  Au+Au Collisions at RHIC}}.
\bibinfo{journal}{{\em Phys. Rev. Lett.}} \bibinfo{volume}{135}
  (\bibinfo{number}{14}): \bibinfo{pages}{142301}.
  \bibinfo{doi}{\doi{10.1103/9l69-2d7p}}.
\eprint{2504.00817}.

\bibtype{Article}%
\bibitem[Adamczewski-Musch et al.(2020)]{HADES:2020wpc}
\bibinfo{author}{Adamczewski-Musch J} and  et al.
  (\bibinfo{collaboration}{HADES}) (\bibinfo{year}{2020}).
\bibinfo{title}{{Proton-number fluctuations in $\sqrt {s_{NN}}$ =2.4 GeV Au +
  Au collisions studied with the High-Acceptance DiElectron Spectrometer
  (HADES)}}.
\bibinfo{journal}{{\em Phys. Rev. C}} \bibinfo{volume}{102}
  (\bibinfo{number}{2}): \bibinfo{pages}{024914}.
  \bibinfo{doi}{\doi{10.1103/PhysRevC.102.024914}}.
\eprint{2002.08701}.

\bibtype{Article}%
\bibitem[Aliberti et al.(2025)]{Aliberti:2025beg}
\bibinfo{author}{Aliberti R} and  et al. (\bibinfo{year}{2025}).
\bibinfo{title}{{The anomalous magnetic moment of the muon in the Standard
  Model: an update}}.
\bibinfo{journal}{{\em Phys. Rept.}} \bibinfo{volume}{1143}:
  \bibinfo{pages}{1--158}. \bibinfo{doi}{\doi{10.1016/j.physrep.2025.08.002}}.
\eprint{2505.21476}.

\bibtype{Article}%
\bibitem[Alkofer and Greensite(2007)]{Alkofer:2006fu}
\bibinfo{author}{Alkofer R} and  \bibinfo{author}{Greensite J}
  (\bibinfo{year}{2007}).
\bibinfo{title}{Quark confinement: The hard problem of hadron physics}.
\bibinfo{journal}{{\em J. Phys. G}} \bibinfo{volume}{34}: \bibinfo{pages}{S3}.
  \bibinfo{doi}{\doi{10.1088/0954-3899/34/7/S02}}.
\eprint{hep-ph/0610365}.

\bibtype{Article}%
\bibitem[Alkofer and von Smekal(2001)]{Alkofer:2000wg}
\bibinfo{author}{Alkofer R} and  \bibinfo{author}{von Smekal L}
  (\bibinfo{year}{2001}).
\bibinfo{title}{{The Infrared behavior of QCD Green's functions: Confinement
  dynamical symmetry breaking, and hadrons as relativistic bound states}}.
\bibinfo{journal}{{\em Phys. Rept.}} \bibinfo{volume}{353}:
  \bibinfo{pages}{281}. \bibinfo{doi}{\doi{10.1016/S0370-1573(01)00010-2}}.
\eprint{hep-ph/0007355}.

\bibtype{Article}%
\bibitem[Alkofer et al.(2010)]{Alkofer:2008jy}
\bibinfo{author}{Alkofer R}, \bibinfo{author}{Huber MQ} and
  \bibinfo{author}{Schwenzer K} (\bibinfo{year}{2010}).
\bibinfo{title}{{Infrared singularities in Landau gauge Yang-Mills theory}}.
\bibinfo{journal}{{\em Phys. Rev. D}} \bibinfo{volume}{81}:
  \bibinfo{pages}{105010}. \bibinfo{doi}{\doi{10.1103/PhysRevD.81.105010}}.
\eprint{0801.2762}.

\bibtype{Article}%
\bibitem[Andronic et al.(2018)]{Andronic:2017pug}
\bibinfo{author}{Andronic A}, \bibinfo{author}{Braun-Munzinger P},
  \bibinfo{author}{Redlich K} and  \bibinfo{author}{Stachel J}
  (\bibinfo{year}{2018}).
\bibinfo{title}{{Decoding the phase structure of QCD via particle production at
  high energy}}.
\bibinfo{journal}{{\em Nature}} \bibinfo{volume}{561} (\bibinfo{number}{7723}):
  \bibinfo{pages}{321--330}. \bibinfo{doi}{\doi{10.1038/s41586-018-0491-6}}.
\eprint{1710.09425}.

\bibtype{Article}%
\bibitem[Aoki et al.(2006)]{Aoki:2006we}
\bibinfo{author}{Aoki Y}, \bibinfo{author}{Endrodi G}, \bibinfo{author}{Fodor
  Z}, \bibinfo{author}{Katz SD} and  \bibinfo{author}{Szabo KK}
  (\bibinfo{year}{2006}).
\bibinfo{title}{{The Order of the quantum chromodynamics transition predicted
  by the standard model of particle physics}}.
\bibinfo{journal}{{\em Nature}} \bibinfo{volume}{443}:
  \bibinfo{pages}{675--678}. \bibinfo{doi}{\doi{10.1038/nature05120}}.
\eprint{hep-lat/0611014}.

\bibtype{Article}%
\bibitem[Aoyama et al.(2020)]{Aoyama:2020ynm}
\bibinfo{author}{Aoyama T} and  et al. (\bibinfo{year}{2020}).
\bibinfo{title}{{The anomalous magnetic moment of the muon in the Standard
  Model}}.
\bibinfo{journal}{{\em Phys. Rept.}} \bibinfo{volume}{887}:
  \bibinfo{pages}{1--166}. \bibinfo{doi}{\doi{10.1016/j.physrep.2020.07.006}}.
\eprint{2006.04822}.

\bibtype{Article}%
\bibitem[Asakawa et al.(2000)]{Asakawa:2000wh}
\bibinfo{author}{Asakawa M}, \bibinfo{author}{Heinz UW} and
  \bibinfo{author}{Muller B} (\bibinfo{year}{2000}).
\bibinfo{title}{Fluctuation probes of quark deconfinement}.
\bibinfo{journal}{{\em Phys. Rev. Lett.}} \bibinfo{volume}{85}:
  \bibinfo{pages}{2072--2075}.
  \bibinfo{doi}{\doi{10.1103/PhysRevLett.85.2072}}.
\eprint{hep-ph/0003169}.

\bibtype{Article}%
\bibitem[Bazavov et al.(2012)]{Bazavov:2011nk}
\bibinfo{author}{Bazavov A} and  et al. (\bibinfo{year}{2012}).
\bibinfo{title}{{The chiral and deconfinement aspects of the QCD transition}}.
\bibinfo{journal}{{\em Phys. Rev. D}} \bibinfo{volume}{85}:
  \bibinfo{pages}{054503}. \bibinfo{doi}{\doi{10.1103/PhysRevD.85.054503}}.
\eprint{1111.1710}.

\bibtype{Article}%
\bibitem[Bazavov et al.(2019)]{Bazavov:2018mes}
\bibinfo{author}{Bazavov A} and  et al. (\bibinfo{collaboration}{HotQCD})
  (\bibinfo{year}{2019}).
\bibinfo{title}{{Chiral crossover in QCD at zero and non-zero chemical
  potentials}}.
\bibinfo{journal}{{\em Phys. Lett.}} \bibinfo{volume}{B795}:
  \bibinfo{pages}{15--21}. \bibinfo{doi}{\doi{10.1016/j.physletb.2019.05.013}}.
\eprint{1812.08235}.

\bibtype{Article}%
\bibitem[Berges(2004)]{Berges:2004yj}
\bibinfo{author}{Berges J} (\bibinfo{year}{2004}).
\bibinfo{title}{{Introduction to nonequilibrium quantum field theory}}.
\bibinfo{journal}{{\em AIP Conf. Proc.}} \bibinfo{volume}{739}
  (\bibinfo{number}{1}): \bibinfo{pages}{3--62}.
  \bibinfo{doi}{\doi{10.1063/1.1843591}}.
\eprint{hep-ph/0409233}.

\bibtype{Article}%
\bibitem[Berges and Hoffmeister(2009)]{Berges:2008sr}
\bibinfo{author}{Berges J} and  \bibinfo{author}{Hoffmeister G}
  (\bibinfo{year}{2009}).
\bibinfo{title}{{Nonthermal fixed points and the functional renormalization
  group}}.
\bibinfo{journal}{{\em Nucl. Phys. B}} \bibinfo{volume}{813}:
  \bibinfo{pages}{383--407}.
  \bibinfo{doi}{\doi{10.1016/j.nuclphysb.2008.12.017}}.
\eprint{0809.5208}.

\bibtype{Article}%
\bibitem[Berges et al.(2002)]{Berges:2000ew}
\bibinfo{author}{Berges J}, \bibinfo{author}{Tetradis N} and
  \bibinfo{author}{Wetterich C} (\bibinfo{year}{2002}).
\bibinfo{title}{{Nonperturbative renormalization flow in quantum field theory
  and statistical physics}}.
\bibinfo{journal}{{\em Phys. Rept.}} \bibinfo{volume}{363}:
  \bibinfo{pages}{223--386}.
  \bibinfo{doi}{\doi{10.1016/S0370-1573(01)00098-9}}.
\eprint{hep-ph/0005122}.

\bibtype{Article}%
\bibitem[Bernhardt and Fischer(2023{\natexlab{a}})]{Bernhardt:2023ezo}
\bibinfo{author}{Bernhardt J} and  \bibinfo{author}{Fischer CS}
  (\bibinfo{year}{2023}{\natexlab{a}}).
\bibinfo{title}{{From imaginary to real chemical potential QCD with functional
  methods}}.
\bibinfo{journal}{{\em Eur. Phys. J. A}} \bibinfo{volume}{59}
  (\bibinfo{number}{8}): \bibinfo{pages}{181}.
  \bibinfo{doi}{\doi{10.1140/epja/s10050-023-01098-1}}.
\eprint{2305.01434}.

\bibtype{Article}%
\bibitem[Bernhardt and Fischer(2023{\natexlab{b}})]{Bernhardt:2023hpr}
\bibinfo{author}{Bernhardt J} and  \bibinfo{author}{Fischer CS}
  (\bibinfo{year}{2023}{\natexlab{b}}).
\bibinfo{title}{Qcd phase transitions in the light quark chiral limit}.
\bibinfo{journal}{{\em Phys. Rev. D}} \bibinfo{volume}{108}
  (\bibinfo{number}{11}): \bibinfo{pages}{114018}.
  \bibinfo{doi}{\doi{10.1103/PhysRevD.108.114018}}.
\eprint{2309.06737}.

\bibtype{Article}%
\bibitem[Bernhardt and Fischer(2025)]{Bernhardt:2025fvk}
\bibinfo{author}{Bernhardt J} and  \bibinfo{author}{Fischer CS}
  (\bibinfo{year}{2025}), \bibinfo{month}{7}.
\bibinfo{title}{{Quark mass dependence of a QCD critical point and structure of
  the Columbia plot}} \eprint{2507.21680}.

\bibtype{Article}%
\bibitem[Bernhardt et al.(2021)]{Bernhardt:2021iql}
\bibinfo{author}{Bernhardt J}, \bibinfo{author}{Fischer CS},
  \bibinfo{author}{Isserstedt P} and  \bibinfo{author}{Schaefer BJ}
  (\bibinfo{year}{2021}).
\bibinfo{title}{{Critical endpoint of QCD in a finite volume}}.
\bibinfo{journal}{{\em Phys. Rev. D}} \bibinfo{volume}{104}
  (\bibinfo{number}{7}): \bibinfo{pages}{074035}.
  \bibinfo{doi}{\doi{10.1103/PhysRevD.104.074035}}.
\eprint{2107.05504}.

\bibtype{Article}%
\bibitem[Bhattacharya et al.(2014)]{Bhattacharya:2014ara}
\bibinfo{author}{Bhattacharya T} and  et al. (\bibinfo{year}{2014}).
\bibinfo{title}{{QCD Phase Transition with Chiral Quarks and Physical Quark
  Masses}}.
\bibinfo{journal}{{\em Phys. Rev. Lett.}} \bibinfo{volume}{113}
  (\bibinfo{number}{8}): \bibinfo{pages}{082001}.
  \bibinfo{doi}{\doi{10.1103/PhysRevLett.113.082001}}.
\eprint{1402.5175}.

\bibtype{Article}%
\bibitem[Bluhm et al.(2019)]{Bluhm:2018qkf}
\bibinfo{author}{Bluhm M}, \bibinfo{author}{Jiang Y}, \bibinfo{author}{Nahrgang
  M}, \bibinfo{author}{Pawlowski JM}, \bibinfo{author}{Rennecke F} and
  \bibinfo{author}{Wink N} (\bibinfo{year}{2019}).
\bibinfo{title}{{Time-evolution of fluctuations as signal of the phase
  transition dynamics in a QCD-assisted transport approach}}.
\bibinfo{journal}{{\em Nucl. Phys.}} \bibinfo{volume}{A982}:
  \bibinfo{pages}{871--874}.
  \bibinfo{doi}{\doi{10.1016/j.nuclphysa.2018.09.058}}.
\eprint{1808.01377}.

\bibtype{Article}%
\bibitem[Borsanyi et al.(2010)]{Borsanyi:2010bp}
\bibinfo{author}{Borsanyi S}, \bibinfo{author}{Fodor Z},
  \bibinfo{author}{Hoelbling C}, \bibinfo{author}{Katz SD},
  \bibinfo{author}{Krieg S}, \bibinfo{author}{Ratti C} and
  \bibinfo{author}{Szabo KK} (\bibinfo{collaboration}{Wuppertal-Budapest})
  (\bibinfo{year}{2010}).
\bibinfo{title}{{Is there still any $T_c$ mystery in lattice QCD? Results with
  physical masses in the continuum limit III}}.
\bibinfo{journal}{{\em JHEP}} \bibinfo{volume}{09}: \bibinfo{pages}{073}.
  \bibinfo{doi}{\doi{10.1007/JHEP09(2010)073}}.
\eprint{1005.3508}.

\bibtype{Article}%
\bibitem[Borsanyi et al.(2020)]{Borsanyi:2020fev}
\bibinfo{author}{Borsanyi S}, \bibinfo{author}{Fodor Z},
  \bibinfo{author}{Guenther JN}, \bibinfo{author}{Kara R},
  \bibinfo{author}{Katz SD}, \bibinfo{author}{Parotto P},
  \bibinfo{author}{Pasztor A}, \bibinfo{author}{Ratti C} and
  \bibinfo{author}{Szabo KK} (\bibinfo{year}{2020}).
\bibinfo{title}{{The QCD crossover at finite chemical potential from lattice
  simulations}}.
\bibinfo{journal}{{\em Phys. Rev. Lett.}} \bibinfo{volume}{125}
  (\bibinfo{number}{5}): \bibinfo{pages}{052001}.
  \bibinfo{doi}{\doi{10.1103/PhysRevLett.125.052001}}.
\eprint{2002.02821}.

\bibtype{Article}%
\bibitem[Braun(2012)]{Braun:2011pp}
\bibinfo{author}{Braun J} (\bibinfo{year}{2012}).
\bibinfo{title}{{Fermion Interactions and Universal Behavior in Strongly
  Interacting Theories}}.
\bibinfo{journal}{{\em J. Phys.}} \bibinfo{volume}{G39}:
  \bibinfo{pages}{033001}. \bibinfo{doi}{\doi{10.1088/0954-3899/39/3/033001}}.
\eprint{1108.4449}.

\bibtype{Article}%
\bibitem[Braun et al.(2010{\natexlab{a}})]{Braun:2010cy}
\bibinfo{author}{Braun J}, \bibinfo{author}{Eichhorn A}, \bibinfo{author}{Gies
  H} and  \bibinfo{author}{Pawlowski JM} (\bibinfo{year}{2010}{\natexlab{a}}).
\bibinfo{title}{{On the Nature of the Phase Transition in SU(N), Sp(2) and E(7)
  Yang-Mills theory}}.
\bibinfo{journal}{{\em Eur.Phys.J.}} \bibinfo{volume}{C70}:
  \bibinfo{pages}{689--702}.
  \bibinfo{doi}{\doi{10.1140/epjc/s10052-010-1485-1}}.
\eprint{1007.2619}.

\bibtype{Article}%
\bibitem[Braun et al.(2010{\natexlab{b}})]{Braun:2007bx}
\bibinfo{author}{Braun J}, \bibinfo{author}{Gies H} and
  \bibinfo{author}{Pawlowski JM} (\bibinfo{year}{2010}{\natexlab{b}}).
\bibinfo{title}{{Quark Confinement from Color Confinement}}.
\bibinfo{journal}{{\em Phys.Lett.}} \bibinfo{volume}{B684}:
  \bibinfo{pages}{262--267}.
  \bibinfo{doi}{\doi{10.1016/j.physletb.2010.01.009}}.
\eprint{0708.2413}.

\bibtype{Article}%
\bibitem[Braun et al.(2011)]{Braun:2009gm}
\bibinfo{author}{Braun J}, \bibinfo{author}{Haas LM},
  \bibinfo{author}{Marhauser F} and  \bibinfo{author}{Pawlowski JM}
  (\bibinfo{year}{2011}).
\bibinfo{title}{{Phase Structure of Two-Flavor QCD at Finite Chemical
  Potential}}.
\bibinfo{journal}{{\em Phys. Rev. Lett.}} \bibinfo{volume}{106}:
  \bibinfo{pages}{022002}. \bibinfo{doi}{\doi{10.1103/PhysRevLett.106.022002}}.
\eprint{0908.0008}.

\bibtype{Article}%
\bibitem[Braun et al.(2020)]{Braun:2020ada}
\bibinfo{author}{Braun J}, \bibinfo{author}{Fu Wj}, \bibinfo{author}{Pawlowski
  JM}, \bibinfo{author}{Rennecke F}, \bibinfo{author}{Rosenbl\"uh D} and
  \bibinfo{author}{Yin S} (\bibinfo{year}{2020}).
\bibinfo{title}{{Chiral susceptibility in ( 2+1 )-flavor QCD}}.
\bibinfo{journal}{{\em Phys. Rev. D}} \bibinfo{volume}{102}
  (\bibinfo{number}{5}): \bibinfo{pages}{056010}.
  \bibinfo{doi}{\doi{10.1103/PhysRevD.102.056010}}.
\eprint{2003.13112}.

\bibtype{Article}%
\bibitem[Braun et al.(2023)]{Braun:2022mgx}
\bibinfo{author}{Braun J} and  et al. (\bibinfo{year}{2023}).
\bibinfo{title}{{Renormalised spectral flows}}.
\bibinfo{journal}{{\em SciPost Phys. Core}} \bibinfo{volume}{6}:
  \bibinfo{pages}{061}. \bibinfo{doi}{\doi{10.21468/SciPostPhysCore.6.3.061}}.
\eprint{2206.10232}.

\bibtype{Article}%
\bibitem[Braun et al.(2025)]{Braun:2023qak}
\bibinfo{author}{Braun J} and  et al. (\bibinfo{year}{2025}).
\bibinfo{title}{{Soft modes in hot QCD matter}}.
\bibinfo{journal}{{\em Phys. Rev. D}} \bibinfo{volume}{111}
  (\bibinfo{number}{9}): \bibinfo{pages}{094010}.
  \bibinfo{doi}{\doi{10.1103/PhysRevD.111.094010}}.
\eprint{2310.19853}.

\bibtype{Misc}%
\bibitem[Braun et al.(2026)]{fQCD}
\bibinfo{author}{Braun J}, \bibinfo{author}{Chen Yr}, \bibinfo{author}{Fu Wj},
  \bibinfo{author}{Gao F}, \bibinfo{author}{Huang C}, \bibinfo{author}{Ihssen
  F}, \bibinfo{author}{Kockler K}, \bibinfo{author}{Lu Y},
  \bibinfo{author}{Pawlowski JM}, \bibinfo{author}{Rennecke F},
  \bibinfo{author}{Sattler FR}, \bibinfo{author}{Stoll J}, \bibinfo{author}{Tan
  Yy}, \bibinfo{author}{Wang Zn}, \bibinfo{author}{Wen R},
  \bibinfo{author}{Wessely J}, \bibinfo{author}{Yin S}, \bibinfo{author}{Zheng
  Hw} and  \bibinfo{author}{Zorbach N} (\bibinfo{year}{2026}).
\bibinfo{title}{fqcd collaboration}.
\bibinfo{note}{\url{https://fqcd-collaboration.github.io/}}.

\bibtype{Article}%
\bibitem[Buballa and Carignano(2015)]{Buballa:2014tba}
\bibinfo{author}{Buballa M} and  \bibinfo{author}{Carignano S}
  (\bibinfo{year}{2015}).
\bibinfo{title}{{Inhomogeneous chiral condensates}}.
\bibinfo{journal}{{\em Prog. Part. Nucl. Phys.}} \bibinfo{volume}{81}:
  \bibinfo{pages}{39--96}. \bibinfo{doi}{\doi{10.1016/j.ppnp.2014.11.001}}.
\eprint{1406.1367}.

\bibtype{Article}%
\bibitem[Chen et al.(2025)]{Chen:2024lzz}
\bibinfo{author}{Chen Yr}, \bibinfo{author}{Tan Yy} and  \bibinfo{author}{Fu
  Wj} (\bibinfo{year}{2025}).
\bibinfo{title}{{Critical dynamics of model H within the real-time FRG
  approach}}.
\bibinfo{journal}{{\em Phys. Rev. D}} \bibinfo{volume}{111}
  (\bibinfo{number}{9}): \bibinfo{pages}{094025}.
  \bibinfo{doi}{\doi{10.1103/PhysRevD.111.094025}}.
\eprint{2406.00679}.

\bibtype{Article}%
\bibitem[Christiansen et al.(2015)]{Christiansen:2014ypa}
\bibinfo{author}{Christiansen N}, \bibinfo{author}{Haas M},
  \bibinfo{author}{Pawlowski JM} and  \bibinfo{author}{Strodthoff N}
  (\bibinfo{year}{2015}).
\bibinfo{title}{{Transport Coefficients in Yang--Mills Theory and QCD}}.
\bibinfo{journal}{{\em Phys. Rev. Lett.}} \bibinfo{volume}{115}
  (\bibinfo{number}{11}): \bibinfo{pages}{112002}.
  \bibinfo{doi}{\doi{10.1103/PhysRevLett.115.112002}}.
\eprint{1411.7986}.

\bibtype{Article}%
\bibitem[Connelly et al.(2020)]{Connelly:2020gwa}
\bibinfo{author}{Connelly A}, \bibinfo{author}{Johnson G},
  \bibinfo{author}{Rennecke F} and  \bibinfo{author}{Skokov V}
  (\bibinfo{year}{2020}).
\bibinfo{title}{{Universal Location of the Yang-Lee Edge Singularity in $O(N)$
  Theories}}.
\bibinfo{journal}{{\em Phys. Rev. Lett.}} \bibinfo{volume}{125}
  (\bibinfo{number}{19}): \bibinfo{pages}{191602}.
  \bibinfo{doi}{\doi{10.1103/PhysRevLett.125.191602}}.
\eprint{2006.12541}.

\bibtype{Article}%
\bibitem[Cuteri et al.(2021{\natexlab{a}})]{Cuteri:2020yke}
\bibinfo{author}{Cuteri F}, \bibinfo{author}{Philipsen O},
  \bibinfo{author}{Sch{\"o}n A} and  \bibinfo{author}{Sciarra A}
  (\bibinfo{year}{2021}{\natexlab{a}}).
\bibinfo{title}{Deconfinement critical point of lattice qcd with $n_f=2$ wilson
  fermions}.
\bibinfo{journal}{{\em Phys. Rev. D}} \bibinfo{volume}{103}
  (\bibinfo{number}{1}): \bibinfo{pages}{014513}.
  \bibinfo{doi}{\doi{10.1103/PhysRevD.103.014513}}.
\eprint{2009.14033}.

\bibtype{Article}%
\bibitem[Cuteri et al.(2021{\natexlab{b}})]{Cuteri:2021ikv}
\bibinfo{author}{Cuteri F}, \bibinfo{author}{Philipsen O} and
  \bibinfo{author}{Sciarra A} (\bibinfo{year}{2021}{\natexlab{b}}).
\bibinfo{title}{On the order of the qcd chiral phase transition for different
  numbers of quark flavours}.
\bibinfo{journal}{{\em JHEP}} \bibinfo{volume}{11}: \bibinfo{pages}{141}.
  \bibinfo{doi}{\doi{10.1007/JHEP11(2021)141}}.
\eprint{2107.12739}.

\bibtype{Article}%
\bibitem[Cuteri et al.(2022)]{Cuteri:2022vwk}
\bibinfo{author}{Cuteri F}, \bibinfo{author}{Goswami J},
  \bibinfo{author}{Karsch F}, \bibinfo{author}{Lahiri A},
  \bibinfo{author}{Neumann M}, \bibinfo{author}{Philipsen O},
  \bibinfo{author}{Schmidt C} and  \bibinfo{author}{Sciarra A}
  (\bibinfo{year}{2022}).
\bibinfo{title}{Toward the chiral phase transition in the roberge-weiss plane}.
\bibinfo{journal}{{\em Phys. Rev. D}} \bibinfo{volume}{106}
  (\bibinfo{number}{1}): \bibinfo{pages}{014510}.
  \bibinfo{doi}{\doi{10.1103/PhysRevD.106.014510}}.
\eprint{2205.12707}.

\bibtype{Article}%
\bibitem[D'Ambrosio et al.(2023)]{DAmbrosio:2022kig}
\bibinfo{author}{D'Ambrosio A}, \bibinfo{author}{Philipsen O} and
  \bibinfo{author}{Kaiser R} (\bibinfo{year}{2023}).
\bibinfo{title}{The chiral phase transition at non-zero imaginary baryon
  chemical potential for different numbers of quark flavours}.
\bibinfo{journal}{{\em PoS}} \bibinfo{volume}{LATTICE2022}:
  \bibinfo{pages}{172}. \bibinfo{doi}{\doi{10.22323/1.430.0172}}.
\eprint{2212.03655}.

\bibtype{Article}%
\bibitem[D'Ambrosio et al.(2025)]{DAmbrosio:2025ldv}
\bibinfo{author}{D'Ambrosio A}, \bibinfo{author}{Fromm M},
  \bibinfo{author}{Kaiser R} and  \bibinfo{author}{Philipsen O}
  (\bibinfo{year}{2025}), \bibinfo{month}{12}.
\bibinfo{title}{{On the nature of the QCD chiral phase transition with
  imaginary chemical potential}} \eprint{2512.15418}.

\bibtype{Article}%
\bibitem[de~Forcrand and Philipsen(2002)]{deForcrand:2002hgr}
\bibinfo{author}{de~Forcrand P} and  \bibinfo{author}{Philipsen O}
  (\bibinfo{year}{2002}).
\bibinfo{title}{{The QCD phase diagram for small densities from imaginary
  chemical potential}}.
\bibinfo{journal}{{\em Nucl. Phys.}} \bibinfo{volume}{B642}:
  \bibinfo{pages}{290--306}.
  \bibinfo{doi}{\doi{10.1016/S0550-3213(02)00626-0}}.
\eprint{hep-lat/0205016}.

\bibtype{Article}%
\bibitem[de~Forcrand and Philipsen(2010)]{deForcrand:2010he}
\bibinfo{author}{de~Forcrand P} and  \bibinfo{author}{Philipsen O}
  (\bibinfo{year}{2010}).
\bibinfo{title}{Constraining the qcd phase diagram by tricritical lines at
  imaginary chemical potential}.
\bibinfo{journal}{{\em Phys. Rev. Lett.}} \bibinfo{volume}{105}:
  \bibinfo{pages}{152001}. \bibinfo{doi}{\doi{10.1103/PhysRevLett.105.152001}}.
\eprint{1004.3144}.

\bibtype{Article}%
\bibitem[Ding et al.(2019)]{HotQCD:2019xnw}
\bibinfo{author}{Ding HT} and  et al. (\bibinfo{collaboration}{HotQCD})
  (\bibinfo{year}{2019}).
\bibinfo{title}{Chiral phase transition temperature in ( 2+1 )-flavor qcd}.
\bibinfo{journal}{{\em Phys. Rev. Lett.}} \bibinfo{volume}{123}
  (\bibinfo{number}{6}): \bibinfo{pages}{062002}.
  \bibinfo{doi}{\doi{10.1103/PhysRevLett.123.062002}}.
\eprint{1903.04801}.

\bibtype{Article}%
\bibitem[Ding et al.(2024)]{Ding:2024sux}
\bibinfo{author}{Ding HT}, \bibinfo{author}{Kaczmarek O},
  \bibinfo{author}{Karsch F}, \bibinfo{author}{Petreczky P},
  \bibinfo{author}{Sarkar M}, \bibinfo{author}{Schmidt C} and
  \bibinfo{author}{Sharma S} (\bibinfo{year}{2024}).
\bibinfo{title}{{Curvature of the chiral phase transition line from the
  magnetic equation of state of (2+1)-flavor QCD}}.
\bibinfo{journal}{{\em Phys. Rev. D}} \bibinfo{volume}{109}
  (\bibinfo{number}{11}): \bibinfo{pages}{114516}.
  \bibinfo{doi}{\doi{10.1103/PhysRevD.109.114516}}.
\eprint{2403.09390}.

\bibtype{Article}%
\bibitem[Dudal and Vercauteren(2023)]{Dudal:2023nbt}
\bibinfo{author}{Dudal D} and  \bibinfo{author}{Vercauteren D}
  (\bibinfo{year}{2023}).
\bibinfo{title}{{Gap equations of background field invariant refined
  Gribov-Zwanziger action proposals and the deconfinement transition}}.
\bibinfo{journal}{{\em Phys. Rev. D}} \bibinfo{volume}{107}
  (\bibinfo{number}{7}): \bibinfo{pages}{074020}.
  \bibinfo{doi}{\doi{10.1103/PhysRevD.107.074020}}.
\eprint{2302.03230}.

\bibtype{Article}%
\bibitem[Dupuis et al.(2021)]{Dupuis:2020fhh}
\bibinfo{author}{Dupuis N}, \bibinfo{author}{Canet L},
  \bibinfo{author}{Eichhorn A}, \bibinfo{author}{Metzner W},
  \bibinfo{author}{Pawlowski JM}, \bibinfo{author}{Tissier M} and
  \bibinfo{author}{Wschebor N} (\bibinfo{year}{2021}).
\bibinfo{title}{{The nonperturbative functional renormalization group and its
  applications}}.
\bibinfo{journal}{{\em Phys. Rept.}} \bibinfo{volume}{910}:
  \bibinfo{pages}{1--114}. \bibinfo{doi}{\doi{10.1016/j.physrep.2021.01.001}}.
\eprint{2006.04853}.

\bibtype{Article}%
\bibitem[Eichmann(2025)]{Eichmann:2025wgs}
\bibinfo{author}{Eichmann G} (\bibinfo{year}{2025}), \bibinfo{month}{3}.
\bibinfo{title}{{Hadron physics with functional methods}} \eprint{2503.10397}.

\bibtype{Article}%
\bibitem[Eichmann et al.(2016)]{Eichmann:2016yit}
\bibinfo{author}{Eichmann G}, \bibinfo{author}{Sanchis-Alepuz H},
  \bibinfo{author}{Williams R}, \bibinfo{author}{Alkofer R} and
  \bibinfo{author}{Fischer CS} (\bibinfo{year}{2016}).
\bibinfo{title}{{Baryons as relativistic three-quark bound states}}.
\bibinfo{journal}{{\em Prog. Part. Nucl. Phys.}} \bibinfo{volume}{91}:
  \bibinfo{pages}{1--100}. \bibinfo{doi}{\doi{10.1016/j.ppnp.2016.07.001}}.
\eprint{1606.09602}.

\bibtype{Article}%
\bibitem[Ejiri et al.(2006)]{Ejiri:2005wq}
\bibinfo{author}{Ejiri S}, \bibinfo{author}{Karsch F} and
  \bibinfo{author}{Redlich K} (\bibinfo{year}{2006}).
\bibinfo{title}{Hadronic fluctuations at the qcd phase transition}.
\bibinfo{journal}{{\em Phys. Lett. B}} \bibinfo{volume}{633}:
  \bibinfo{pages}{275--282}.
  \bibinfo{doi}{\doi{10.1016/j.physletb.2005.11.083}}.
\eprint{hep-ph/0509051}.

\bibtype{Article}%
\bibitem[Ejiri et al.(2020)]{Ejiri:2019csa}
\bibinfo{author}{Ejiri S}, \bibinfo{author}{Itagaki S}, \bibinfo{author}{Iwami
  R}, \bibinfo{author}{Kanaya K}, \bibinfo{author}{Kitazawa M},
  \bibinfo{author}{Kiyohara A}, \bibinfo{author}{Shirogane M} and
  \bibinfo{author}{Umeda T} (\bibinfo{collaboration}{WHOT-QCD})
  (\bibinfo{year}{2020}).
\bibinfo{title}{End point of the first-order phase transition of qcd in the
  heavy quark region by reweighting from quenched qcd}.
\bibinfo{journal}{{\em Phys. Rev. D}} \bibinfo{volume}{101}
  (\bibinfo{number}{5}): \bibinfo{pages}{054505}.
  \bibinfo{doi}{\doi{10.1103/PhysRevD.101.054505}}.
\eprint{1912.10500}.

\bibtype{Article}%
\bibitem[Fejos(2022)]{Fejos:2022mso}
\bibinfo{author}{Fejos G} (\bibinfo{year}{2022}).
\bibinfo{title}{Second-order chiral phase transition in three-flavor quantum
  chromodynamics?}
\bibinfo{journal}{{\em Phys. Rev. D}} \bibinfo{volume}{105}
  (\bibinfo{number}{7}): \bibinfo{pages}{L071506}.
  \bibinfo{doi}{\doi{10.1103/PhysRevD.105.L071506}}.
\eprint{2201.07909}.

\bibtype{Article}%
\bibitem[Fejos and Hatsuda(2024)]{Fejos:2024bgl}
\bibinfo{author}{Fejos G} and  \bibinfo{author}{Hatsuda T}
  (\bibinfo{year}{2024}).
\bibinfo{title}{{Order of the SU(Nf){\texttimes}SU(Nf) chiral transition via
  the functional renormalization group}}.
\bibinfo{journal}{{\em Phys. Rev. D}} \bibinfo{volume}{110}
  (\bibinfo{number}{1}): \bibinfo{pages}{016021}.
  \bibinfo{doi}{\doi{10.1103/PhysRevD.110.016021}}.
\eprint{2404.00554}.

\bibtype{Article}%
\bibitem[Ferreira et al.(2025)]{Ferreira:2025tzo}
\bibinfo{author}{Ferreira MN}, \bibinfo{author}{Papavassiliou J},
  \bibinfo{author}{Pawlowski JM} and  \bibinfo{author}{Wink N}
  (\bibinfo{year}{2025}).
\bibinfo{title}{{Physics of the gluon mass gap}}.
\bibinfo{journal}{{\em Eur. Phys. J. C}} \bibinfo{volume}{85}
  (\bibinfo{number}{11}): \bibinfo{pages}{1339}.
  \bibinfo{doi}{\doi{10.1140/epjc/s10052-025-15027-7}}.
\eprint{2508.20568}.

\bibtype{Article}%
\bibitem[Fischer(2006)]{Fischer:2006ub}
\bibinfo{author}{Fischer CS} (\bibinfo{year}{2006}).
\bibinfo{title}{{Infrared properties of QCD from Dyson-Schwinger equations}}.
\bibinfo{journal}{{\em J. Phys.}} \bibinfo{volume}{G32}:
  \bibinfo{pages}{R253--R291}. \bibinfo{doi}{\doi{10.1088/0954-3899/32/8/R02}}.
\eprint{hep-ph/0605173}.

\bibtype{Article}%
\bibitem[Fischer(2009)]{Fischer:2009wc}
\bibinfo{author}{Fischer CS} (\bibinfo{year}{2009}).
\bibinfo{title}{Deconfinement phase transition and the quark condensate}.
\bibinfo{journal}{{\em Phys. Rev. Lett.}} \bibinfo{volume}{103}:
  \bibinfo{pages}{052003}. \bibinfo{doi}{\doi{10.1103/PhysRevLett.103.052003}}.
\eprint{0904.2700}.

\bibtype{Article}%
\bibitem[Fischer(2019)]{Fischer:2018sdj}
\bibinfo{author}{Fischer CS} (\bibinfo{year}{2019}).
\bibinfo{title}{{QCD at finite temperature and chemical potential from
  Dyson-Schwinger equations}}.
\bibinfo{journal}{{\em Prog. Part. Nucl. Phys.}} \bibinfo{volume}{105}:
  \bibinfo{pages}{1--60}. \bibinfo{doi}{\doi{10.1016/j.ppnp.2019.01.002}}.
\eprint{1810.12938}.

\bibtype{Article}%
\bibitem[Fischer and Mueller(2009)]{Fischer:2009gk}
\bibinfo{author}{Fischer CS} and  \bibinfo{author}{Mueller JA}
  (\bibinfo{year}{2009}).
\bibinfo{title}{Chiral and deconfinement transition from dyson-schwinger
  equations}.
\bibinfo{journal}{{\em Phys. Rev. D}} \bibinfo{volume}{80}:
  \bibinfo{pages}{074029}. \bibinfo{doi}{\doi{10.1103/PhysRevD.80.074029}}.
\eprint{0908.0007}.

\bibtype{Article}%
\bibitem[Fischer and Pawlowski(2007)]{Fischer:2006vf}
\bibinfo{author}{Fischer CS} and  \bibinfo{author}{Pawlowski JM}
  (\bibinfo{year}{2007}).
\bibinfo{title}{{Uniqueness of infrared asymptotics in Landau gauge Yang-Mills
  theory}}.
\bibinfo{journal}{{\em Phys. Rev. D}} \bibinfo{volume}{75}:
  \bibinfo{pages}{025012}. \bibinfo{doi}{\doi{10.1103/PhysRevD.75.025012}}.
\eprint{hep-th/0609009}.

\bibtype{Article}%
\bibitem[Fischer and Pawlowski(2009)]{Fischer:2009tn}
\bibinfo{author}{Fischer CS} and  \bibinfo{author}{Pawlowski JM}
  (\bibinfo{year}{2009}).
\bibinfo{title}{{Uniqueness of infrared asymptotics in Landau gauge Yang-Mills
  theory II}}.
\bibinfo{journal}{{\em Phys. Rev. D}} \bibinfo{volume}{80}:
  \bibinfo{pages}{025023}. \bibinfo{doi}{\doi{10.1103/PhysRevD.80.025023}}.
\eprint{0903.2193}.

\bibtype{Article}%
\bibitem[Fischer and Pawlowski(2026)]{Fischer:2026uni}
\bibinfo{author}{Fischer CS} and  \bibinfo{author}{Pawlowski JM}
  (\bibinfo{year}{2026}), \bibinfo{month}{3}.
\bibinfo{title}{{Phase structure and observables at high densities from first
  principles QCD}} \eprint{2603.11135}.

\bibtype{Article}%
\bibitem[Fischer et al.(2014{\natexlab{a}})]{Fischer:2013eca}
\bibinfo{author}{Fischer CS}, \bibinfo{author}{Fister L},
  \bibinfo{author}{Luecker J} and  \bibinfo{author}{Pawlowski JM}
  (\bibinfo{year}{2014}{\natexlab{a}}).
\bibinfo{title}{{Polyakov loop potential at finite density}}.
\bibinfo{journal}{{\em Phys. Lett.}} \bibinfo{volume}{B732}:
  \bibinfo{pages}{273--277}.
  \bibinfo{doi}{\doi{10.1016/j.physletb.2014.03.057}}.
\eprint{1306.6022}.

\bibtype{Article}%
\bibitem[Fischer et al.(2014{\natexlab{b}})]{Fischer:2014ata}
\bibinfo{author}{Fischer CS}, \bibinfo{author}{Luecker J} and
  \bibinfo{author}{Welzbacher CA} (\bibinfo{year}{2014}{\natexlab{b}}).
\bibinfo{title}{{Phase structure of three and four flavor QCD}}.
\bibinfo{journal}{{\em Phys. Rev.}} \bibinfo{volume}{D90}
  (\bibinfo{number}{3}): \bibinfo{pages}{034022}.
  \bibinfo{doi}{\doi{10.1103/PhysRevD.90.034022}}.
\eprint{1405.4762}.

\bibtype{Article}%
\bibitem[Fischer et al.(2015)]{Fischer:2014vxa}
\bibinfo{author}{Fischer CS}, \bibinfo{author}{Luecker J} and
  \bibinfo{author}{Pawlowski JM} (\bibinfo{year}{2015}).
\bibinfo{title}{{Phase structure of QCD for heavy quarks}}.
\bibinfo{journal}{{\em Phys. Rev.}} \bibinfo{volume}{D91}
  (\bibinfo{number}{1}): \bibinfo{pages}{014024}.
  \bibinfo{doi}{\doi{10.1103/PhysRevD.91.014024}}.
\eprint{1409.8462}.

\bibtype{Article}%
\bibitem[Fister and Pawlowski(2013)]{Fister:2013bh}
\bibinfo{author}{Fister L} and  \bibinfo{author}{Pawlowski JM}
  (\bibinfo{year}{2013}).
\bibinfo{title}{{Confinement from Correlation Functions}}.
\bibinfo{journal}{{\em Phys.Rev.}} \bibinfo{volume}{D88}:
  \bibinfo{pages}{045010}. \bibinfo{doi}{\doi{10.1103/PhysRevD.88.045010}}.
\eprint{1301.4163}.

\bibtype{Article}%
\bibitem[Friman et al.(2011)]{Friman:2011pf}
\bibinfo{author}{Friman B}, \bibinfo{author}{Karsch F},
  \bibinfo{author}{Redlich K} and  \bibinfo{author}{Skokov V}
  (\bibinfo{year}{2011}).
\bibinfo{title}{Fluctuations as probe of the qcd phase transition and
  freeze-out in heavy ion collisions at lhc and rhic}.
\bibinfo{journal}{{\em Eur. Phys. J. C}} \bibinfo{volume}{71}:
  \bibinfo{pages}{1694}. \bibinfo{doi}{\doi{10.1140/epjc/s10052-011-1694-2}}.
\eprint{1103.3511}.

\bibtype{Article}%
\bibitem[Fromm et al.(2012)]{Fromm:2011qi}
\bibinfo{author}{Fromm M}, \bibinfo{author}{Langelage J},
  \bibinfo{author}{Lottini S} and  \bibinfo{author}{Philipsen O}
  (\bibinfo{year}{2012}).
\bibinfo{title}{The qcd deconfinement transition for heavy quarks and all
  baryon chemical potentials}.
\bibinfo{journal}{{\em JHEP}} \bibinfo{volume}{01}: \bibinfo{pages}{042}.
  \bibinfo{doi}{\doi{10.1007/JHEP01(2012)042}}.
\eprint{1111.4953}.

\bibtype{Article}%
\bibitem[Fu(2022)]{Fu:2022gou}
\bibinfo{author}{Fu Wj} (\bibinfo{year}{2022}).
\bibinfo{title}{{QCD at finite temperature and density within the fRG approach:
  an overview}}.
\bibinfo{journal}{{\em Commun. Theor. Phys.}} \bibinfo{volume}{74}
  (\bibinfo{number}{9}): \bibinfo{pages}{097304}.
  \bibinfo{doi}{\doi{10.1088/1572-9494/ac86be}}.
\eprint{2205.00468}.

\bibtype{Article}%
\bibitem[Fu et al.(2020)]{Fu:2019hdw}
\bibinfo{author}{Fu Wj}, \bibinfo{author}{Pawlowski JM} and
  \bibinfo{author}{Rennecke F} (\bibinfo{year}{2020}).
\bibinfo{title}{{QCD phase structure at finite temperature and density}}.
\bibinfo{journal}{{\em Phys. Rev. D}} \bibinfo{volume}{101}
  (\bibinfo{number}{5}): \bibinfo{pages}{054032}.
  \bibinfo{doi}{\doi{10.1103/PhysRevD.101.054032}}.
\eprint{1909.02991}.

\bibtype{Article}%
\bibitem[Fu et al.(2025{\natexlab{a}})]{Fu:2023lcm}
\bibinfo{author}{Fu Wj}, \bibinfo{author}{Luo X}, \bibinfo{author}{Pawlowski
  JM}, \bibinfo{author}{Rennecke F} and  \bibinfo{author}{Yin S}
  (\bibinfo{year}{2025}{\natexlab{a}}).
\bibinfo{title}{{Ripples of the QCD critical point}}.
\bibinfo{journal}{{\em Phys. Rev. D}} \bibinfo{volume}{111}
  (\bibinfo{number}{3}): \bibinfo{pages}{L031502}.
  \bibinfo{doi}{\doi{10.1103/PhysRevD.111.L031502}}.
\eprint{2308.15508}.

\bibtype{Article}%
\bibitem[Fu et al.(2025{\natexlab{b}})]{Fu:2024rto}
\bibinfo{author}{Fu Wj}, \bibinfo{author}{Pawlowski JM},
  \bibinfo{author}{Pisarski RD}, \bibinfo{author}{Rennecke F},
  \bibinfo{author}{Wen R} and  \bibinfo{author}{Yin S}
  (\bibinfo{year}{2025}{\natexlab{b}}).
\bibinfo{title}{{QCD moat regime and its real-time properties}}.
\bibinfo{journal}{{\em Phys. Rev. D}} \bibinfo{volume}{111}
  (\bibinfo{number}{9}): \bibinfo{pages}{094026}.
  \bibinfo{doi}{\doi{10.1103/PhysRevD.111.094026}}.
\eprint{2412.15949}.

\bibtype{Article}%
\bibitem[Fu et al.(2026)]{Fu:2026qnl}
\bibinfo{author}{Fu Wj}, \bibinfo{author}{Huang C}, \bibinfo{author}{Pawlowski
  JM}, \bibinfo{author}{Rennecke F}, \bibinfo{author}{Wen R} and
  \bibinfo{author}{Yin S} (\bibinfo{year}{2026}), \bibinfo{month}{3}.
\bibinfo{title}{{Strangeness neutrality and the QCD phase diagram}}
  \eprint{2603.13455}.

\bibtype{Article}%
\bibitem[Fukushima and Kashiwa(2013)]{Fukushima:2012qa}
\bibinfo{author}{Fukushima K} and  \bibinfo{author}{Kashiwa K}
  (\bibinfo{year}{2013}).
\bibinfo{title}{{Polyakov loop and QCD thermodynamics from the gluon and ghost
  propagators}}.
\bibinfo{journal}{{\em Phys. Lett.}} \bibinfo{volume}{B723}:
  \bibinfo{pages}{360--364}.
  \bibinfo{doi}{\doi{10.1016/j.physletb.2013.05.037}}.
\eprint{1206.0685}.

\bibtype{Article}%
\bibitem[Fukushima and Skokov(2017)]{Fukushima:2017csk}
\bibinfo{author}{Fukushima K} and  \bibinfo{author}{Skokov V}
  (\bibinfo{year}{2017}).
\bibinfo{title}{{Polyakov loop modeling for hot QCD}}.
\bibinfo{journal}{{\em Prog. Part. Nucl. Phys.}} \bibinfo{volume}{96}:
  \bibinfo{pages}{154--199}. \bibinfo{doi}{\doi{10.1016/j.ppnp.2017.05.002}}.
\eprint{1705.00718}.

\bibtype{Article}%
\bibitem[Fukushima et al.(2024{\natexlab{a}})]{Fukushima:2023tpv}
\bibinfo{author}{Fukushima K}, \bibinfo{author}{Hidaka Y},
  \bibinfo{author}{Inoue K}, \bibinfo{author}{Shigaki K} and
  \bibinfo{author}{Yamaguchi Y} (\bibinfo{year}{2024}{\natexlab{a}}).
\bibinfo{title}{{Hanbury-Brown{\textendash}Twiss signature for clustered
  substructures probing primordial inhomogeneity in hot and dense QCD matter}}.
\bibinfo{journal}{{\em Phys. Rev. C}} \bibinfo{volume}{109}
  (\bibinfo{number}{5}): \bibinfo{pages}{L051903}.
  \bibinfo{doi}{\doi{10.1103/PhysRevC.109.L051903}}.
\eprint{2306.17619}.

\bibtype{Article}%
\bibitem[Fukushima et al.(2024{\natexlab{b}})]{Fukushima:2023wnl}
\bibinfo{author}{Fukushima K}, \bibinfo{author}{Horak J},
  \bibinfo{author}{Pawlowski JM}, \bibinfo{author}{Wink N} and
  \bibinfo{author}{Zelle CP} (\bibinfo{year}{2024}{\natexlab{b}}).
\bibinfo{title}{{Nuclear liquid-gas transition in QCD}}.
\bibinfo{journal}{{\em Phys. Rev. D}} \bibinfo{volume}{110}
  (\bibinfo{number}{7}): \bibinfo{pages}{076022}.
  \bibinfo{doi}{\doi{10.1103/PhysRevD.110.076022}}.
\eprint{2308.16594}.

\bibtype{Article}%
\bibitem[Gao and Oldengott(2022)]{Gao:2021nwz}
\bibinfo{author}{Gao F} and  \bibinfo{author}{Oldengott IM}
  (\bibinfo{year}{2022}).
\bibinfo{title}{{Cosmology Meets Functional QCD: First-Order Cosmic QCD
  Transition Induced by Large Lepton Asymmetries}}.
\bibinfo{journal}{{\em Phys. Rev. Lett.}} \bibinfo{volume}{128}
  (\bibinfo{number}{13}): \bibinfo{pages}{131301}.
  \bibinfo{doi}{\doi{10.1103/PhysRevLett.128.131301}}.
\eprint{2106.11991}.

\bibtype{Article}%
\bibitem[Gao and Pawlowski(2021)]{Gao:2020fbl}
\bibinfo{author}{Gao F} and  \bibinfo{author}{Pawlowski JM}
  (\bibinfo{year}{2021}).
\bibinfo{title}{{Chiral phase structure and critical end point in QCD}}.
\bibinfo{journal}{{\em Phys. Lett. B}} \bibinfo{volume}{820}:
  \bibinfo{pages}{136584}. \bibinfo{doi}{\doi{10.1016/j.physletb.2021.136584}}.
\eprint{2010.13705}.

\bibtype{Article}%
\bibitem[Gao and Pawlowski(2022)]{Gao:2021vsf}
\bibinfo{author}{Gao F} and  \bibinfo{author}{Pawlowski JM}
  (\bibinfo{year}{2022}).
\bibinfo{title}{{Phase structure of (2+1)-flavor QCD and the magnetic equation
  of state}}.
\bibinfo{journal}{{\em Phys. Rev. D}} \bibinfo{volume}{105}
  (\bibinfo{number}{9}): \bibinfo{pages}{094020}.
  \bibinfo{doi}{\doi{10.1103/PhysRevD.105.094020}}.
\eprint{2112.01395}.

\bibtype{Article}%
\bibitem[Gasenzer and Pawlowski(2008)]{Gasenzer:2007za}
\bibinfo{author}{Gasenzer T} and  \bibinfo{author}{Pawlowski JM}
  (\bibinfo{year}{2008}).
\bibinfo{title}{{Towards far-from-equilibrium quantum field dynamics: A
  functional renormalisation-group approach}}.
\bibinfo{journal}{{\em Phys. Lett. B}} \bibinfo{volume}{670}:
  \bibinfo{pages}{135--140}.
  \bibinfo{doi}{\doi{10.1016/j.physletb.2008.10.049}}.
\eprint{0710.4627}.

\bibtype{Article}%
\bibitem[Gies(2012)]{Gies:2006wv}
\bibinfo{author}{Gies H} (\bibinfo{year}{2012}).
\bibinfo{title}{{Introduction to the functional RG and applications to gauge
  theories}}.
\bibinfo{journal}{{\em Lect. Notes Phys.}} \bibinfo{volume}{852}:
  \bibinfo{pages}{287--348}. \bibinfo{doi}{\doi{10.1007/978-3-642-27320-9_6}}.
\eprint{hep-ph/0611146}.

\bibtype{Article}%
\bibitem[Greensite(2003)]{Greensite:2003bk}
\bibinfo{author}{Greensite J} (\bibinfo{year}{2003}).
\bibinfo{title}{The confinement problem in lattice gauge theory}.
\bibinfo{journal}{{\em Prog. Part. Nucl. Phys.}} \bibinfo{volume}{51}:
  \bibinfo{pages}{1}. \bibinfo{doi}{\doi{10.1016/S0146-6410(03)90012-3}}.
\eprint{hep-lat/0301023}.

\bibtype{Book}%
\bibitem[Greensite(2011)]{Greensite:2011zz}
\bibinfo{author}{Greensite J} (\bibinfo{year}{2011}).
\bibinfo{title}{An introduction to the confinement problem},
  \bibinfo{volume}{821}.
\bibinfo{doi}{\doi{10.1007/978-3-642-14382-3}}.

\bibtype{Article}%
\bibitem[Gross et al.(1981)]{Gross:1980br}
\bibinfo{author}{Gross DJ}, \bibinfo{author}{Pisarski RD} and
  \bibinfo{author}{Yaffe LG} (\bibinfo{year}{1981}).
\bibinfo{title}{{QCD and Instantons at Finite Temperature}}.
\bibinfo{journal}{{\em Rev. Mod. Phys.}} \bibinfo{volume}{53}:
  \bibinfo{pages}{43}. \bibinfo{doi}{\doi{10.1103/RevModPhys.53.43}}.

\bibtype{Article}%
\bibitem[Gunkel and Fischer(2021)]{Gunkel:2021oya}
\bibinfo{author}{Gunkel PJ} and  \bibinfo{author}{Fischer CS}
  (\bibinfo{year}{2021}).
\bibinfo{title}{{Locating the critical endpoint of QCD: Mesonic backcoupling
  effects}}.
\bibinfo{journal}{{\em Phys. Rev. D}} \bibinfo{volume}{104}
  (\bibinfo{number}{5}): \bibinfo{pages}{054022}.
  \bibinfo{doi}{\doi{10.1103/PhysRevD.104.054022}}.
\eprint{2106.08356}.

\bibtype{Article}%
\bibitem[Haas et al.(2013)]{Haas:2013qwp}
\bibinfo{author}{Haas LM}, \bibinfo{author}{Stiele R}, \bibinfo{author}{Braun
  J}, \bibinfo{author}{Pawlowski JM} and  \bibinfo{author}{Schaffner-Bielich J}
  (\bibinfo{year}{2013}).
\bibinfo{title}{{Improved Polyakov-loop potential for effective models from
  functional calculations}}.
\bibinfo{journal}{{\em Phys. Rev.}} \bibinfo{volume}{D87}
  (\bibinfo{number}{7}): \bibinfo{pages}{076004}.
  \bibinfo{doi}{\doi{10.1103/PhysRevD.87.076004}}.
\eprint{1302.1993}.

\bibtype{Article}%
\bibitem[Haas et al.(2014)]{Haas:2013hpa}
\bibinfo{author}{Haas M}, \bibinfo{author}{Fister L} and
  \bibinfo{author}{Pawlowski JM} (\bibinfo{year}{2014}).
\bibinfo{title}{{Gluon spectral functions and transport coefficients in
  Yang--Mills theory}}.
\bibinfo{journal}{{\em Phys. Rev.}} \bibinfo{volume}{D90}:
  \bibinfo{pages}{091501}. \bibinfo{doi}{\doi{10.1103/PhysRevD.90.091501}}.
\eprint{1308.4960}.

\bibtype{Article}%
\bibitem[Haensch et al.(2024)]{Haensch:2023sig}
\bibinfo{author}{Haensch M}, \bibinfo{author}{Rennecke F} and
  \bibinfo{author}{von Smekal L} (\bibinfo{year}{2024}).
\bibinfo{title}{{Medium induced mixing, spatial modulations, and critical modes
  in QCD}}.
\bibinfo{journal}{{\em Phys. Rev. D}} \bibinfo{volume}{110}
  (\bibinfo{number}{3}): \bibinfo{pages}{036018}.
  \bibinfo{doi}{\doi{10.1103/PhysRevD.110.036018}}.
\eprint{2308.16244}.

\bibtype{Article}%
\bibitem[Halasz et al.(1998)]{Halasz:1998qr}
\bibinfo{author}{Halasz AM}, \bibinfo{author}{Jackson AD},
  \bibinfo{author}{Shrock RE}, \bibinfo{author}{Stephanov MA} and
  \bibinfo{author}{Verbaarschot JJM} (\bibinfo{year}{1998}).
\bibinfo{title}{On the phase diagram of qcd}.
\bibinfo{journal}{{\em Phys. Rev. D}} \bibinfo{volume}{58}:
  \bibinfo{pages}{096007}. \bibinfo{doi}{\doi{10.1103/PhysRevD.58.096007}}.
\eprint{hep-ph/9804290}.

\bibtype{Article}%
\bibitem[Herbst et al.(2015)]{Herbst:2015ona}
\bibinfo{author}{Herbst TK}, \bibinfo{author}{Luecker J} and
  \bibinfo{author}{Pawlowski JM} (\bibinfo{year}{2015}).
\bibinfo{title}{{Confinement order parameters and fluctuations}}
  \eprint{1510.03830}.

\bibtype{Article}%
\bibitem[Horak et al.(2020)]{Horak:2020eng}
\bibinfo{author}{Horak J}, \bibinfo{author}{Pawlowski JM} and
  \bibinfo{author}{Wink N} (\bibinfo{year}{2020}).
\bibinfo{title}{{Spectral functions in the $\phi^4$-theory from the spectral
  DSE}}.
\bibinfo{journal}{{\em Phys. Rev. D}} \bibinfo{volume}{102}:
  \bibinfo{pages}{125016}. \bibinfo{doi}{\doi{10.1103/PhysRevD.102.125016}}.
\eprint{2006.09778}.

\bibtype{Article}%
\bibitem[Horak et al.(2023)]{Horak:2022aza}
\bibinfo{author}{Horak J}, \bibinfo{author}{Pawlowski JM} and
  \bibinfo{author}{Wink N} (\bibinfo{year}{2023}).
\bibinfo{title}{{On the quark spectral function in QCD}}.
\bibinfo{journal}{{\em SciPost Phys.}} \bibinfo{volume}{15}
  (\bibinfo{number}{4}): \bibinfo{pages}{149}.
  \bibinfo{doi}{\doi{10.21468/SciPostPhys.15.4.149}}.
\eprint{2210.07597}.

\bibtype{Article}%
\bibitem[Horak et al.(2025)]{Horak:2022myj}
\bibinfo{author}{Horak J}, \bibinfo{author}{Pawlowski JM} and
  \bibinfo{author}{Wink N} (\bibinfo{year}{2025}).
\bibinfo{title}{{On the complex structure of Yang-Mills theory}}.
\bibinfo{journal}{{\em SciPost Phys. Core}} \bibinfo{volume}{8}:
  \bibinfo{pages}{048}. \bibinfo{doi}{\doi{10.21468/SciPostPhysCore.8.3.048}}.
\eprint{2202.09333}.

\bibtype{Article}%
\bibitem[Huang et al.(2026)]{Huang:2026zfv}
\bibinfo{author}{Huang Yz}, \bibinfo{author}{Yin S}, \bibinfo{author}{Han Sn},
  \bibinfo{author}{Wu J}, \bibinfo{author}{Li F} and  \bibinfo{author}{Fu Wj}
  (\bibinfo{year}{2026}), \bibinfo{month}{5}.
\bibinfo{title}{{Functional renormalization group study of the jet quenching
  parameter near the QCD critical end point}} \eprint{2605.30816}.

\bibtype{Article}%
\bibitem[Huber(2025)]{Huber:2025cbd}
\bibinfo{author}{Huber MQ} (\bibinfo{year}{2025}), \bibinfo{month}{10}.
\bibinfo{title}{{A beginner's guide to functional methods in particle physics}}
  \eprint{2510.18960}.

\bibtype{Article}%
\bibitem[Huber et al.(2020)]{Huber:2020ngt}
\bibinfo{author}{Huber MQ}, \bibinfo{author}{Fischer CS} and
  \bibinfo{author}{Sanchis-Alepuz H} (\bibinfo{year}{2020}).
\bibinfo{title}{{Spectrum of scalar and pseudoscalar glueballs from functional
  methods}}.
\bibinfo{journal}{{\em Eur. Phys. J. C}} \bibinfo{volume}{80}
  (\bibinfo{number}{11}): \bibinfo{pages}{1077}.
  \bibinfo{doi}{\doi{10.1140/epjc/s10052-020-08649-6}}.
\eprint{2004.00415}.

\bibtype{Article}%
\bibitem[Huber et al.(2021)]{Huber:2021yfy}
\bibinfo{author}{Huber MQ}, \bibinfo{author}{Fischer CS} and
  \bibinfo{author}{Sanchis-Alepuz H} (\bibinfo{year}{2021}).
\bibinfo{title}{{Higher spin glueballs from functional methods}}.
\bibinfo{journal}{{\em Eur. Phys. J. C}} \bibinfo{volume}{81}
  (\bibinfo{number}{12}): \bibinfo{pages}{1083}.
  \bibinfo{doi}{\doi{10.1140/epjc/s10052-021-09864-5}}.
\bibinfo{note}{[Erratum: Eur.Phys.J.C 82, 38 (2022)]}, \eprint{2110.09180}.

\bibtype{Article}%
\bibitem[Huber et al.(2025)]{Huber:2025kwy}
\bibinfo{author}{Huber MQ}, \bibinfo{author}{Fischer CS} and
  \bibinfo{author}{Sanchis-Alepuz H} (\bibinfo{year}{2025}).
\bibinfo{title}{{Apparent convergence in functional glueball calculations}}.
\bibinfo{journal}{{\em Eur. Phys. J. C}} \bibinfo{volume}{85}
  (\bibinfo{number}{8}): \bibinfo{pages}{859}.
  \bibinfo{doi}{\doi{10.1140/epjc/s10052-025-14590-3}}.
\eprint{2503.03821}.

\bibtype{Article}%
\bibitem[Ihssen et al.(2024)]{Ihssen:2024miv}
\bibinfo{author}{Ihssen F}, \bibinfo{author}{Pawlowski JM},
  \bibinfo{author}{Sattler FR} and  \bibinfo{author}{Wink N}
  (\bibinfo{year}{2024}), \bibinfo{month}{8}.
\bibinfo{title}{{Towards quantitative precision in functional QCD I}}
  \eprint{2408.08413}.

\bibtype{Article}%
\bibitem[Isserstedt et al.(2021)]{Isserstedt:2020qll}
\bibinfo{author}{Isserstedt P}, \bibinfo{author}{Fischer CS} and
  \bibinfo{author}{Steinert T} (\bibinfo{year}{2021}).
\bibinfo{title}{{Thermodynamics from the quark condensate}}.
\bibinfo{journal}{{\em Phys. Rev. D}} \bibinfo{volume}{103}
  (\bibinfo{number}{5}): \bibinfo{pages}{054012}.
  \bibinfo{doi}{\doi{10.1103/PhysRevD.103.054012}}.
\eprint{2012.04991}.

\bibtype{Article}%
\bibitem[Jeon and Koch(2000)]{Jeon:2000wg}
\bibinfo{author}{Jeon S} and  \bibinfo{author}{Koch V} (\bibinfo{year}{2000}).
\bibinfo{title}{Charged particle ratio fluctuation as a signal for qgp}.
\bibinfo{journal}{{\em Phys. Rev. Lett.}} \bibinfo{volume}{85}:
  \bibinfo{pages}{2076--2079}.
  \bibinfo{doi}{\doi{10.1103/PhysRevLett.85.2076}}.
\eprint{hep-ph/0003168}.

\bibtype{Article}%
\bibitem[Johnson et al.(2023)]{Johnson:2022cqv}
\bibinfo{author}{Johnson G}, \bibinfo{author}{Rennecke F} and
  \bibinfo{author}{Skokov VV} (\bibinfo{year}{2023}).
\bibinfo{title}{{Universal location of Yang-Lee edge singularity in classic
  O(N) universality classes}}.
\bibinfo{journal}{{\em Phys. Rev. D}} \bibinfo{volume}{107}
  (\bibinfo{number}{11}): \bibinfo{pages}{116013}.
  \bibinfo{doi}{\doi{10.1103/PhysRevD.107.116013}}.
\eprint{2211.00710}.

\bibtype{Article}%
\bibitem[Kamikado et al.(2014)]{Kamikado:2013sia}
\bibinfo{author}{Kamikado K}, \bibinfo{author}{Strodthoff N},
  \bibinfo{author}{von Smekal L} and  \bibinfo{author}{Wambach J}
  (\bibinfo{year}{2014}).
\bibinfo{title}{{Real-time correlation functions in the $O(N)$ model from the
  functional renormalization group}}.
\bibinfo{journal}{{\em Eur. Phys. J.}} \bibinfo{volume}{C74}
  (\bibinfo{number}{3}): \bibinfo{pages}{2806}.
  \bibinfo{doi}{\doi{10.1140/epjc/s10052-014-2806-6}}.
\eprint{1302.6199}.

\bibtype{Article}%
\bibitem[Karsch(2019)]{Karsch:2019mbv}
\bibinfo{author}{Karsch F} (\bibinfo{year}{2019}).
\bibinfo{title}{Critical behavior and net-charge fluctuations from lattice
  qcd}.
\bibinfo{journal}{{\em PoS}} \bibinfo{volume}{CORFU2018}: \bibinfo{pages}{163}.
  \bibinfo{doi}{\doi{10.22323/1.347.0163}}.
\eprint{1905.03936}.

\bibtype{Article}%
\bibitem[Kashiwa and Maezawa(2012)]{Kashiwa:2012td}
\bibinfo{author}{Kashiwa K} and  \bibinfo{author}{Maezawa Y}
  (\bibinfo{year}{2012}), \bibinfo{month}{12}.
\bibinfo{title}{{Quark back reaction to deconfinement transition via gluon
  propagators}} \eprint{1212.2184}.

\bibtype{Article}%
\bibitem[Kiyohara et al.(2021)]{Kiyohara:2021smr}
\bibinfo{author}{Kiyohara A}, \bibinfo{author}{Kitazawa M},
  \bibinfo{author}{Ejiri S} and  \bibinfo{author}{Kanaya K}
  (\bibinfo{year}{2021}).
\bibinfo{title}{Finite-size scaling around the critical point in the heavy
  quark region of qcd}.
\bibinfo{journal}{{\em Phys. Rev. D}} \bibinfo{volume}{104}
  (\bibinfo{number}{11}): \bibinfo{pages}{114509}.
  \bibinfo{doi}{\doi{10.1103/PhysRevD.104.114509}}.
\eprint{2108.00118}.

\bibtype{Article}%
\bibitem[Koch et al.(2005)]{Koch:2005vg}
\bibinfo{author}{Koch V}, \bibinfo{author}{Majumder A} and
  \bibinfo{author}{Randrup J} (\bibinfo{year}{2005}).
\bibinfo{title}{Baryon-strangeness correlations: A diagnostic of strongly
  interacting matter}.
\bibinfo{journal}{{\em Phys. Rev. Lett.}} \bibinfo{volume}{95}:
  \bibinfo{pages}{182301}. \bibinfo{doi}{\doi{10.1103/PhysRevLett.95.182301}}.
\eprint{nucl-th/0505052}.

\bibtype{Article}%
\bibitem[Landsman and van Weert(1987)]{Landsman:1986uw}
\bibinfo{author}{Landsman NP} and  \bibinfo{author}{van Weert CG}
  (\bibinfo{year}{1987}).
\bibinfo{title}{Real and imaginary time field theory at finite temperature and
  density}.
\bibinfo{journal}{{\em Phys. Rept.}} \bibinfo{volume}{145}:
  \bibinfo{pages}{141}. \bibinfo{doi}{\doi{10.1016/0370-1573(87)90121-9}}.

\bibtype{Inproceedings}%
\bibitem[Litim and Pawlowski(1998)]{Litim:1998nf}
\bibinfo{author}{Litim DF} and  \bibinfo{author}{Pawlowski JM}
  (\bibinfo{year}{1998}), \bibinfo{month}{9}, \bibinfo{title}{{On gauge
  invariant Wilsonian flows}}, \bibinfo{booktitle}{{Workshop on the Exact
  Renormalization Group}},  \bibinfo{pages}{168--185}, \eprint{hep-th/9901063}.

\bibtype{Article}%
\bibitem[Lu et al.(2024)]{Lu:2023mkn}
\bibinfo{author}{Lu Y}, \bibinfo{author}{Gao F}, \bibinfo{author}{Liu YX} and
  \bibinfo{author}{Pawlowski JM} (\bibinfo{year}{2024}).
\bibinfo{title}{{QCD equation of state and thermodynamic observables from
  computationally minimal Dyson-Schwinger equations}}.
\bibinfo{journal}{{\em Phys. Rev. D}} \bibinfo{volume}{110}
  (\bibinfo{number}{1}): \bibinfo{pages}{014036}.
  \bibinfo{doi}{\doi{10.1103/PhysRevD.110.014036}}.
\eprint{2310.18383}.

\bibtype{Article}%
\bibitem[Lu et al.(2025)]{Lu:2025cls}
\bibinfo{author}{Lu Y}, \bibinfo{author}{Gao F}, \bibinfo{author}{Liu Yx} and
  \bibinfo{author}{Pawlowski JM} (\bibinfo{year}{2025}), \bibinfo{month}{4}.
\bibinfo{title}{{Finite density signatures of confining and chiral dynamics in
  QCD thermodynamics and fluctuations of conserved charges}}
  \eprint{2504.05099}.

\bibtype{Article}%
\bibitem[Lu et al.(2026)]{Lu:2026ezr}
\bibinfo{author}{Lu Y}, \bibinfo{author}{Fischer CS}, \bibinfo{author}{Gao F},
  \bibinfo{author}{Liu Yx} and  \bibinfo{author}{Pawlowski JM}
  (\bibinfo{year}{2026}), \bibinfo{month}{3}.
\bibinfo{title}{{Extracting freeze-out conditions in beam energy scan via
  functional QCD}} \eprint{2603.09336}.

\bibtype{Article}%
\bibitem[Luo and Xu(2017)]{Luo:2017faz}
\bibinfo{author}{Luo X} and  \bibinfo{author}{Xu N} (\bibinfo{year}{2017}).
\bibinfo{title}{{Search for the QCD Critical Point with Fluctuations of
  Conserved Quantities in Relativistic Heavy-Ion Collisions at RHIC : An
  Overview}}.
\bibinfo{journal}{{\em Nucl. Sci. Tech.}} \bibinfo{volume}{28}
  (\bibinfo{number}{8}): \bibinfo{pages}{112}.
  \bibinfo{doi}{\doi{10.1007/s41365-017-0257-0}}.
\eprint{1701.02105}.

\bibtype{Article}%
\bibitem[Marhauser and Pawlowski(2008)]{Marhauser:2008fz}
\bibinfo{author}{Marhauser F} and  \bibinfo{author}{Pawlowski JM}
  (\bibinfo{year}{2008}).
\bibinfo{title}{{Confinement in Polyakov Gauge}} \eprint{0812.1144}.

\bibtype{Article}%
\bibitem[Mari~Surkau and Reinosa(2026)]{MariSurkau:2026irs}
\bibinfo{author}{Mari~Surkau VT} and  \bibinfo{author}{Reinosa U}
  (\bibinfo{year}{2026}), \bibinfo{month}{1}.
\bibinfo{title}{{Interplay between the chiral and deconfinement transitions
  from a Curci-Ferrari-based Polyakov loop potential}} \eprint{2601.15839}.

\bibtype{Article}%
\bibitem[Motta et al.(2023)]{Motta:2023pks}
\bibinfo{author}{Motta TF}, \bibinfo{author}{Bernhardt J},
  \bibinfo{author}{Buballa M} and  \bibinfo{author}{Fischer CS}
  (\bibinfo{year}{2023}).
\bibinfo{title}{{Toward a stability analysis of inhomogeneous phases in QCD}}.
\bibinfo{journal}{{\em Phys. Rev. D}} \bibinfo{volume}{108}
  (\bibinfo{number}{11}): \bibinfo{pages}{114019}.
  \bibinfo{doi}{\doi{10.1103/PhysRevD.108.114019}}.
\eprint{2306.09749}.

\bibtype{Article}%
\bibitem[Motta et al.(2025)]{Motta:2024rvk}
\bibinfo{author}{Motta TF}, \bibinfo{author}{Bernhardt J},
  \bibinfo{author}{Buballa M} and  \bibinfo{author}{Fischer CS}
  (\bibinfo{year}{2025}).
\bibinfo{title}{{New tool to detect inhomogeneous chiral-symmetry breaking}}.
\bibinfo{journal}{{\em Phys. Rev. D}} \bibinfo{volume}{111}
  (\bibinfo{number}{7}): \bibinfo{pages}{074030}.
  \bibinfo{doi}{\doi{10.1103/PhysRevD.111.074030}}.
\eprint{2411.02285}.

\bibtype{Article}%
\bibitem[Mukherjee et al.(2022)]{Mukherjee:2021tyg}
\bibinfo{author}{Mukherjee S}, \bibinfo{author}{Rennecke F} and
  \bibinfo{author}{Skokov VV} (\bibinfo{year}{2022}).
\bibinfo{title}{{Analytical structure of the equation of state at finite
  density: Resummation versus expansion in a low energy model}}.
\bibinfo{journal}{{\em Phys. Rev. D}} \bibinfo{volume}{105}
  (\bibinfo{number}{1}): \bibinfo{pages}{014026}.
  \bibinfo{doi}{\doi{10.1103/PhysRevD.105.014026}}.
\eprint{2110.02241}.

\bibtype{Article}%
\bibitem[Pawlowski(2014)]{Pawlowski:2014aha}
\bibinfo{author}{Pawlowski JM} (\bibinfo{year}{2014}).
\bibinfo{title}{{Equation of state and phase diagram of strongly interacting
  matter}}.
\bibinfo{journal}{{\em Nucl. Phys.}} \bibinfo{volume}{A931}:
  \bibinfo{pages}{113--124}.
  \bibinfo{doi}{\doi{10.1016/j.nuclphysa.2014.09.074}}.

\bibtype{Article}%
\bibitem[Pawlowski and Strodthoff(2015)]{Pawlowski:2015mia}
\bibinfo{author}{Pawlowski JM} and  \bibinfo{author}{Strodthoff N}
  (\bibinfo{year}{2015}).
\bibinfo{title}{{Real time correlation functions and the functional
  renormalization group}}.
\bibinfo{journal}{{\em Phys. Rev.}} \bibinfo{volume}{D92}
  (\bibinfo{number}{9}): \bibinfo{pages}{094009}.
  \bibinfo{doi}{\doi{10.1103/PhysRevD.92.094009}}.
\eprint{1508.01160}.

\bibtype{Article}%
\bibitem[Pawlowski et al.(2023)]{Pawlowski:2022zhh}
\bibinfo{author}{Pawlowski JM}, \bibinfo{author}{Schneider CS},
  \bibinfo{author}{Turnwald J}, \bibinfo{author}{Urban JM} and
  \bibinfo{author}{Wink N} (\bibinfo{year}{2023}).
\bibinfo{title}{{Yang-Mills glueball masses from spectral reconstruction}}.
\bibinfo{journal}{{\em Phys. Rev. D}} \bibinfo{volume}{108}
  (\bibinfo{number}{7}): \bibinfo{pages}{076018}.
  \bibinfo{doi}{\doi{10.1103/PhysRevD.108.076018}}.
\eprint{2212.01113}.

\bibtype{Article}%
\bibitem[Pawlowski et al.(2025)]{Pawlowski:2025jpg}
\bibinfo{author}{Pawlowski JM}, \bibinfo{author}{Rennecke F} and
  \bibinfo{author}{Sattler FR} (\bibinfo{year}{2025}), \bibinfo{month}{12}.
\bibinfo{title}{{Inhomogeneous instabilities in high-density QCD}}
  \eprint{2512.20510}.

\bibtype{Article}%
\bibitem[Pisarski and Rennecke(2021)]{Pisarski:2021qof}
\bibinfo{author}{Pisarski RD} and  \bibinfo{author}{Rennecke F}
  (\bibinfo{year}{2021}).
\bibinfo{title}{{Signatures of Moat Regimes in Heavy-Ion Collisions}}.
\bibinfo{journal}{{\em Phys. Rev. Lett.}} \bibinfo{volume}{127}
  (\bibinfo{number}{15}): \bibinfo{pages}{152302}.
  \bibinfo{doi}{\doi{10.1103/PhysRevLett.127.152302}}.
\eprint{2103.06890}.

\bibtype{Article}%
\bibitem[Pisarski and Rennecke(2024)]{Pisarski:2024esv}
\bibinfo{author}{Pisarski RD} and  \bibinfo{author}{Rennecke F}
  (\bibinfo{year}{2024}).
\bibinfo{title}{{Conjectures about the Chiral Phase Transition in QCD from
  Anomalous Multi-Instanton Interactions}}.
\bibinfo{journal}{{\em Phys. Rev. Lett.}} \bibinfo{volume}{132}
  (\bibinfo{number}{25}): \bibinfo{pages}{251903}.
  \bibinfo{doi}{\doi{10.1103/PhysRevLett.132.251903}}.
\eprint{2401.06130}.

\bibtype{Article}%
\bibitem[Quandt and Reinhardt(2022)]{Quandt:2022lhe}
\bibinfo{author}{Quandt M} and  \bibinfo{author}{Reinhardt H}
  (\bibinfo{year}{2022}).
\bibinfo{title}{{Effective potential of the Polyakov loop in the Hamiltonian
  approach to QCD}}.
\bibinfo{journal}{{\em Phys. Rev. D}} \bibinfo{volume}{106}
  (\bibinfo{number}{11}): \bibinfo{pages}{114001}.
  \bibinfo{doi}{\doi{10.1103/PhysRevD.106.114001}}.
\eprint{2209.04967}.

\bibtype{Article}%
\bibitem[Reinhardt and Heffner(2012)]{Reinhardt:2012qe}
\bibinfo{author}{Reinhardt H} and  \bibinfo{author}{Heffner J}
  (\bibinfo{year}{2012}).
\bibinfo{title}{{The effective potential of the confinement order parameter in
  the Hamilton approach}}.
\bibinfo{journal}{{\em Phys. Lett. B}} \bibinfo{volume}{718}:
  \bibinfo{pages}{672--677}.
  \bibinfo{doi}{\doi{10.1016/j.physletb.2012.10.084}}.
\eprint{1210.1742}.

\bibtype{Article}%
\bibitem[Reinosa(2025)]{Reinosa:2024njc}
\bibinfo{author}{Reinosa U} (\bibinfo{year}{2025}).
\bibinfo{title}{{Aspects of confinement within non-abelian gauge theories}}.
\bibinfo{journal}{{\em Eur. Phys. J. C}} \bibinfo{volume}{85}
  (\bibinfo{number}{2}): \bibinfo{pages}{199}.
  \bibinfo{doi}{\doi{10.1140/epjc/s10052-025-13801-1}}.
\eprint{2404.06118}.

\bibtype{Article}%
\bibitem[Reinosa et al.(2015{\natexlab{a}})]{Reinosa:2014zta}
\bibinfo{author}{Reinosa U}, \bibinfo{author}{Serreau J},
  \bibinfo{author}{Tissier M} and  \bibinfo{author}{Wschebor N}
  (\bibinfo{year}{2015}{\natexlab{a}}).
\bibinfo{title}{{Deconfinement transition in SU(2) Yang-Mills theory: A
  two-loop study}}.
\bibinfo{journal}{{\em Phys. Rev.}} \bibinfo{volume}{D91}:
  \bibinfo{pages}{045035}. \bibinfo{doi}{\doi{10.1103/PhysRevD.91.045035}}.
\eprint{1412.5672}.

\bibtype{Article}%
\bibitem[Reinosa et al.(2015{\natexlab{b}})]{Reinosa:2014ooa}
\bibinfo{author}{Reinosa U}, \bibinfo{author}{Serreau J},
  \bibinfo{author}{Tissier M} and  \bibinfo{author}{Wschebor N}
  (\bibinfo{year}{2015}{\natexlab{b}}).
\bibinfo{title}{{Deconfinement transition in SU($N$) theories from perturbation
  theory}}.
\bibinfo{journal}{{\em Phys. Lett.}} \bibinfo{volume}{B742}:
  \bibinfo{pages}{61--68}. \bibinfo{doi}{\doi{10.1016/j.physletb.2015.01.006}}.
\eprint{1407.6469}.

\bibtype{Article}%
\bibitem[Rennecke(2026)]{Rennecke:2025bcw}
\bibinfo{author}{Rennecke F} (\bibinfo{year}{2026}).
\bibinfo{title}{{QCD phase structure {\&} equation of state: A functional
  perspective}}.
\bibinfo{journal}{{\em EPJ Web Conf.}} \bibinfo{volume}{364}:
  \bibinfo{pages}{01018}. \bibinfo{doi}{\doi{10.1051/epjconf/202636401018}}.
\eprint{2510.11270}.

\bibtype{Article}%
\bibitem[Rennecke and Skokov(2022)]{Rennecke:2022ohx}
\bibinfo{author}{Rennecke F} and  \bibinfo{author}{Skokov VV}
  (\bibinfo{year}{2022}).
\bibinfo{title}{{Universal location of Yang{\textendash}Lee edge singularity
  for a one-component field theory in
  1{\ensuremath{\leq}}d{\ensuremath{\leq}}4}}.
\bibinfo{journal}{{\em Annals Phys.}} \bibinfo{volume}{444}:
  \bibinfo{pages}{169010}. \bibinfo{doi}{\doi{10.1016/j.aop.2022.169010}}.
\eprint{2203.16651}.

\bibtype{Article}%
\bibitem[Rennecke et al.(2023)]{Rennecke:2023xhc}
\bibinfo{author}{Rennecke F}, \bibinfo{author}{Pisarski RD} and
  \bibinfo{author}{Rischke DH} (\bibinfo{year}{2023}).
\bibinfo{title}{{Particle interferometry in a moat regime}}.
\bibinfo{journal}{{\em Phys. Rev. D}} \bibinfo{volume}{107}
  (\bibinfo{number}{11}): \bibinfo{pages}{116011}.
  \bibinfo{doi}{\doi{10.1103/PhysRevD.107.116011}}.
\eprint{2301.11484}.

\bibtype{Article}%
\bibitem[Resch et al.(2019)]{Resch:2017vjs}
\bibinfo{author}{Resch S}, \bibinfo{author}{Rennecke F} and
  \bibinfo{author}{Schaefer BJ} (\bibinfo{year}{2019}).
\bibinfo{title}{{Mass sensitivity of the three-flavor chiral phase
  transition}}.
\bibinfo{journal}{{\em Phys. Rev.}} \bibinfo{volume}{D99}
  (\bibinfo{number}{7}): \bibinfo{pages}{076005}.
  \bibinfo{doi}{\doi{10.1103/PhysRevD.99.076005}}.
\eprint{1712.07961}.

\bibtype{Article}%
\bibitem[Roberge and Weiss(1986)]{Roberge:1986mm}
\bibinfo{author}{Roberge A} and  \bibinfo{author}{Weiss N}
  (\bibinfo{year}{1986}).
\bibinfo{title}{{Gauge Theories With Imaginary Chemical Potential and the
  Phases of {QCD}}}.
\bibinfo{journal}{{\em Nucl. Phys.}} \bibinfo{volume}{B275}:
  \bibinfo{pages}{734--745}. \bibinfo{doi}{\doi{10.1016/0550-3213(86)90582-1}}.

\bibtype{Article}%
\bibitem[Roberts and Schmidt(2000)]{Roberts:2000aa}
\bibinfo{author}{Roberts CD} and  \bibinfo{author}{Schmidt SM}
  (\bibinfo{year}{2000}).
\bibinfo{title}{{Dyson-Schwinger equations: Density, temperature and continuum
  strong QCD}}.
\bibinfo{journal}{{\em Prog. Part. Nucl. Phys.}} \bibinfo{volume}{45}:
  \bibinfo{pages}{S1--S103}.
  \bibinfo{doi}{\doi{10.1016/S0146-6410(00)90011-5}}.
\eprint{nucl-th/0005064}.

\bibtype{Article}%
\bibitem[Roberts and Williams(1994)]{Roberts:1994dr}
\bibinfo{author}{Roberts CD} and  \bibinfo{author}{Williams AG}
  (\bibinfo{year}{1994}).
\bibinfo{title}{{Dyson-Schwinger equations and their application to hadronic
  physics}}.
\bibinfo{journal}{{\em Prog. Part. Nucl. Phys.}} \bibinfo{volume}{33}:
  \bibinfo{pages}{477--575}. \bibinfo{doi}{\doi{10.1016/0146-6410(94)90049-3}}.
\eprint{hep-ph/9403224}.

\bibtype{Article}%
\bibitem[Roth et al.(2022)]{Roth:2021nrd}
\bibinfo{author}{Roth JV}, \bibinfo{author}{Schweitzer D},
  \bibinfo{author}{Sieke LJ} and  \bibinfo{author}{von Smekal L}
  (\bibinfo{year}{2022}).
\bibinfo{title}{{Real-time methods for spectral functions}}.
\bibinfo{journal}{{\em Phys. Rev. D}} \bibinfo{volume}{105}
  (\bibinfo{number}{11}): \bibinfo{pages}{116017}.
  \bibinfo{doi}{\doi{10.1103/PhysRevD.105.116017}}.
\eprint{2112.12568}.

\bibtype{Article}%
\bibitem[Roth et al.(2025)]{Roth:2024hcu}
\bibinfo{author}{Roth JV}, \bibinfo{author}{Ye Y}, \bibinfo{author}{Schlichting
  S} and  \bibinfo{author}{von Smekal L} (\bibinfo{year}{2025}).
\bibinfo{title}{{Universal critical dynamics near the chiral phase transition
  and the QCD critical point}}.
\bibinfo{journal}{{\em Phys. Rev. D}} \bibinfo{volume}{111}
  (\bibinfo{number}{11}): \bibinfo{pages}{L111901}.
  \bibinfo{doi}{\doi{10.1103/PhysRevD.111.L111901}}.
\eprint{2409.14470}.

\bibtype{Article}%
\bibitem[Sagun et al.(2018)]{Sagun:2017eye}
\bibinfo{author}{Sagun VV}, \bibinfo{author}{Bugaev KA},
  \bibinfo{author}{Ivanytskyi AI}, \bibinfo{author}{Yakimenko IP},
  \bibinfo{author}{Nikonov EG}, \bibinfo{author}{Taranenko AV},
  \bibinfo{author}{Greiner C}, \bibinfo{author}{Blaschke DB} and
  \bibinfo{author}{Zinovjev GM} (\bibinfo{year}{2018}).
\bibinfo{title}{{Hadron Resonance Gas Model with Induced Surface Tension}}.
\bibinfo{journal}{{\em Eur. Phys. J.}} \bibinfo{volume}{A54}
  (\bibinfo{number}{6}): \bibinfo{pages}{100}.
  \bibinfo{doi}{\doi{10.1140/epja/i2018-12535-1}}.
\eprint{1703.00049}.

\bibtype{Article}%
\bibitem[Saito et al.(2011)]{Saito:2011fs}
\bibinfo{author}{Saito H}, \bibinfo{author}{Ejiri S}, \bibinfo{author}{Aoki S},
  \bibinfo{author}{Hatsuda T}, \bibinfo{author}{Kanaya K},
  \bibinfo{author}{Maezawa Y}, \bibinfo{author}{Ohno H} and
  \bibinfo{author}{Umeda T} (\bibinfo{collaboration}{WHOT-QCD})
  (\bibinfo{year}{2011}).
\bibinfo{title}{Phase structure of finite temperature qcd in the heavy quark
  region}.
\bibinfo{journal}{{\em Phys. Rev. D}} \bibinfo{volume}{84}:
  \bibinfo{pages}{054502}. \bibinfo{doi}{\doi{10.1103/PhysRevD.85.079902}}.
\bibinfo{note}{[Erratum: Phys.Rev.D 85, 079902 (2012)]}, \eprint{1106.0974}.

\bibtype{Article}%
\bibitem[Santowsky et al.(2020)]{Santowsky:2020pwd}
\bibinfo{author}{Santowsky N}, \bibinfo{author}{Eichmann G},
  \bibinfo{author}{Fischer CS}, \bibinfo{author}{Wallbott PC} and
  \bibinfo{author}{Williams R} (\bibinfo{year}{2020}).
\bibinfo{title}{{$\sigma$-meson: Four-quark versus two-quark components and
  decay width in a Bethe-Salpeter approach}}.
\bibinfo{journal}{{\em Phys. Rev. D}} \bibinfo{volume}{102}
  (\bibinfo{number}{5}): \bibinfo{pages}{056014}.
  \bibinfo{doi}{\doi{10.1103/PhysRevD.102.056014}}.
\eprint{2007.06495}.

\bibtype{Article}%
\bibitem[Schaefer and Wambach(2005)]{Schaefer:2004en}
\bibinfo{author}{Schaefer BJ} and  \bibinfo{author}{Wambach J}
  (\bibinfo{year}{2005}).
\bibinfo{title}{{The Phase diagram of the quark meson model}}.
\bibinfo{journal}{{\em Nucl. Phys.}} \bibinfo{volume}{A757}:
  \bibinfo{pages}{479--492}.
  \bibinfo{doi}{\doi{10.1016/j.nuclphysa.2005.04.012}}.
\eprint{nucl-th/0403039}.

\bibtype{Article}%
\bibitem[Schwarz and Stuke(2009)]{Schwarz:2009ii}
\bibinfo{author}{Schwarz DJ} and  \bibinfo{author}{Stuke M}
  (\bibinfo{year}{2009}).
\bibinfo{title}{{Lepton asymmetry and the cosmic QCD transition}}.
\bibinfo{journal}{{\em JCAP}} \bibinfo{volume}{11}: \bibinfo{pages}{025}.
  \bibinfo{doi}{\doi{10.1088/1475-7516/2009/11/025}}.
\bibinfo{note}{[Erratum: JCAP 10, E01 (2010)]}, \eprint{0906.3434}.

\bibtype{Article}%
\bibitem[Stephanov et al.(1998)]{Stephanov:1998dy}
\bibinfo{author}{Stephanov MA}, \bibinfo{author}{Rajagopal K} and
  \bibinfo{author}{Shuryak EV} (\bibinfo{year}{1998}).
\bibinfo{title}{Signatures of the tricritical point in qcd}.
\bibinfo{journal}{{\em Phys. Rev. Lett.}} \bibinfo{volume}{81}:
  \bibinfo{pages}{4816--4819}.
  \bibinfo{doi}{\doi{10.1103/PhysRevLett.81.4816}}.
\eprint{hep-ph/9806219}.

\bibtype{Article}%
\bibitem[Stephanov et al.(1999)]{Stephanov:1999zu}
\bibinfo{author}{Stephanov MA}, \bibinfo{author}{Rajagopal K} and
  \bibinfo{author}{Shuryak EV} (\bibinfo{year}{1999}).
\bibinfo{title}{{Event-by-event fluctuations in heavy ion collisions and the
  QCD critical point}}.
\bibinfo{journal}{{\em Phys. Rev.}} \bibinfo{volume}{D60}:
  \bibinfo{pages}{114028}. \bibinfo{doi}{\doi{10.1103/PhysRevD.60.114028}}.
\eprint{hep-ph/9903292}.

\bibtype{Article}%
\bibitem[Tan et al.(2022)]{Tan:2021zid}
\bibinfo{author}{Tan Yy}, \bibinfo{author}{Chen Yr} and  \bibinfo{author}{Fu
  Wj} (\bibinfo{year}{2022}).
\bibinfo{title}{{Real-time dynamics of the $O(4)$ scalar theory within the fRG
  approach}}.
\bibinfo{journal}{{\em SciPost Phys.}} \bibinfo{volume}{12}
  (\bibinfo{number}{1}): \bibinfo{pages}{026}.
  \bibinfo{doi}{\doi{10.21468/SciPostPhys.12.1.026}}.
\eprint{2107.06482}.

\bibtype{Article}%
\bibitem[Tan et al.(2025)]{Tan:2025bsv}
\bibinfo{author}{Tan Yy}, \bibinfo{author}{Yin S}, \bibinfo{author}{Chen Yr},
  \bibinfo{author}{Huang C} and  \bibinfo{author}{Fu Wj}
  (\bibinfo{year}{2025}), \bibinfo{month}{12}.
\bibinfo{title}{{Real-time evolution of critical modes in the QCD phase
  diagram}} \eprint{2512.03614}.

\bibtype{Article}%
\bibitem[Vovchenko et al.(2020)]{Vovchenko:2020tsr}
\bibinfo{author}{Vovchenko V}, \bibinfo{author}{Savchuk O},
  \bibinfo{author}{Poberezhnyuk RV}, \bibinfo{author}{Gorenstein MI} and
  \bibinfo{author}{Koch V} (\bibinfo{year}{2020}).
\bibinfo{title}{{Connecting fluctuation measurements in heavy-ion collisions
  with the grand-canonical susceptibilities}}.
\bibinfo{journal}{{\em Phys. Lett. B}} \bibinfo{volume}{811}:
  \bibinfo{pages}{135868}. \bibinfo{doi}{\doi{10.1016/j.physletb.2020.135868}}.
\eprint{2003.13905}.

\bibtype{Article}%
\bibitem[Wan et al.(2025)]{Wan:2025wdg}
\bibinfo{author}{Wan Zy}, \bibinfo{author}{Lu Y}, \bibinfo{author}{Gao F} and
  \bibinfo{author}{Liu Yx} (\bibinfo{year}{2025}).
\bibinfo{title}{{Lee Yang edge singularities of QCD in association with the
  Roberge-Weiss and chiral phase transitions}}.
\bibinfo{journal}{{\em Phys. Rev. D}} \bibinfo{volume}{112}
  (\bibinfo{number}{9}): \bibinfo{pages}{094007}.
  \bibinfo{doi}{\doi{10.1103/p7y6-hq15}}.
\eprint{2504.12964}.

\bibtype{Article}%
\bibitem[Weil et al.(2017)]{Weil:2017knt}
\bibinfo{author}{Weil E}, \bibinfo{author}{Eichmann G},
  \bibinfo{author}{Fischer CS} and  \bibinfo{author}{Williams R}
  (\bibinfo{year}{2017}).
\bibinfo{title}{{Electromagnetic decays of the neutral pion}}.
\bibinfo{journal}{{\em Phys. Rev. D}} \bibinfo{volume}{96}
  (\bibinfo{number}{1}): \bibinfo{pages}{014021}.
  \bibinfo{doi}{\doi{10.1103/PhysRevD.96.014021}}.
\eprint{1704.06046}.

\bibtype{Article}%
\bibitem[Weiss(1981)]{Weiss:1980rj}
\bibinfo{author}{Weiss N} (\bibinfo{year}{1981}).
\bibinfo{title}{{The Effective Potential for the Order Parameter of Gauge
  Theories at Finite Temperature}}.
\bibinfo{journal}{{\em Phys. Rev. D}} \bibinfo{volume}{24}:
  \bibinfo{pages}{475}. \bibinfo{doi}{\doi{10.1103/PhysRevD.24.475}}.

\bibtype{Article}%
\bibitem[Williams(2019)]{Williams:2018adr}
\bibinfo{author}{Williams R} (\bibinfo{year}{2019}).
\bibinfo{title}{{Vector mesons as dynamical resonances in the
  Bethe{\textendash}Salpeter framework}}.
\bibinfo{journal}{{\em Phys. Lett. B}} \bibinfo{volume}{798}:
  \bibinfo{pages}{134943}. \bibinfo{doi}{\doi{10.1016/j.physletb.2019.134943}}.
\eprint{1804.11161}.

\bibtype{Inproceedings}%
\bibitem[Zhang et al.(2026)]{Zhang:2026dny}
\bibinfo{author}{Zhang Y}, \bibinfo{author}{Wang Z}, \bibinfo{author}{Luo X}
  and  \bibinfo{author}{Xu N} (\bibinfo{year}{2026}), \bibinfo{month}{2},
  \bibinfo{title}{{Search for the QCD Critical Point in High Energy Nuclear
  Collisions: A Status Report}}, \eprint{2602.08356}.

\end{thebibliography*}

\end{document}